%Paper: adap-org/9411001
%From: borgs@math.ias.edu (Christian H Borgs)
%Date: Mon, 14 Nov 94 14:13:39 EST
%Date (revised): Mon, 14 Nov 94 14:39:56 EST

%%%%%%%%%%%%%%%%%%%%%%%%%%%%%%%%%%%%%%%%%%
%%                                      %%
%%        C.Borgs, J.T.Chayes           %%
%%   The covariance matrix of the       %%
%%          Potts model:                %%
%%      A random cluster analysis       %%
%%                                      %%
%%%%%%%%%%%%%%%%%%%%%%%%%%%%%%%%%%%%%%%%%%

\input amstex
\documentstyle{amsppt}

\magnification=1200

%\vcorrection{-.4in}
%\vcorrection{.4in}
%\hcorrection{-.37in}
%\addto\tenpoint
{\normalbaselineskip15pt}
%\normalbaselines
\pagewidth{15.25truecm}
\pageheight{23truecm}

\TagsOnRight
\CenteredTagsOnSplits
%\NoBlackBoxes
%\NoRunningHeads
%\nopagenumbers

\font\cal=cmsy10
\font\headerfont=cmcsc8

\def\half{\hbox{$\textstyle{1\over 2}$}}

\def\splus{\text{\fivebf +}}

\def\supp{{\text{supp}}}

\def\diam{{\text{diam}}}
\def\dist{{\text{dist}}}

\def\free{{\text{free}}}
\def\cyl{{\text{cyl}}}

\def\wir{{\text{wir}}}
\def\fin{{\text{fin}}}

\def\picture #1 by #2 (#3)
{\vbox to #2{
    \hrule width #1 height 0pt
    depth 0pt
    \vfill
    \special{picture #3}}}

\leftheadtext\nofrills
{\headerfont C. Borgs, J. T. Chayes}
\rightheadtext\nofrills
{\headerfont Covariance matrix of the Potts model}
\pageno=0
\rightline{UCLA preprint}
\rightline{September 1994}
\bigskip

\topmatter

\title
The covariance matrix of the  Potts model:\\
A random cluster analysis
\vskip -.5cm
\endtitle

\author
 C\. Borgs \footnotemark"${}^{1,\dag}$" \,\,
\, and \,J\. T\. Chayes\footnotemark"${}^{2,\ddag}$"
\vskip -.1
cm
\endauthor

\footnotetext"${}^\dag$"{DFG Heisenberg Fellow}
\footnotetext"${}^\ddag$"{Partialy supported by
NSF grant DMS-9403842}

\affil
%{\sevenit
{\eightit
${}^1$Center for Theoretical Study,
Charles University,
Prague
and Department of Physics, Freie Universit\"at, Berlin\\
${}^2$Department of Mathematics,
University of California at Los Angeles
}
\endaffil

\address
Christian Borgs
\hfill\newline
Institut f\"ur Theoretische Physik
\hfill\newline
Freie Universit\"at Berlin
\hfill\newline
Arnimallee 14, 14195 Berlin
\endaddress
\email
borgs\@dirac.physik.fu-berlin.de
\endemail

\address
Jennifer T. Chayes
\hfill\newline
Department of Mathematics
\hfill\newline
University of
California at Los Angeles
\hfill\newline
Los Angles, CA 90024
\endaddress
\email
jchayes\@math.ucla.edu
\endemail

%\keywords
%\endkeywords

\abstract
We consider the covariance matrix
$$
G^{mn}(x-y)
=
\langle q\delta(\sigma_x,m)\,q\delta(\sigma_y,n)\rangle
-
\langle q\delta(\sigma_x,m)\rangle\,
\langle q\delta(\sigma_y,n)\rangle
$$
of the $d$-dimensional $q$-states Potts model,
rewriting it in terms of the connectivity, the finite-cluster
connectivity and the infinite-cluster covariance in
the random cluster representation
of Fortuin and Kasteleyn.  In any of the $q$ ordered
phases, we show that --
in addition to the trivial eigenvalue $0$ --
the matrix $G^{mn}(x-y)$ has
one simple eigenvalue \hbox{$G_{\wir}^{(1)}(x-y)$} and
one ($q-2$)-fold degenerate eigenvalue
\hbox{$G_{\wir}^{(2)}(x-y)$}.  Furthermore, we identify
the eigenvalues both in terms of representations
of the unbroken
symmetry group of the model, and in terms of connectivities
and cluster covariances, thereby attributing
algebraic significance to these
stochastic geometric quantities.
In addition to establishing
the existence of the correlation
lengths $\xi_{\wir}^{(1)}$ and $\xi_{\wir}^{(2)}$
corresponding to $G_{\wir}^{(1)}(x-y)$ and
$G_{\wir}^{(2)}(x-y)$,
we show that
$\xi_{\wir}^{(1)}(\beta)\geq \xi_{\wir}^{(2)}(\beta)$
for all
inverse temperatures $\beta$. We also prove that
$\xi_{\wir}^{(2)}$ coincides with the decay rate of
the diameter of finite clusters.

For dimension $d=2$ and $q \geq 1$,
we establish a duality relation between
$\xi_{\wir}^{(2)}$ and
$\xi_{\free}$,
the correlation length of the two-point function
with free boundary conditions:
We show
$\xi_{\wir}^{(2)}(\beta) = \frac{1}{2} \xi_{\free}(\beta^\ast)$
for all
$\beta \geq \beta_o$,
where $\beta^\ast$ is the dual inverse temperature and
$\beta_o$ is the self-dual point.
For systems with first-order transitions, this
relation helps to resolve certain inconsistencies
between recent exact and numerical work on correlation lengths
at
$\beta = \beta_o$.  For systems with
second order transitions,
this relation implies Widom scaling.  Namely,
asssuming that $\xi_{\free}(\beta) \sim |\beta - \beta_o|^{-\nu}$
as $\beta \uparrow \beta_o$, the duality relation gives
$\xi_{\wir}^{(2)}(\beta^{\ast}) \sim |\beta^{\ast}
- \beta_o|^{-\tilde\nu}$ as $\beta^{\ast} \downarrow \beta_o$
with $\tilde\nu = \nu$.

In the course of proving the above results, we establish
several properties of independent interest for the random
cluster model, including left continuity of the inverse
correlation length $1/\xi_{\free}(\beta)$
and upper semicontinuity of the
inverse length
$1/\xi_{\wir}^{(2)}(\beta)$ in all dimensions,
and left continuity of the two-dimensional
free boundary condition percolation probability at $\beta_o$.
We also introduce DLR equations for the random cluster model
and use them to establish ergodicity of the free
measure.

In order to prove the above results, we introduce a new class
of events which we call decoupling events and two
inequalities for these events.  The first is similar to
the FKG inequality, but holds
for events which are
neither increasing nor decreasing, and replaces independence in
the standard percolation model; the second replaces
the van den Berg - Kesten inequality.

\endabstract

\endtopmatter

\document

\bigskip
\head{1. Introduction:  Background and Discussion of Results}
\endhead

The $q$-state Potts model has been the subject of increasing interest
in recent years.  On the one hand, it
has been studied by probabilists and statistical mechanicists due to
its relationship to the random cluster model (see \cite{FK} and
\cite{ACCN}), where many of the known results for percolation are open
and interesting problems.  On the other hand, the phase transitions
in the Potts model provide a paradigm for
testing numerical methods developed for more complex transitions,
such as deconfinement in lattice QCD:\quad  The Potts model is
relatively easy to simulate with efficient algorithms (see
e.g\. \cite{SW}), it can be tuned from a second-order through a
weakly first-order to a strong first-order transition by varying
the number of states $q$, and many quantities of interest are
explicitly known for dimension $d=2$, thus allowing for a direct
test of numerical methods.  Finally, many of the exact results on
the Potts model
have recently been shown to have fascinating algebraic interpretations
(see e.g\. section VII.B in the review of \cite{W}).

Motivated by discrepancies between recent exact and numerical
results on the correlation length of the Potts model, we have
undertaken to identify and study the relevant length scales
in the problem.  We relate these scales both
to the algebraic structure
of the unbroken symmetry group and to stochastic geometric
quantities in the random cluster representation of
the Potts model.  In the process, we show that
the some of the natural stochastic geometric quantities
one defines in the random cluster representation --
e.g\. the finite-cluster connectivity -- have
independent algebraic significance.  In two dimensions,
we prove a relation between various scales which is an extension
of known relations for percolation and the Ising magnet, and which
establishes a strong form of Widom scaling for Potts models with
continuous transitions.  We also prove an analogue of this
relation for two-dimensional Potts models with discontinuous
transitions; this analogue helps to explain the apparent discrepancy
between the exact and numerical results.

Adopting a field theoretic perspective, we identify
the relevant lengths in the model by studying the
eigenvalues of the
covariance matrix
$$
G_b^{mn}(x-y)
=
{\langle} q\delta(\sigma_x,m)\,;\,q\delta(\sigma_y,n){\rangle}_b
\,.
\tag 1.1
$$
Here $\sigma_x\in\{0,\cdots,q-1\}$ are the usual spins of
the Potts model, $\delta(\cdot,\cdot)$ is the Kronecker delta,
${\langle} \cdot{\rangle}_b$ is the expectation with
respect to the infinite-volume
state obtained from finite-volume
states with ``$b$'' boundary conditions, and
${\langle} A;B{\rangle}_b \equiv {\langle} AB{\rangle}_b
-{\langle} A{\rangle}_b {\langle}B{\rangle}_b$ is the
truncated expectation of the functions $A$ and $B$.

In the disordered phase, we consider the
covariance matrix with free boundary conditions,
$G_{\free}^{mn}(x-y)$.  We find that this is
proportional
to the standard two-point function, which
in turn is equal to the connectivity
function in the random cluster representation:
$$
G_{\free}^{mn}(x-y)=
(q\delta(m,n)-1)
{\langle}\frac{1}{q-1} (q\delta(\sigma_x,\sigma_y)-1)
{\rangle} _{\free}=
(q\delta(m,n)-1)\tau_{\free}(x-y)
\,.
\tag 1.2
$$
Here the connectivity, $\tau_{\free}(x-y)$, is the probabililty
with respect to the free boundary condition
random cluster measure that $x$
and $y$ lie in the same component.  That $\tau_{\free}(x-y)$ is equal
to the two-point function in finite volume is well known
both to probabilists and to numerical physicists, the latter
of whom
use this equivalence to measure the two-point function
according to the ``improved estimators'' approach.
Our only contribution here is to verify the
equivalence in infinite volume.
We note that, in the disordered phase, the covariance matrix
contains no more information than the standard
two-point function, or equivalently, the connectivity function.

The problem is more subtle in the ordered phase, where
we consider the matrix
$G_c^{mn}(x-y)$
with fixed constant boundary
conditions, $c \in S = \{0, \cdots, q-1\}$.
Defining the finite-cluster connectivity,
$\tau^{\fin}_{\wir}(x-y)$, to be the
probability, in the so-called wired random cluster measure,
that $x$ and $y$
lie in the same finite component, and the infinite-cluster
covariance, $C_{\wir}(x-y)$, to be the covariance, again in the
wired measure, of the events
that $x$ and $y$ lie in the infinite component, we prove
that the matrix elements $G_{c}^{mn}(x-y)$ are linear
combinations of $\tau^{\fin}_{\wir}(x-y)$ and $C_{\wir}(x-y)$, namely
$$
\align
G_{c}^{mn}(x-y)
=
(q\delta(m,n)-1)\,
\tau^{\fin}_{\wir}(x-y)
+
(q\delta(m,c)-1) (q\delta(n,c)-1)
C_{\wir}(x-y)
\,.
\tag 1.3
\endalign
$$
We remark that such a relation was not known previously, either
in finite or infinite volume; furthermore, the proof of the
infinite-volume limit involves some subtleties related to how
the infinite cluster emerges from large finite clusters in
the wired problem (for more details, see the remark following
Proposition 3.4).

Percolation analogues of $\tau^{\fin}_{\wir}(x-y)$ and $C_{\wir}(x-y)$
-- in the absence of boundary conditions --
have arisen previously in \cite{CCGKS}, where they appeared
as a natural decomposition of the
truncated percolation connectivity in the ordered phase.
There, however, they did
not have independent signficance, appearing only as a sum.
The question naturally arises whether they have independent
significance here.  Obviously, this is not the case for
$q=2$, for which (1.3) can be rewritten as
$$
G_{c}^{mn}(x-y)
=
(2\delta(m,n)-1)\,
(\tau^{\fin}_{\wir}(x-y)
+ C_{\wir}(x-y))
\,,
$$
involving again only the sum $\tau^{\fin}_{\wir}(x-y) + C_{\wir}(x-y)$.

For $q \geq 3$, however, the fixed boundary
condition covariance matrix $G_{c}^{mn}(x-y)$
has a richer structure.
We prove that it has a simple eigenvalue zero and
a nontrivial simple eigenvalue
$$
G_{\wir}^{(1)}(x-y)=q\,\tau^{\fin}_{\wir}(x-y)+q(q-1)C_{\wir}(x-y)
\,,
\tag 1.4
$$
both corresponding to the trival
representation of the unbroken subgroup $S_{q-1}$ of permutations
of $S \setminus \{c\}$, as well as one ($q-2$)-fold degenerate
eigenvalue
$$
G_{\wir}^{(2)}(x-y)=q\,\tau^{\fin}_{\wir}(x-y)
\,,
\tag 1.5
$$
corresponding to the remaining orthogonal
subspace.\footnote{For the Ising model ($q=2$),
$G_{c}^{mn}(x-y)$ has
only the trivial eigenvalue
zero and the eigenvalue $G_{\wir}^{(1)}(x-y)$.}
Thus we see
that for $q \geq 3$,
the finite-cluster cluster connectivity, $\tau^{\fin}_{\wir}(x-y)$,
has independent algebraic significance as an eigenvalue
of the covariance matrix, and hence also physical significance
in terms of the associated one-particle spectrum.
As for the infinite cluster covariance $C_{\wir}(x-y)$, we will
show in Theorem 4.3 that its {\it decay rate} is equal to the
{\it decay rate} of the eigenvalue $G_{\wir}^{(1)}$ whenever
the magnetization is positive.  Thus although $C_{\wir}(x-y)$
does not have independent algebraic significance, its decay
rate does.

For completeness, we note that the free boundary
condition matrix, $G_{\free}^{mn}(x-y)$, can be diagonalized
as well, yielding a simple eigenvalue zero and
a $(q-1)$-fold degenerate eigenvalue
$$
G_{\free}(x-y) = q\,\tau_{\free}(x-y)\,.
\tag 1.6
$$

Given the eigenvalues (1.4)--(1.6), one naturally defines the
 inverse correlation lengths:
$$
\frac{1}{\xi_{\free}(\beta)}
=-\lim_{|x|\to\infty}
\frac{1}{|x|} \log G_{\free}(x)
\,,
\tag 1.7
$$
$$
\frac{1}{\xi_{\wir}^{(1)}(\beta)}
=-\lim_{|x|\to\infty}
\frac{1}{|x|} \log G_{\wir}^{(1)}(x)
\tag 1.8
$$
and
$$
\frac{1}{\xi_{\wir}^{(2)}(\beta)}
=-\lim_{|x|\to\infty}
\frac{1}{|x|} \log G_{\wir}^{(2)}(x)
\,.
\tag 1.9
$$
Here, as usual, $\beta$ is the inverse temperature
of the model.  In all cases, the limits are taken
so that $x$ lies along a coordinate axis.
In order
to establish the existence of the limits,
we return
to the spin representation and use reflection
positivity.  We also give alternative subadditive proofs of
the existence of the limits (1.7) and (1.9),
which though more complicated
than the reflection positivity arguments, have the advantage that
they hold for non-integer
$q \geq 1$
\comment
\footnote{Although the
covariance matrices and hence the eigenvalues are defined
only for integer $q$,
 the random cluster expressions (1.4)--(1.6)
and the limits (1.7)  and (1.9)
make sense for non-integer $q \geq 1$ as well.
It therefore turns out that most of our results
hold also in this case.}
\endcomment
and can be used
to establish additional properties.
In particular, we use subadditivity to show
left continuity of the inverse
correlation length $1/\xi_{\free}(\beta)$
and upper semicontinuity of the inverse correlation
length $1/\xi_{\wir}^{(2)}(\beta)$.
We also use subadditivity arguments to prove
that $\xi_{\wir}^{(2)}(\beta)$ is equal
to several other geometrical correlation
lengths in the problem, one of which is the decay rate
of the diameter of finite clusters in the wired measure
--- a
quantity which
should be easiy accessible to numerical measurement.

All three correlation lengths coincide in the
high-temper\-a\-ture regime,
where their common value is often denoted by
$\xi_{\text{dis}}(\beta)$.
In the low-temper\-a\-ture regime, we expect
$\xi_{\free}(\beta) \equiv \infty$.
The non-trivial correlation length in this regime is often
denoted by $\xi_{\text{ord}}(\beta)$.  Here, however, we see that
for $q \geq 3$, there are two {\it a priori} different
non-trivial lengths, $\xi_{\wir}^{(1)}(\beta)$ and
$\xi_{\wir}^{(2)}(\beta)$.
Equations (1.4) and (1.5) immediately imply that
$$
\xi_{\wir}^{(1)}(\beta) \,\geq \, \xi_{\wir}^{(2)}(\beta) \,,
\tag 1.10
$$
so that the correlation length $\xi_{\wir}^{(1)}$ of the
symmetric state (i.e\. symmetric with respect to
$S_{q-1}$) is not smaller than those of the unsymmetric
states.
An interesting open question is whether or not the
inequality is strict.  It is worth noting that in
percolation, analogues of
$C_{\wir}(x-y)$ and $\tau^{\fin}_{\wir}(x-y)$
in the absence of boundary conditions
have equal exponential decay rates
\cite{CCGKS}, which here would imply equality of
$\xi_{\wir}^{(1)}(\beta)$ and $\xi_{\wir}^{(2)}(\beta)$.
However, it is not at all clear whether
the Potts models for $q \geq 3$ should have
analogous behavior.

We return finally to our original question,
namely the discrepancy between the exact and
numerical correlation lengths of two-dimensional Potts models
with discontinuous transitions.
Explicit calculations based on a mapping to the
six-vertex model
yielded a correlation
length $\xi_{\text{dis}}(\beta_o)$ of the disordered phase
at the self-dual point $\beta_o$
(\cite{BW},
\cite{KSZ}, \cite{Kl})
which disagreed with previous numerical measurements (\cite{PL},
\cite{GI})
of the ordered correlation length at the transition point
$\xi_{\text{ord}}(\beta_o)$ by roughly a factor of 2,
suggesting the possible relation
$\xi_{\text{ord}}(\beta_o) = \frac12\xi_{\text{dis}}(\beta_o)$
\cite{BJ}.
A continuous transition analogue of this
relation
is already known for both two-dimensional bond
percolation, where $\xi(p) = \frac12\xi(1-p)$ has been
rig\-orously established for all $p>p_c$ \cite{CCGKS}, and
the two-dimensional
Ising magnet, where $\xi(\beta) = \frac12\xi(\beta^{\ast})$
has been been established via exact solution for all
$\beta > \beta_o$ \cite{MW}.  Here, as usual,
$\beta^{\ast}$ is the dual inverse temperature.
However, from our
results discussed above, we now know
that the situation is more complicated in the
$q$-state Potts model, $q \geq 3$, than it is in either
percolation or the Ising magnet, since in the ordered phase
the Potts model has two {\it a priori} different
correlation lengths.  One of our principal results
is a relation of the conjectured form in terms of the
smaller ordered correlation length, $\xi_{\wir}^{(2)}$.

Our result follows from a dichotomy which we prove for all
two-dimensional random cluster models with $q \geq 1$. In addition to
the conjectured relation, the dichotomy
implies Widom scaling for Potts models
with continuous transitions.
Let $P_{\infty}^{\free}(\beta)$ be the percolation
probabilty in the free boundary condition random cluster measure.
Our dichotomy is:\quad  If $P_{\infty}^{\free}(\beta^\ast) = 0$, then
$$
\xi_{\wir}^{(2)}(\beta) = \frac12\xi_{\free}(\beta^\ast)\,,
\tag 1.11
$$
whereas if $P_{\infty}^{\free}(\beta^\ast) > 0$, then
$$
\xi_{\free}(\beta) \, =
\xi_{\wir}^{(1)}(\beta) =
\, \xi_{\wir}^{(2)}(\beta) \,.
\tag 1.12
$$
In order to interpret the dichotomy, we supplement it with the
two-dimensional relation
$$
P_{\infty}^{\wir}(\beta)\,P_{\infty}^{\free}(\beta^\ast) = 0 \,,
\tag 1.13
$$
where $P_{\infty}^{\wir}(\beta)$ is the percolation probability
in the wired measure, which is of course equal to the spontaneous
magnetization $M(\beta)$.  Note that (1.13) shows that
$P_{\infty}^{\free}(\beta^\ast) > 0$ implies $M(\beta) = 0$,
so that (1.12) is simply the equality of the three correlation
lengths in the high-temperature regime, as mentioned earlier.

Our more interesting corollaries follow from the first branch of
the dichotomy, i.e\. the duality relation (1.11).
In order to see this,
we combine (1.13) with the obvious bound
$P_{\infty}^{\wir}(\beta) \geq P_{\infty}^{\free}(\beta)$ to obtain
$
P_{\infty}^{\free}(\beta)\,P_{\infty}^{\free}(\beta^\ast) = 0,
$
so that $P_{\infty}^{\free}(\beta_o) = 0$.
Since $P_{\infty}^{\free}(\beta^\ast)$ is an increasing function
of $\beta^\ast$,
this in turn implies
$$
P_{\infty}^{\free}(\beta^\ast) = 0
\quad\text{for all}\,\,\,\, \beta \geq \beta_o\,.
\tag 1.14
$$
Equation (1.14) implies in particular that
$P_{\infty}^{\free}(\beta)$
is left continuous at
the self-dual point $\beta_o$. Moreover, it means that
that the first branch of the dichotomy (i\.e\. equation (1.11))
holds throughout the low-temperature phase
$\beta \geq \beta_o$.  For systems with
first-order transitions, this implies the conjectured relation
at $\beta_o$:
$$
\xi_{\wir}^{(2)}(\beta_o) = \frac12\xi_{\free}(\beta_o)
\,.
\tag 1.15
$$
For systems with second-order transitions, (1.11)
is a generalization of the aforementioined results
on two-dimensional
percolation \cite{CCGKS} and the Ising magnet \cite{MW}.
In particular, it gives
a strong form of Widom scaling as $\beta \rightarrow \beta_o$:\,\,\,
If
$\xi_{\free}(\beta^{\ast})$ diverges
 with critical exponent
$\nu$, $\xi_{\free}(\beta^{\ast}) \sim
|\beta^{\ast} - \beta_o|^{-\nu}$
as $\beta^{\ast} \uparrow \beta_o$, (1.11) implies that
$\xi_{\wir}^{(2)}(\beta)$ diverges with the same exponent:
$\xi_{\wir}^{(2)}(\beta) \sim |\beta
- \beta_t|^{-\tilde\nu}$ as $\beta \downarrow \beta_o$
with $\tilde\nu = \nu$.

As noted above, the interpretation (and in fact, the
proof) of the dichotomy (1.11) and (1.12) requires
the relation (1.13), which we obtain as a special
case of a general two-dimensional result of
Gandolfi, Keane and Russo \cite{GKR}.  However,
in order to apply the \cite{GKR} theorem, we
need to know that the free random
measure is ergodic, a result which we establish in
all dimensions.  We prove ergodicity
by introducing suitable DLR equations
(\cite{D}, \cite{LR}) for the random cluster problem.
Here the justification of the DLR equations is  much more
delicate than in standard spin systems due to the
nonlocal nature of the random cluster weights:\,\,\,
because of this nonlocality, the specification
used to construct the DLR states is not quasilocal,
and thus standard theorems do not apply.

Before reviewing the organization of the paper,
let us briefly discuss our methods.
These methods are necessarily quite different
from those used in the analysis of the
Bernoulli percolation model, since
the random cluster model lacks several
properties which are used extensively in
percolation -- namely, independence of events
occurring on fixed disjoint sets and the
van den Berg - Kesten \cite{BK} inequality
for events occurring on random disjoint sets.
Moreover, the random cluster model has an
additional feature -- boundary conditions --
which significantly complicates its analysis
relative to the independent model. However,
it is by actually focusing on the boundary
conditions that we are
able to circumvent the other difficulties
and in fact derive two correlation inequalities
which we expect will be useful in many other
contexts.  We do this by noting that in many
cases, the events of interest carry with them
boundary conditions which {\it decouple} them
from other events and thus effectively overcome
the coupling of the random cluster weights.
This idea is formalized by introducing the
notion of {\it decoupling events}.  We use
our decoupling events in formulating and proving
two sets of inequalities which effectively
replace independence and the BK inequality.
The independence is replaced by a relation
which resembles the FKG inequality, but
contains two decoupling events and holds
for a much larger class of events than
the original FKG inequality -- in particular,
for events which are neither increasing
nor decreasing.   The BK inequality is
replaced by a relation which resembles
the independent BK inequality but contains
a decoupling event.  Both inequalities
hold for {\it any} FKG measure,
and thus in particular for
the free and wired random cluster measures with
$q \geq 1$.

\bigskip

The organization of this paper is as follows. In first two
parts of Section 2,
we review the necessary properties of the standard spin
and random cluster representations of the Potts model.
The third part of Section 2 contains our
inequalities for decoupling events.
In the last part of the section, we derive the DLR equation
and establish ergodicity of the free measure.
Section 3 is concerned with the covariance matrix.
In the first part of the section, we derive the
finite- and infinite-volume
representations of the matrix with free and
constant boundary conditions, in particular establishing
the infinite-volume limits of $\tau_{\free}(x-y)$,
$\tau^{\fin}_{\wir}(x-y)$
and $C_{\wir}(x-y)$ from their finite-volume analogues.
The matrix is diagonalized in the second part of
Section 3.  Section 4 concerns the correlation lengths
 $\xi_{\free}$, $\xi_{\wir}^{(1)}$
and $\xi_{\wir}^{(2)}$.  In the first part of the section,
we establish existence of the lengths using
reflection positivity, as reviewed in the Appendix.
The second part of the section concerns alternative
characterizations of $\xi_{\free}$, $\xi_{\wir}^{(1)}$
and $\xi_{\wir}^{(2)}$, proved via subadditivity arguments
and our
inequalities for decoupling events.
In particular, we show that
$1/\xi_{\free}$ is left continuous
and $1/\xi_{\wir}^{(2)}$ is upper semicontinuous; we prove that
$\xi_{\wir}^{(1)}$ is the decay rate of
$C_{\wir}$ whenever the magnetization is positive; and
we establish that
$\xi_{\wir}^{(2)}$ coincides with the decay rate of
the diameter of finite clusters, as well as with
the limiting decay rate of connectivity functions for clusters
in boxes.
Section 5 contains our proof of the two-dimensional
dichotomy (1.11) and (1.12), as well as derivations of a few
results on two-dimensional
percolation probabilities.  The first part of this section
contains a discussion of the heuristics of the duality relation
(1.11) in terms of the behavior of interfaces in the system.
In the second and third parts of the section, we
prove upper and lower bounds
of the form needed for the duality relation (1.11).  Finally, in the
fourth part of the section, we
combine these upper and lower bounds with several
results from Section
4 and the relation (1.13) to obtain our dichotomy.

\bigskip
\head{2. Preliminaries}
\endhead

\medskip
\subhead{2.1. Definition of the Spin Model}
\endsubhead
\smallskip

We consider the $q$-states Potts ferromagnet,
a model with spins
$\sigma_x$ in the set $S=\{0,1,\cdots,q-1\}$,
$q \geq 1$. In
a finite volume
$
\Lambda\subset \Bbb Z^d
$,
the Hamiltonian with free boundary conditions
is
$$
H_\free (\sigma_{\Lambda})=-
\sum_{{\langle}x,y{\rangle} \in B(\Lambda)}
(\delta(\sigma_x, \sigma_y) - 1)
\tag 2.1
$$
where the sum goes over
the set $ B(\Lambda)$
of all
nearest neighbor pairs
${\langle}x,y{\rangle}$ for which
both $x$ and $y$ lie in $\Lambda$. The
Hamiltonian with $c$-boundary conditions,
$c\in S=\{0,1,\cdots,q-1\}$, is
$$
H_c(\sigma_\Lambda)=
H_\free (\sigma_{\Lambda})
-
\sum_{x\in\Lambda,\atop {y\in\partial\Lambda}}
(\delta(\sigma_x, \sigma_y) - 1) \,,
\tag2.2
$$
where $\partial \Lambda=
\{x\notin\Lambda|\dist(x,\Lambda)=1\}$
is the (outer) boundary of $\Lambda$.
Using the label $b$ for ``free'' or $c\in S$,
one introduces the partition function
with boundary condition $b$ as
$$
Z_b(\Lambda) =
\sum_{\sigma_{\Lambda}}
e^{-\beta H_b(\sigma_{\Lambda})}
\tag2.3
$$
where the sum runs over all configurations
$\sigma_\Lambda:\Lambda\to S,x\mapsto \sigma_x$
and $\beta$ is the inverse temperature
$\beta=1/k_BT$.

As usual, an observable $A$ with support $\supp A$
is a function
$A:\sigma_\Lambda\to\Bbb C$ which does not
depend on the spin variables
$\sigma_x$ for $x\notin \supp A$.
A local observable is an observable with a support
$\supp A$ not depending on $\Lambda$. Expectation values
of a local observable $A$ are defined as
$$
{\langle}A{\rangle}_{b,\Lambda} =
{1\over Z_b(\Lambda)}
\sum_{\sigma_{\Lambda}}
A(\sigma_\Lambda)e^{-\beta H_b(\sigma_{\Lambda})}.
\tag2.4
$$
If $A$ and $\tilde A$ are two local
observables, one also
considers the truncated expectation value
defined by
$$
{\langle}A;\tilde A{\rangle}_{b,\Lambda}=
{\langle}A\tilde A{\rangle}_{b,\Lambda}
-{\langle}A{\rangle}_{b,\Lambda}{\langle}\tilde A{\rangle}_{b,\Lambda} .
\tag 2.5
$$
For observables of the form
$$
A_p=\exp({i(\sigma,p)})=\exp(i\sum_{x\in\supp A_p}\sigma_x p_x)
\,,
$$
where $p$ is a function of finite support from $\Bbb Z^d$ into
$\hat S=\{0,2\pi /q,\cdots,2\pi (q-1)/q\}$,
the expectation values
$
{\langle}A_p{\rangle}_{\free, \Lambda}
$
are monotone increasing (i.e. non-decreasing)
in $\Lambda$, while the expectation values
$
{\langle}A_p{\rangle}_{0, \Lambda}
$
are monotone decreasing in $\Lambda$ by Griffiths'
second inequality \cite{Grf}, as generalized by
Ginibre
%(\cite{Grf1}, \cite{Grf2},\cite{Grf3}, \cite{KeS},
\cite{Gi}.
As a consequence, for $b=\text{``free''}$
or $b=0$,
the thermodynamic limit
$$
{\langle}A_p{\rangle}_b =\lim_{\Lambda\to\Bbb Z^d}
{\langle}A_p{\rangle}_{b,\Lambda}
\tag 2.6
$$
exists for all local observables of the form
$
A_p=e^{i(\sigma,p)}
$
and hence for all local observables $A$.  In
(2.6), the limit may be taken through any increasing
sequence of sets.
Using the permutation symmetry of the Hamitonian
(2.2), one concludes that the limit (2.6)
exists for all boundary conditons $b$ considered here
(i.e. free or any constant boundary conditions).
Also by Griffiths' second inequality
(\cite {Grf}, \cite{Gi}), the limit (2.6) is translation invariant.

The order parameter of the Potts model
is the magnetization
$$
M(\beta)=
{1\over q-1}{\langle}q \delta(\sigma_{x_0},0)-1{\rangle}_0
\,,
\tag 2.7
$$
where $x_0$ is an arbitrary point in $\Bbb Z^d$
(recall that the infinite-volume states ${\langle}\cdot{\rangle}_b$
are translation invariant). It is known that
$M(\beta)$ is increasing in $\beta$
(\cite{Gi}, \cite{ACCN}),
decreasing in $q$ \cite{ACCN}, and that the infinite-volume
states
${\langle}\cdot{\rangle}_c$, $c\in S$ are
equal to ${\langle}\cdot{\rangle}_\free$ if and only if $M(\beta)=0$
\cite{ACCN}\footnote{Actually,
\cite{ACCN} proved that {\it all} infinite-volume Gibbs states
are equal to ${\langle}\cdot{\rangle}_\free$ if and only if $M(\beta)=0$.}.
Defining the transition point
$$
\beta_t=\inf\{\beta | M(\beta) > 0\},
\tag 2.8
$$
we  remark that it is believed
that $M(\beta_t)$ is increasing in $q$,
and that
$$
q_c=\max\{q\in\Bbb N | M(\beta_t)=0\}
\tag 2.9
$$
is $4$ for $d=2$ and $2$ for $d > 2$. The fact that
$M(\beta_t) > 0$, i.e\. the existence of a
first-order phase transition, has been
rigorously established for all $d\geq 2$
provided $q$ is sufficiently large
(\cite{KoS}, see also \cite{LMR} and \cite{LMMRS}).

\bigskip
\subhead{2.2. The Random Cluster Representation:
Review of Basic Properties}
\endsubhead
\smallskip

It is often useful to reexpress the $q$-state Potts model
as an integer value of a two-parameter interacting
percolation model, the so-called random cluster model
of Fortuin and Kasteleyn \cite{FK}.
In order to set our notation and state
the results we will use in the rest of
this paper, we briefly review
the derivation and some basic properties
of the FK representation.  The representation
is defined in terms of configurations
$\omega\in\Omega\equiv\{0,1\}^{\Bbb B_d}$,
where
$
\Bbb B_d=
\{{\langle}x,y{\rangle}|x,y\in\Bbb Z^d\}
$
is the nearest-neighbor bond lattice.  For subsets
$B\subset\Bbb B_d$, the configuration
space is denoted by
$\Omega_B\equiv\{0,1\}^B$.

Let us
start with the finite-volume partition funtion with free
boundary conditions.
We write the Gibbs factor
$e^{-\beta H_\free(\sigma_\Lambda)}$
as
$$
\prod_{{\langle}x,y{\rangle}\in B(\Lambda)}
e^{\beta(\delta(\sigma_x,\sigma_y)-1)}
\,
$$
and expand the product with the help of the
identity
$$
e^{\beta(\delta(\sigma_x,\sigma_y)-1)}
=(1-p)+p\delta(\sigma_x,\sigma_y)
\quad\text{where}\quad p=1-e^{-\beta}
\,.
\tag 2.10
$$
We identify each term of this expansion
with a configuration $\omega\in\Omega_{ B(\Lambda)}$; $\omega$
is chosen so that it is zero
on those bonds for which the
factor in the product is $p$,
and one on those bonds for which the
factor is $p\delta(\sigma_x,\sigma_y)$. Geometrically,
we think of the bonds $b={\langle}x,y{\rangle}$ for which
$\omega(b)=1$ as occupied or ordered,
and
those for which
$\omega(b)=0$ as vacant or disordered.
With a slight abuse of notation,
we sometimes use the symbol
$\omega$ to denote the set of occupied
bonds in $ B(\Lambda)$,
and
$\omega^c$ to denote the set of empty
bonds in $ B(\Lambda)$, see e.g\. (2.11) below.

Rewriting the Gibbs factor in expanded form,
we obtain
$$
Z_\free(\Lambda)=
\sum_{\omega\in\Omega_{ B(\Lambda)}}
\sum_{\sigma_\Lambda}
(1-p)^{|\omega^c|}
p^{|\omega|}
\prod_{{\langle}x,y{\rangle}\in\omega}
\delta(\sigma_x,\sigma_y)
\,.
\tag 2.11
$$
Evaluating  the sum over
$\sigma_\Lambda$,
we pick up a factor $q$ for each connected
component of the graph $(\Lambda,\omega)$
(regarding isolated points as separate clusters).
Denoting the number of clusters in this graph
by $\#(\omega)$,
we find
$$
Z_\free(\Lambda)=
\sum_{\omega\in\Omega_{ B(\Lambda)}}
(1-p)^{|\omega^c|}
p^{|\omega|}
q^{\#(\omega)}
\,.
\tag 2.12
$$
It is an easy exercise to generalize (2.12)
to the expectations of local
observables $A=A(\sigma)$. One obtains
$$
{\langle}A{\rangle}_{\free,\Lambda}=
\sum_{\omega\in\Omega_{ B(\Lambda)}}
G_{\free,\Lambda}(\omega)
E_\free(A|\omega)
\tag 2.13
$$
where
$$
G_{\free,\Lambda}(\omega)=
\frac 1{Z_\free(\Lambda)}
(1-p)^{|\omega^c|}
p^{|\omega|}
q^{\#(\omega)}
\,
\tag 2.14
$$
is the weight of the configuration
$\omega$,
while
$E_\free(\cdot|\omega)$ is an average over
spins with the spins constrained to be
constant on each connected cluster of $\omega$ and
with values for different
clusters being chosen uniformly from
$\{0,1,\cdots,q-1\}$.  We remark that for
the purposes of interpreting expectations
of this sort, it is often convenient to consider
the joint distribution on the spin and bond
variables with weights given by the terms in
(2.11), as introduced implicitly in \cite{SW}
and explicitly in \cite{ES}.  In terms of this
distribution, the expectation $E_\free(\cdot|\omega)$ is an
average over the conditional distribution
of spins, given the bond variables.

For constant boundary conditions, one obtains a
similar representation, with the following
differences (as noted in
\cite{ACCN}):

\item{i)} The set $ B(\Lambda)$ is
replaced by the set
$B^{\splus}(\Lambda)$
of all nearest neighbor
pairs ${\langle}x,y{\rangle}$ for which at least one of the
two points $x$ and $y$ lies in $\Lambda$.

\item{ii)}  The points of the boundary
$\partial \Lambda$
are regarded as {\it preconnected} or {\it wired},
in the sense that these points
are taken to be lying in one cluster.
This of course modifies the value
of $\#(\omega)$.

\item{iii)} The expectation $E_\free(A|\omega)$
in (2.14) is replaced by
$E_c(A|\omega)$, where the average is
computed with the additional constraint
that spins in clusters connected to the
boundary now only assume the value $\sigma_x=c$.

\noindent
We have:
$$
{\langle}A{\rangle}_{c,\Lambda}=
\sum_{\omega\in\Omega_{ B^{+}(\Lambda)}}
G_{\wir,\Lambda}(\omega)
E_c(A|\omega)\,,
\tag 2.15
$$
where
$$
G_{\wir,\Lambda}(\omega)=
\frac 1{ Z_{\wir}(\Lambda)}
(1-p)^{|\omega^c|}
p^{|\omega|}
q^{\#(\omega)}
\tag 2.16
$$
and $Z_{\wir} = \sum_{c\in S}{Z_c} = qZ_0$.

We denote by
$\mu_{\free,\Lambda}(\cdot)$
and
$\mu_{\wir,\Lambda}(\cdot)$
the finite-volume measures
defined by the weights (2.14)
and (2.16), respectively.

\remark{Remark} The measures
$\mu_{\free,\Lambda}(\cdot)$
and
$\mu_{\wir,\Lambda}(\cdot)$
are defined on the probability
spaces
$(\Omega_{ B(\Lambda)},\Cal F_{ B(\Lambda)})$
and
$(\Omega_{ B^{+}(\Lambda)},\Cal F_{ B^{+}(\Lambda)})$,
respectively. (In general, we use $\Cal F_B$
to denote the $\sigma$-algebra
generated by cylinder events $A\subset\Omega_B$.)
It is sometimes convenient
to extend these to measures on the
full space $(\Omega,\Cal F)$
by declaring all bonds in
$\Bbb B_d\setminus B(\Lambda)$
to be vacant for
$\mu_{\free,\Lambda}(\cdot)$,
and all bonds in
$\Bbb B_d\setminus B^{\splus}(\Lambda)$
to be occupied for
$\mu_{\wir,\Lambda}(\cdot)$.
\endremark

\smallskip

An important property of the
FK representation is that it
obeys the Harris-FKG inequality.
This inequality, first proved for
percolation in \cite{H} and proved
for a large class of models in
\cite{FKG}, was established for
the $q\geq 1$ random cluster
representation in \cite{F}
(see also \cite{ACCN}).  We begin
with the standard:

\definition{Definition 2.1}  Consider the
natural partial order on bond configurations
$\omega\in\Omega_{ B}$,
$ B\subset \Bbb B_d$, namely
$\omega\prec\omega^{\prime}$ if
$\omega(b) = 1 \Rightarrow \omega^{\prime}(b) = 1$.
A function $f:\Omega_{ B}\rightarrow\Bbb R$ is said to
be {\it increasing} if it is nondecreasing with respect to
this partial order, i.e. $f(\omega)\leq f(\omega^\prime)$
for all $\omega\prec\omega^{\prime}$. An event is
said to be
increasing if its indicator is an
increasing function. Similarly, a function $f$ is
{\it decreasing} if the function $-f$ is increasing,
and an event is decreasing if its complement
is increasing.

A measure $\mu$ on  $(\Omega_{ B},\Cal F_{ B})$
is said to be an
{\it FKG measure} if it obeys the so-called Harris-FKG
inequality
$$
\mu(A_1\cap A_2)\geq
\mu(A_1)\mu(A_2)
\tag 2.17
$$
for all pairs of increasing events
$A_1,A_2\in\Cal F_{ B}$.
It is said to be a {\it strong FKG measure}
if for each cylinder event $C \in \Cal F_{ B}$,
the conditional measure $\mu(\cdot \mid C)$
is an FKG measure.
Finally, a measure
$\mu$ on $(\Omega_{ B},\Cal F_{ B})$ is said
to {\it FKG dominate} a measure
$\nu$ on $(\Omega_{ B},\Cal F_{ B})$,
denoted by $\nu
\underset\text{FKG} \to\leq
\mu$, if $\nu (A) \leq \mu (A)$ for all increasing
events $A\in\Cal F_{ B}$.
\enddefinition

\proclaim{Proposition 2.2} {\rm (\cite{F}, \cite{ACCN})}
Let $q\geq 1$.  Then the finite-volume free and wired FK measures,
$\mu_{\free,\Lambda}$ and $\mu_{\wir,\Lambda}$,
are strong FKG measures.
\endproclaim

\remark{Consequences {\rm (See e.g. \cite{ACCN})}}
\item{1)} The finite-volume measures are monotonic in the volume:
$$
\mu_{\free,\Lambda}
\underset\text{FKG} \to\leq \mu_{\free,\Lambda^\prime}
\quad\text{if}\quad
\Lambda\subset\Lambda^\prime
\tag 2.18
$$
and
$$
\mu_{\wir,\Lambda}
\underset\text{FKG} \to\geq \mu_{\wir,\Lambda^\prime}
\quad\text{if}\quad
\Lambda\subset\Lambda^\prime
\,,
\tag 2.19
$$
from which it follows that the
infinite-volume measures
$$
\mu_\free(\cdot)=
\lim_{\Lambda\to\Bbb Z^d}
\mu_{\free,\Lambda}(\cdot)
\tag 2.20
$$
and
$$
\mu_\wir(\cdot)=
\lim_{\Lambda\to\Bbb Z^d}
\mu_{\wir,\Lambda}(\cdot)
\tag 2.21
$$
exist for all monotone local functions,
and hence for all local functions.
Furthermore, these infinite-volume
measures are translation invariant and
inherit the strong FKG property.

\item{2)} The wired measures FKG dominate the
free measures, i\.e\.
$$
\mu_{\free,\Lambda}
\underset\text{FKG} \to\leq \mu_{\wir,\Lambda}
\tag 2.22
$$
and
$$
\mu_\free
\underset\text{FKG} \to\leq \mu_\wir\,.
\tag 2.23
$$

\endremark

\smallskip
Another useful property of these measures is that they have
{\it finite energy}, a notion introduced by Newman and Schulman
\cite{NS}.

\definition{Definition 2.3}  Let $ B\subset \Bbb B_d$,
$| B| < \infty$, and $\phi\in\Omega_{ B}$
a configuration on $ B$.  If $\omega\in\Omega$ is a
configuration on the full space, let $\phi(\omega)$ be the
configuration which agrees with $\phi$ on $ B$ and
with $\omega$ on $ B^c$:
$$\phi(\omega)(b) =
\cases
\phi(b) & b\in B \\
\omega(b) &  b\in B^c \\
\endcases
{}.
$$
Finally, if $A\subset\Omega$ is an event, let $\phi(A) =
\{\phi(\omega)\, |\, \omega\in A \}$.  The measure $\mu$ on
$\Omega$ is said to have {\it finite energy} if for every
finite $B\subset \Bbb B_d$ and for every $\phi\in\Omega_{ B}$,
$$
\mu (A) > 0 \,\implies\, \mu (\phi (A)) > 0 \,.
$$
\enddefinition
It is easy to see that finite energy is equivalent to the
statement:  \quad For each bond $b$, the conditional probability
of the event that $b$ is occupied, given the configuration on
all the other bonds, is non-trivial:
$$
0 \,<\, \mu(\omega(b) = 1\, |\,
\omega(\tilde b) , \tilde b \neq b) \,<\, 1 \,.
$$
For the free and wired measures, it was observed in \cite{ACCN}
that this probability can be explicitly calculated:
$$
\mu(\omega(b) = 1\, |\,
\omega(\tilde b) , \tilde b \neq b)
=
\cases
p &   {\text{if the endpoints of $b$ are connected,}
  \atop\text{regardless of $\omega(b)$
\phantom{ssssssssssssssssssi}}}\\
\frac{p}{p+q(1-p)} & \text{otherwise} \\
\endcases
\,,
\tag 2.24
$$
where $\mu= \mu_\free$ or $\mu_\wir$.  Thus
for all $q \geq 1$ and all $p \neq 0,1$, the random cluster
measures $\mu_\free$ and $\mu_\wir$ have finite energy.
Note that this is not true in all random cluster measures:
\, Boundary conditions can impose constraints which exclude
certain con\-figurations.

Given stationarity and finite energy, it follows immediately
from a general result of Burton and Keane \cite{BuK}
that the infinite cluster is unique:

\proclaim{Proposition 2.4}  For any $q \geq 1$ and any
$p\in (0,1)$, the free and wired random cluster
states have at most one infinite cluster with probability
one.
\endproclaim

Since the Burton and Keane theorem requires only stationarity,
it applies also to non-extremal states, and therefore allows
the possibility of a convex combination of states with zero
and one infinite cluster.
If, in addition, the measures are ergodic,
then
at any given value of $p$,
there is {\it either} zero {\it or} one infinite cluster
with probability one.  This is presumably the case for
both the free and wired measures, although we only prove
it for the free state (see subsection 2.4).
Of course, ergodicity does not exclude the possibility that,
for a fixed value of $p$, the wired state has an
infinite cluster and the free state does not -- indeed,
for $q$ large enough, this is exactly what happens
at the transition point.

\bigskip
\subhead{2.3. Two Useful Inequalities}
\endsubhead
\smallskip

There are three main technical tools
for factoring intersections of events in standard Bernoulli
percolation:  the FKG inequality for monotone events,
independence for events which occur on nonrandom
disjoint sets, and the van den Berg - Kesten \cite {BK} inequality
for events which occur on random disjoint sets.
As discussed in the last section, the free and
wired random cluster
measures obey an FKG inequality.  However, due to the nonlocality
of the weights (2.14) and (2.16), they satisfy neither an
independence condition nor a BK inequality.  Indeed, it is
clear from (2.24) that the probability of even a simple bond
occupation event can be enhanced by the occurrence of some
other event at an arbitrarily long distance from the bond
in question.  In this subsection, we provide alternatives
to independence and the BK inequality for many events of interest
in a general setting.

As a substitute for independence of events
occuring on nonrandom disjoint sets, we might try to use the
FKG inequality as a bound, provided that the desired
events are monotone.  However, many of the events we care about
-- especially in the low-temperature phase -- are not monotone.
For example, the probability of a connection via finite
clusters is the {\it intersection} of an increasing and a
decreasing event.  The presence of boundary conditions,
which very often complicates proofs in the random cluster
model, can be used to our advantage here.  Certain boundary
conditions {\it decouple} a set from its exterior.  Many
events of interest carry with them decoupling boundary
conditions for the (random) sets on which they occur.  We
make this notion precise
by introducing the definition of a {\it decoupling
event} below.  It turns out that, given this definition,
it is possible to prove a general inequality which is similar
to the FKG inequality and which replaces independence for
events whose random boundaries occur within disjoint
nonrandom sets.  Our inequality holds for any FKG measure
and for events which are intersections of {\it arbitrary} events
with monotone decoupling events.

As explained above, the BK inequality is certainly not true
in general
for the random cluster model -- there are numerous examples
in which the occurrence of one event enhances the occurrence of
another.  However, this enhancement cannot take place if the two
events are {\it decoupled} from one another, in a sense to
be made precise in the definition below.  Thus we prove a
second inequality, which replaces the BK inequality of
Bernoulli percolation, and which holds for the intersection of
an arbitrary event, an increasing event and a decreasing
decoupling event.

In Proposition 2.6 below, we actually present two
versions of each of our inequalities:  one which
is easy to formulate
(but not that useful),
and a  more involved one
which is of the form
needed for our applications.  All of these inequalities
hold for general FKG measures.
We also give a useful corollary that concerns
monotonicity in the volume and FKG domination
in the random cluster model.  We begin with the
definition of a decoupling event.

\definition{Definition 2.5}
Given a probability space
$(\Omega, \Cal F, \mu)$
and events $A_1, A_2, D\in \Cal F$, we say that
$D$ is a {\it decoupling event} for $A_1$ and $A_2$,
if
$$
\mu(A_1\cap A_2\mid D)=
\mu(A_1\mid D)\,
\mu(A_2\mid D)\,.
\tag 2.25
$$
For brevity, we will
sometimes say $D$ decouples $A_1$ from $A_2$.
\enddefinition

While this definition makes sense in any probability
space, it may be useful to illustrate it with a typical
example from the random cluster model.  Consider a set
$B\subset\Bbb B_d$ such that $\Bbb B_d \setminus B
= B_1 \cup B_2$, $B_1 \cap B_2
= \emptyset$.  The event that the bonds of $B$ are vacant
then decouples any event $A_1 \in
\Cal F_{B_1\cup B}$ from any event $A_2 \in
\Cal F_{B_2\cup B}$.  In this paper, such decoupling
events typically occur when $B$ is the boundary of a finite
occupied cluster.  Returning to the general context of
Definition 2.5, we have:

\proclaim{Proposition 2.6}
Let $(\Omega, \Cal F, \mu)$ be a probability space
with $\Omega$ partially ordered
and $\mu$ an FKG measure with respect to this
order. Then the following inequalities hold:

\noindent
1) {The First Inequality}

\item {}
i) Consider two arbitrary events $A_1, A_2 \in\Cal F$,
and two increasing (or two decreasing)
events
$D_1, D_2\in\Cal F$
such that $D_1$ decouples
$A_1$ from $D_2$
while $D_2$ decouples $A_2$ from $A_1\cap D_1$.
Then $E_1=A_1\cap D_1$ and $E_2=A_2\cap D_2$
obey the inequality
$$
\mu(E_1\cap E_2)
\geq
\mu(E_1)
\,
\mu(E_2).
\tag 2.26
$$

\item{}
ii) More generally, let
$E_i$, $i=1,$ $2$,
be disjoint unions of the form
$$
E_i=\bigcup_{k\in K_i}
A_{i,k}\cap D_{i,k},
\tag 2.27
$$
where $K_i$ are countable index sets,
$A_{i,k}\in \Cal F$ are arbitrary events,
$D_{i,k}\in\Cal F$ are all increasing
(or all decreasing) events, and
$D_{1,k}$
decouples $A_{1,k}$ from $D_{2,k^\prime}$
while
$D_{2,k^\prime}$
decouples $A_{2,k^\prime}$
from $A_{1,k}\cap D_{1,k}$
for all $k\in K_1$ and $k^\prime\in K_2$. Then
$E_1$ and $E_2$ obey the inequality
{\rm (2.26)}.

\noindent 2) {The Second Inequality}

\item {}
i) Let $A_1\in\Cal F$ be an increasing
event, $A_2\in\Cal F$ be arbitrary,
and $D\in\Cal F$ be a decreasing event
which decouples $A_1$ from $A_2$.
Then
$$
\mu(A_1\cap D \cap A_2)
\leq
\mu(A_1)\mu(D\cap A_2)
\leq
\mu(A_1)\mu(A_2).
\tag 2.28
$$

\item{} ii) More generally,
let $A_1\in\Cal F$ be an increasing event,
and let $A_2\in\Cal F$ and $ D\in\Cal F$
be events for which
$D\cap A_2$ can be rewritten as a disjoint
union
of the form {\rm (2.27)},
with
$D_{2,k}$ decreasing events that decouple
$A_1$ from $A_{2,k}$ for all $k\in K_2$.
Then the bound {\rm (2.28)} remains valid.

\endproclaim

\demo{Proof}
Rewriting the left hand side of (2.26)
as
$$
\mu(D_2)\,
\mu(A_1\cap D_1\cap A_2\mid D_2)
$$
and using the fact that $D_2$ decouples
$A_2$ from $A_1\cap D_1$,
we obtain
$$
\mu(A_1\cap D_1\cap A_2\cap D_2)
=
\mu(A_1\cap D_1\cap D_2)\,
\mu(A_2\mid D_2).
$$
Applying the same procedure to the term
$\mu(A_1\cap D_1\cap D_2)$ and using the
decoupling event $D_1$, we get
$$
\mu(A_1\cap D_1\cap A_2\cap D_2)
=
\mu(D_1\cap D_2)\,
\mu(A_1\mid D_1)\,
\mu(A_2\mid D_2)
$$
which by the FKG inequality (2.17)
implies (2.26).
Part 1ii) of the proposition
then follows from the countable additivity
of the measure $\mu$ and the fact that
the events $E_1$ and $E_2$ are
disjoint unions of events
for which (2.26) is valid.

In order to prove 2i), we observe that
$$
\mu(A_1\cap D\cap A_2)=
\mu(D)\,
\mu(A_1\mid D)\,
\mu(A_2\mid D)
$$
by the definition of conditional expectations
and (2.25).
Using the FKG inequality (2.17) to
bound $\mu(A_1\mid D)$
by $\mu(A_1)$, the bound (2.28) now follows.
Again, 2ii) follows from 2i)
and the countable additivity of the
measure.
\qed
\enddemo

\remark{Remark}
It is clear from the above proof
that the inequality (2.26) is reversed
if one of the two decoupling
events $D_1$ and $D_2$
is increasing and the other is decreasing.
Similarly, the first inequality
in (2.28) is reversed if
$A_1$ and $D$ are both
decreasing or both increasing.
\endremark

\proclaim{Corollary}
Let  $q\geq 1$,
$\Lambda\subset\Bbb Z^d$
and  $b=$ ``$\wir$'' or
``$\free$''. Consider the random
cluster measure
$\mu_{b,\Lambda}$
and the corresponding probability
space
$(\Omega_{B_b},\Cal F_{B_b},\mu_{b,\Lambda})$,
where $B_b=B^{\splus}(\Lambda)$
if  $b=$ ``$\wir$''
and $B_b=B(\Lambda)$
if  $b=$ ``$\free$''.
Let $B\subset B_b$, and
let $E$ be an event
of the form (2.27)
where the index set is the set of all
subsets of $B$, i.e\.
$$
E=\bigcup_{S\subset B}
A_S\cap D_S,
$$
with $A_S\in \Cal F_{B_b}$
arbitrary events,
and
$D_S\in \Cal F_{B_b}$
decreasing events that decouple
$A_S$ from all events in
$\Cal F_{B_b\setminus B}$.
If the
events $D_S$ are decoupling events
with respect to the measure
$\mu_{\free,\Lambda}$,
then
$$
\mu_{\free,\Lambda^\prime}(E)
\geq \mu_{\free,\Lambda}(E)
\quad\text{provided}\quad
\Lambda^\prime\subset\Lambda
\quad\text{and}\quad
B\subset B(\Lambda^\prime).
\tag 2.29
$$
If the events $D_S$
are decoupling events
with respect to the measure
$\mu_{\wir,\Lambda}$,
then
$$
\mu_{\wir,\Lambda^\prime}(E)
\leq \mu_{\wir,\Lambda}(E)
\quad\text{provided}\quad
\Lambda^\prime\subset\Lambda
\quad\text{and}\quad
B\subset B^{\splus}(\Lambda^\prime)
\,
\tag 2.30
$$
and
$$
\mu_{\wir,\Lambda}(E)
\leq \mu_{\free,\Lambda}(E)
\quad\text{provided}\quad
B\subset B(\Lambda).
\tag 2.31
$$
\endproclaim

\demo{Proof}
Let $A_1$ be the event that
all bonds in
$B(\Lambda)\setminus B(\Lambda^\prime)$
are vacant. Then
$
A_1\in\Cal F_{B(\Lambda)
\setminus B(\Lambda^\prime)}
\subset
\Cal F_{B(\Lambda)
\setminus B}$
is decoupled from $A_S$
by the event $D_S$.
Since both
$A_1$ and $D_S$
are decreasing events,
$$
\mu_{\free,\Lambda}(E\mid A_1)
\geq \mu_{\free,\Lambda}(E)
$$ by the remark following the proof
of Proposition 2.6.
Observing that
$
\mu_{\free,\Lambda^\prime}(E)=
\mu_{\free,\Lambda}(E\mid A_1)
$,
this proves (2.29).
Defining $A_1$ as the
event that
all bonds in
$B^{\splus}(\Lambda)
\setminus B^{\splus}(\Lambda^\prime)$
are occupied
(all bonds in
$B(\Lambda)^{\splus}\setminus
B(\Lambda)$
are vacant), the  remaining
two inequalities are proved in the same
way.
\qed
\enddemo

In order to illustrate the utility of
Proposition 2.6,
we conclude this subsection with
applications of each of the two inequalities.
These applications  will be needed in our
subsequent analysis
and may be of independent interest.
As usual,
we denote by $C(x) = C(x;\omega)$ the set of occupied
bonds connected to $x$ in the configuration $\omega$,
and define
$\{x\leftrightarrow y\}$
as the event that
$x$ is connected to $y$
by a finite path
of occupied bonds.
We also define,
for each finite set $\Lambda\subset\Bbb Z^d$
and any two points $x, y\in \Lambda$,
the event $R_{x,y}^\fin(\Lambda)$
that $x$ and $y$ are connected by a
cluster $C(x) \subset B(\Lambda)$.

\proclaim{Proposition 2.7}
Let $q\geq 1$, $\Lambda\subset \Bbb Z^d$
be finite or infinite, and let
$\mu=\mu_{\wir,\Lambda}$
or $\mu_{\free,\Lambda}$.
Then for all
finite $\Lambda_1, \Lambda_2\subset\Lambda$
with
$B^{\splus}(\Lambda_1)
\cap
B(\Lambda_2)
=B(\Lambda_1)
\cap
B^{\splus}(\Lambda_2)
=\emptyset$,
and all $x, y\in\Lambda_1$,
$z,w\in\Lambda_2$,
$$
\mu(
    R_{x,y}^\fin(\Lambda_1)
    \cap
    R_{z,w}^\fin(\Lambda_2)
   )
\geq
\mu(R_{x,y}^\fin(\Lambda_1))\,
\mu(R_{z,w}^\fin(\Lambda_2)).
$$
\endproclaim

\proclaim{Proposition 2.8}
Let $q\geq 1$, $\Lambda\subset \Bbb Z^d$
be finite, and let
$\mu=\mu_{\wir,\Lambda}$
or $\mu_{\free,\Lambda}$.
Let $x,y,z,w\in\Lambda$,
and let $D$ be the event
$\{x\not\leftrightarrow z\}
\cap
\{y\not\leftrightarrow w\}
$.
Then
$$
\mu(
\{x\leftrightarrow  y\}
\cap
D
\cap
\{z\leftrightarrow  w\}
)
\leq
\mu(x\leftrightarrow  y)\,
\mu(z\leftrightarrow  w).
$$
\endproclaim

Clearly, Proposition 2.7 is an application of the first
inequality in Proposition 2.6 -- the connections in
question occur on fixed disjoint sets, $B(\Lambda_1)$
and $B(\Lambda_2)$, and due to the finiteness of the
clusters, each connection carries its own decoupling event.
Note that if $B^{\splus}(\Lambda_1)\cap B^{\splus}(\Lambda_2)
=\emptyset$, then in percolation, the probability of
the intersection of the events in Proposition 2.7 would
factor exactly.
Here our first inequality replaces this independence.
In fact, given that the decoupling events
can overlap, Proposition 2.7 gives a new result even
in the case of percolation.  Proposition 2.8 is an
application of the second inequality in Proposition
2.6 -- the connections in question occur on random
disjoint sets separated by the decoupling event $D$.
This obviously replaces the BK inequality.  Note that,
in marked contrast to percolation,
the inequality would fail to hold if we removed the
decoupling event.

\demo{Proof of Proposition 2.7}
Introducing $\Cal B_1$ as
the family of all sets
$B\subset B(\Lambda_1)$
such that $B$ connects $x$
to $y$, and $\Cal B_2$ as the family of all
$B\subset B(\Lambda_2)$ connecting $z$ to $w$,
we decompose
$R_{x,y}^\fin(\Lambda_1)$
and
$R_{z,w}^\fin(\Lambda_2)$
as
$$
R_{x,y}^\fin(\Lambda_1)
=\bigcup_{B\in \Cal B_1}
\{C(x)=B\}
=\bigcup_{B\in \Cal B_1}
\{\omega_B=1\}
\cap
\{\omega_{\partial^* B}=0\}
$$
and
$$
R_{z,w}^\fin(\Lambda_2)
=\bigcup_{B\in \Cal B_2}
\{C(z)=B\}
=\bigcup_{B\in \Cal B_2}
\{\omega_B=1\}
\cap
\{\omega_{\partial^* B}=0\}\,.
$$
Here $\omega_B$ is the
configuration $\omega$ restricted to the
set $B$ and $\partial^\ast B$ is
the set of all bonds in
$\Bbb B_d\setminus B$ which are connected to $B$.
Observing that for all
$B\in \Cal B_1$ and all
$\tilde B\in \Cal B_2$,
$D_{1,B}=\{\omega_{\partial^* B}=0\}$
decouples
$A_{1,B}=\{\omega_B=1\}$
from all events in
$\Cal F_{B^c}\supset\Cal F_{B^{\splus}(\Lambda_2)}$,
while
$D_{2,\tilde B}=\{\omega_{\partial^* \tilde B}=0\}$
decouples
$A_{2,\tilde B}=\{\omega_{\tilde B}=1\}$
from all events in
$\Cal F_{\tilde B^c}
\supset\Cal F_{B^{\splus}(\Lambda_1)}$,
one easily verifies that
$R_{x,y}^\fin(\Lambda_1)$
and
$R_{z,w}^\fin(\Lambda_2)$
are events of the form considered
in part 1ii) of Proposition 2.6.
\qed
\enddemo

\demo{Proof of Proposition 2.8}
Defining $A_1=\{x\leftrightarrow y\}$
and $A_2=\{z\leftrightarrow w\}$,
we rewrite
$A_1$ as the
disjoint union
$A_1^\fin\cup A_1^{\text{inf}}$,
with
$$
A_1^\fin=
A_1
\cap
\{x\not\leftrightarrow\partial\Lambda\}
$$
and
$$
A_1^{\text{inf}}=
A_1
\cap
\{x\leftrightarrow\partial\Lambda\}\,.
$$
Notice that $\mu_{\free ,\Lambda}(A_1^{\text{inf}})
=0$, since with free boundary conditions, $x$ cannot
be connected to the outer boundary
$\partial \Lambda=
\{x\notin\Lambda|\dist(x,\Lambda)=1\}$.
Introducing the family
$\Cal B_1$ of sets $B\subset B(\Lambda)$
that connect
$x$ to $y$ but do not connect
$x$ to $z$ or $y$ to $w$, we then decompose
$A_1^\fin\cap D$
as
$$
A_1^\fin\cap D
=\bigcup_{B\in \Cal B_1}
\{\omega_B=1\}
\cap
\{\omega_{\partial^* B}=0\}\,.
$$
Observing that for all $B\in \Cal B_1$,
the event $\{\omega_{\partial^* B}=0\}$
decouples $A_2$ from the event
$\{\omega_B=1\}$, we obtain
$$
\mu(A_1^\fin\cap D\cap A_2)
\leq
\mu(A_1^\fin\cap D)\,
\mu(A_2)
\leq
\mu(A_1^\fin)\,
\mu(A_2)\,,
$$
where we have used the second inequality of Proposition
2.6 in the first step.  This completes the proof for
the free measure.

In order to complete the proof for the
wired measure, we will
show
$$
\mu(A_1^{\text{inf}}\cap D\cap A_2)
\leq
\mu(A_1^{\text{inf}})\,
\mu(A_2)\,.
$$
To this end, we define
$$
A_2^\fin=
A_2
\cap
\{z\not\leftrightarrow\partial\Lambda\}\,.
$$
Since the wiring would connect two points
if they were both connected to the boundary,
we have
$$
A_1^{\text{inf}}\cap D\cap A_2=
A_1^{\text{inf}}\cap D\cap A_2^\fin
$$
with probability one with respect to the
wired measure.
Applying the same strategy as before,
we then obtain
$$
\mu(A_1^{\text{inf}}\cap D\cap A_2)
=
\mu(A_1^{\text{inf}}\cap D\cap A_2^\fin)
\leq
\mu(A_1^{\text{inf}})
\,
\mu( A_2^\fin)
\leq
\mu(A_1^{\text{inf}})
\,
\mu( A_2)\,,
$$
as claimed.
\qed
\enddemo

\remark{Remarks}
\item{1)}  As can be seen from the above proof, the finite-volume
free measure actually obeys the stronger inequality
$$
\mu_{\free,\Lambda}(
\{x\leftrightarrow  y\}
\cap
D
\cap
\{z\leftrightarrow  w\}
)
\leq
\mu_{\free,\Lambda}(\{x\leftrightarrow  y \} \cap D)\,\,
\mu_{\free,\Lambda}(z\leftrightarrow  w)\,.
$$
\item{2)} Using uniqueness of the infinite cluster
\cite{BuK} (see also Proposition
2.4 above), it follows immediately that
$
A_1^{\text{inf}}\cap D\cap A_2=
A_1^{\text{inf}}\cap D\cap A_2^\fin
$
with probability one with respect to the
infinite-volume measures
$\mu_{\free}$ and $\mu_{\wir}$.
Hence Proposition 2.8 holds for
these measures as well.
\endremark

\bigskip
\subhead{2.4. DLR Equations and States of the Random Cluster Model}
\endsubhead
\smallskip

In this subsection, we introduce the notion of (unconstrained) DLR
states for the random cluster model, prove that the free measure
is such a state, and use this to show that it is ergodic ---
a property we will need in our subsequent analysis.  It is
usually straightforward to establish such results by invoking
the general theory of Gibbs states (see e\.g\. \cite{Pr} and
\cite{Ge}).  However, the general theory requires that the
finite-volume expectations used to construct
the DLR states are {\it
quasilocal} functions
of the boundary conditions,
a property which fails to hold here due to the
nonlocality of the random cluster weights.  Thus the DLR
equation has to be established explicitly.

We start by defining finite-volume measures with general
unconstrained boundary conditions --- conditions which permit
any component to be connected to any other component.
The set of states generated by all such boundary conditions
is quite natural
in the random cluster model.  A larger class including
constrained states will be discussed briefly at the end
of this subsection.
Each measure is defined on an arbitrary finite set of bonds
$B \subset \Bbb B_d$ with boundary
$$
\partial B = \{ x\in \Bbb Z^d \mid  \exists y,z\in \Bbb Z^d
\,\,\,\text{with}\,\,\,
\langle x,y \rangle \in B \,, \langle x,z \rangle \in B^c\}\,.
$$
We specify the boundary condition
by introducing a {\it wiring diagram},
$W$, which is a disjoint partition of
$\partial B$ into $n_W = 1, \cdots, |\partial B|$ components:
$$
W=\{W_1, \cdots, W_{n_W}\} \quad \text{with} \quad
\partial B = \bigcup_{i=1}^{n_W}W_i \,\,,\,\,\,\,
W_i \cap W_j = \emptyset \,\,\, \text{if} \,\, i\neq j \,.
$$
We denote by $\Cal W (\partial B)$ the set of
all such wiring diagrams -- i\.e. the set of all disjoint
partitions of $\partial B$.
Each component, $W_i$, of the wiring diagram $W$ is considered
to be preconnected or wired, so that all bonds $b \in B$
connected to points of $W_i$ are regarded as being connected
to each other.  The number of components $\#(\omega)$ is then
computed as usual.  The random cluster weight
$$
G_{W,B}(\omega)=
\frac 1{ Z_W(B)}
(1-p)^{|\omega^c|}
p^{|\omega|}
q^{\#(\omega)}
\tag 2.32
$$
defines the finite-volume measure $\mu_{W,B}(\cdot)$.
Denoting by $W_{\free}$ the partition with
$n_W = |\partial B|$ components and by $W_{\wir}$
the partition with only a single component,
we see that
$\mu_{\free,\Lambda}(\cdot) = \mu_{W_{\free},B(\Lambda)}(\cdot) $
and
$\mu_{\wir,\Lambda}(\cdot) =
\mu_{W_{\wir},B^+(\Lambda)}(\cdot)$,
so that the free and (fully) wired measures are just special cases
of $\mu_{W,B}(\cdot)$.  Note that among the measures
$\mu_{W,B}(\cdot)$ are some that cannot be obtained
as transforms of any finite-volume states in the spin
system, namely those in which $W$ has more than $q$
components $W_i$ with $|W_i| \geq 2$.

There is a natural partial order on the set
$\Cal W (\partial B)$.  If
$W,W^\prime\in\Cal W (\partial B)$, we
say that $W^\prime$ is {\it coarser} than $W$,
denoted by $W^\prime \succ W$,
if for each $W_i^\prime\in W^\prime$ there
exist $W_{i_1}, W_{i_2}, \cdots, W_{i_m} \in W$ such that
$W_i^\prime = \cup_{j=1}^{m}W_{i_j}$.  Notice
that $W_{\free}$ is the least coarse and
$W_{\wir}$ is the most coarse of all wiring
diagrams.  Moreover if $W^\prime \succ W$,
then $\mu_{W^\prime {\hskip -.05cm},B}$ dominates $\mu_{W,B}$ in
the sense of FKG (see Definition 2.1 above).

Each configuration $\omega\in\Omega$ induces a wiring
diagram on each finite set $B\subset\Bbb B_d$.  The
{\it induced wiring diagram} $W(B,\omega)$ is
a partition into components of $\partial B$,
each of which
is connected using occupied bonds in
$\omega_{ B^c}$.  Thus each $\omega\in\Omega$
gives rise to a sequence of induced finite-volume
measures $\mu_{W(B,\omega),B}$ for any
increasing sequence of sets $B\subset\Bbb B_d$.
Henceforth we will extend the induced
finite-volume measure
$\mu_{W(B,\omega),B}$ to a measure on the full
space $(\Omega, \Cal F)$ by declaring all bonds
in $B^c$ to have the configuration specified by
$\omega$.  (Compare this to our extensions of
of $\mu_{\free,\Lambda}$ and $\mu_{\wir,\Lambda}$
discussed in the remark following equation (2.16).)
Using the form (2.32) of the weights $G_{W,B}$ and our
definition of induced wiring diagrams, it
is straightforward to check that the
(extended) induced finite-volume
measures obey the {\it consistency condition}
$$
\mu_{W(B ,\omega),B}(A) =
\int \mu_{W(B ,\omega), B}(d\tilde\omega)\,\,\,
\mu_{W(\tilde B ,\tilde\omega),\tilde B}(A)
\tag 2.33
$$
for all local events $A\in \Cal F$, any finite
set $B$ and all  $\tilde{B} \subset B$.

For each finite $B$, we may define the function
$\pi_B:(\Cal F,\Omega)\rightarrow\Bbb R$
by $\pi_B(A | \omega) = \mu_{W(B,\omega),B}(A)$.
Since the family $\gamma = \{\pi_B \mid B\subset\Bbb B_d,
|B| < \infty \}$
is a set of proper probability kernels obeying the
consistency condition (2.33), $\gamma$
is a {\it specification} in the sense of \cite{Pr}.

A DLR equation (\cite{D}, \cite{LR}) is just an
infinite-volume analogue of a consistency condition
like (2.33).  Thus we introduce the (unconstrained)
DLR equation for an infinite-volume random cluster
state $\mu$:
$$
\mu(A) = \int \mu(d\omega)\,\,\,\mu_{W(B,\omega),B}(A)
\tag 2.34
$$
where $A\in \Cal F$ is any local observable
and $B\subset\Bbb B_d$ is any finite set.  As usual,
the DLR equation (2.34) --- if it holds --- allows us to
write the infinite-volume expectation of $A$ as an
average over finite-volume expectations.  It is closed
in the sense that the average is computed with respect
to the given measure $\mu$.  Note that this is different
from the equation for states given in \cite{ACCN},
where a random cluster measure was obtained as a transform
of a measure obeying the DLR equation in the spin system.
On the other hand, a DLR equation was implicit in the discussion
of states in \cite{Grm}; there, however, the question of
existence of solutions to the equation was not addressed.

Let us denote the set of states obeying (2.34) by
$\Cal G = \Cal G(\gamma)$, where as above $\gamma$
denotes the specification.  States $\mu \in \Cal G$
will be called {\it DLR states} or {\it Gibbs states}.
{\it A priori} it is not clear whether $\Cal G$
is nonempty, i.e\. whether there exists any $\mu$
satsifying (2.34).  One might try to construct such a
$\mu$ as a subsequential limit of finite-volume
measures $\mu_{W,B}$ --- which clearly exists by
compactness --- but the question of whether such a
limit obeys (2.34) involves a delicate interchange
of limits.  The theory of Gibbs states (\cite{Pr},
\cite{Ge}) provides general conditions under which
(2.34) is satisfied, one
of which is quasilocality
of the specification.

A function $f$ is {\it quasilocal} if it can be approximated in
the supremum norm by local functions, a property
which is equivalent (\cite{Ge}, Remark 2.21) to the
statement
$$
\sup_{{\omega,\eta\,:\,}{\omega_B = \eta_B}}
\mid f(\omega) - f(\eta) \mid
\,\rightarrow 0
\,\,\,\,
\text{as} \,\, B \,\rightarrow \Bbb B_d\,\,.
$$
A specification $\{\pi_B\}$ is quasilocal
if the functions $\pi_B(A,\cdot\,)$ are quasilocal for all
finite $B\subset\Bbb B_d$ and all local events $A\in \Cal F$.

Unfortunately, due to nonlocality of
 the weights $G_{W,B}$, our specification
is not quasilocal.  For
example, the probability of the simple event
$\{\omega(b) = 1\}$, conditioned on the bonds in
$\Bbb B_d \setminus \{b\}$, changes discontinuously
depending on whether or not the endpoints of $b$
are connected by a path (of any length) in
$\Bbb B_d \setminus \{b\}$  (see equation (2.24)).
The general theory of Gibbs states can therefore
not be applied here.  However, we can verify
the DLR equation (2.34)
explicitly in the case of the free measure:

\proclaim{Proposition 2.9}  For all $q \geq 1$ and $0 \leq \beta
\leq \infty$,
$\mu_{\free}\in\Cal G$.
\endproclaim

\demo{Proof}
Let $B\subset\Bbb B_d$ be a finite set and
$A\in\Cal F$ a local event.  We wish to show
$$
\mu_{\free}(A) =
\int \mu_{\free}(d\omega)\,\,\,\mu_{W(B,\omega),B}(A)\,.
\tag 2.35
$$
By the finite-volume consistency condition (2.33)
and convergence of the finite-volume measures (2.20),
it suffices to prove
$$
\lim_{\Lambda\,\rightarrow \Bbb Z_d}
\int \mu_{{\free},\Lambda}(d\omega)\,\,\,\mu_{W(B,\omega),B}(A) =
\int \mu_{\free}(d\omega)\,\,\,\mu_{W(B,\omega),B}(A)\,.
\tag 2.36
$$
Inserting the partition of unity
$
\sum_{W \in \Cal W (\partial B)}
{1{\hskip -.1cm} \text{I}}_{\{W(B,\omega)=W\}} = 1
$
into (2.36) and noting that
$\mu_{W,B}(A)$ is independent
of $\Lambda$, we see that it is enough
to prove
$$
\lim_{\Lambda\,\rightarrow \Bbb Z_d}
\mu_{{\free},\Lambda}(\{W(B,\omega)=W\})
=\mu_{\free}(\{W(B,\omega)=W\})\,,
\tag 2.37
$$
i\.e\. that the probability of a given wiring diagram converges.

Let $R_{W_i}(B^c)$ denote the event that all sites
within the set $W_i$ are connected to each other via
bonds in $B^c$,
let $S_W(B^c) = \bigcap_{W_i\in W}R_{W_i}(B^c)$, and let
$N_{W_i,W_j}(B^c)$ denote the event that none of the
sites in $W_i$ is connected to any of the sites in
$W_j$ via bonds in $B^c$.  Then
$$
\{W(B,\omega)=W\}=
S_W(B^c)\,\cap\,
\bigcap_{{W_i,W_j\in W}\atop{i \ne j}}
N_{W_i,W_j}(B^c)\,.
\tag 2.38
$$
By inclusion-exclusion, it is not hard to show that
if $B(\Lambda)\supset B$, then
$$
\mu_{{\free},\Lambda}(\{W(B,\omega)=W\}) =
\sum_{{\tilde W \in \cal W(\partial B):}\atop{\tilde W \succ W}}
k_{\tilde W}(W)\,\,
\mu_{{\free},\Lambda}(S_{\tilde W}(B^c))\,,
\tag 2.39
$$
where the sum is over $\tilde W$ coarser than $W$ (see
the definition a paragraph below (2.32)), and
$k_{\tilde W}(W)\in\Bbb Z$  are computable coefficients
with $k_W(W) = 1$.  Thus by (2.37)--(2.39), we only
need to show that for all $\tilde W \in \Cal W (\partial B)$
$$
\lim_{\Lambda\,\rightarrow \Bbb Z_d}
\mu_{{\free},\Lambda}(S_{\tilde W}(B^c))
=\mu_{\free}(S_{\tilde W}(B^c))\,.
\tag 2.40
$$

Let $\tilde W\in \Cal W(\partial B)$
and choose $\Lambda$ such that $B(\Lambda)\supset B$.
Due to the
free boundary conditions on $\partial\Lambda$, the argument of
the left hand side of (2.40) can be rewritten as
$$
\mu_{{\free},\Lambda}(S_{\tilde W}(B^c)) =
 \mu_{{\free},\Lambda}(S_{\tilde W}(B(\Lambda)\setminus B)).
\tag 2.41
$$
Approximating the wiring event $S_{\tilde W}(B^c)$
by local events, we see that the right hand side of (2.40)
is actually a double limit:
$$
\align
\mu_{\free}(S_{\tilde W}(B^c))
&=\lim_{\Lambda^\prime\,\rightarrow \Bbb Z_d}
\mu_{\free}(S_{\tilde W}(B(\Lambda^\prime)\setminus B))
\\
&=\lim_{\Lambda^\prime\,\rightarrow \Bbb Z_d}\,\,
\lim_{\Lambda\,\rightarrow \Bbb Z_d}
\mu_{{\free},\Lambda}(S_{\tilde W}(B(\Lambda^\prime)\setminus B))\,.
\tag 2.42
\endalign
$$
Thus by (2.40)--(2.42), we must show
$$
\lim_{\Lambda\,\rightarrow \Bbb Z_d}
\mu_{{\free},\Lambda}(S_{\tilde W}(B(\Lambda)\setminus B))
=
\lim_{\Lambda^\prime\,\rightarrow \Bbb Z_d}\,\,
\lim_{\Lambda\,\rightarrow \Bbb Z_d}
\mu_{{\free},\Lambda}(S_{\tilde W}(B(\Lambda^\prime)\setminus B))\,.
\tag 2.43
$$
In order to prove this, we note that for all
$\Lambda^\prime\subset\Lambda$
$$
\mu_{{\free},\Lambda^\prime}
(S_{\tilde W}(B(\Lambda^\prime)\setminus B))
\leq
\mu_{{\free},\Lambda}(S_{\tilde W}(B(\Lambda^\prime)\setminus B))
\leq
\mu_{{\free},\Lambda}(S_{\tilde W}(B(\Lambda)\setminus B))\,,
\tag 2.44
$$
where the first inequality follows from the monotonicity property
(2.18) and the second is just a consequence of
$S_{\tilde W}(B(\Lambda^\prime)\setminus B)
\subset
S_{\tilde W}(B(\Lambda)\setminus B)$
if $\Lambda^\prime\subset\Lambda$.
Taking the limits
$\Lambda\,\rightarrow \Bbb Z_d$ and
$\Lambda^\prime\,\rightarrow \Bbb Z_d$, equation (2.44)
yields (2.43) and hence (2.35).
\qed
\enddemo

\remark{Remarks}
\item{1)}  The only property of the free measure that was
used to reduce the proposition to equation (2.40) was
convergence of the finite-volume measures.  Thus the wired
analogue of (2.40) --- i\.e. convergence of the probability
of the wiring events $S_{\tilde W}(B^c)$ with respect to the
finite-volume wired measures --- is sufficient to prove
$\mu_{\wir}\in\Cal G$.  Unfortunately, however, $\mu_{\wir}$
does not have nice monotonicity properties like those in
equation (2.44).
\item{2)}  Equation (2.43) of the proof is our first example
of the problem of interchange of limits which arises again
in Proposition 3.4 and in many theorems in Section 4.  Whenever
we deal with the infinite-volume limit of an event which is
not confined to a finite volume, we encounter a double limit
--- one for the construction of the infinite-volume measure
and the other for the approximation of the given event by
local events.  Hence the problem of interchange of limits.
This problem does not arise in percolation because the measure
is defined directly in the infinite-volume limit.  Here, when
we can deal with interchange, it is usually accomplished via
either simple FKG monotonicity (equations (2.18) and (2.19))
or our monotonicity involving decoupling events (corollary
to Proposition 2.6).
\endremark
\medskip

It is now straightforward to show that $\mu_{\free}$ is ergodic.
We have:

\proclaim{Theorem 2.10}
Let $H$ be any nontrivial subgroup of the translation group and
let $\Cal G_o \subset \Cal G$ be the set of all $H$-invariant
DLR states. Then for all $q \geq 1$,
$\mu_{\free}$ is extremal in $\Cal G_o$
and hence is $H$-ergodic.
\endproclaim

\demo{Proof}
As noted earlier, $W_{\free}$ is the least coarse of all wiring
diagrams, so that
$$
\mu_{W_{\free},B}
\underset\text{FKG} \to\leq
\mu_{W,B}
\qquad
\text{for all}\,\,\, W\in\Cal W (\partial B)\,,
\tag 2.45
$$
and thus by convergence of the measure (2.20)
$$
\mu_{\free}
\underset\text{FKG} \to\leq
\mu
\qquad
\text{for all}\,\,\,\mu\in\Cal G \,\,.
\tag 2.46
$$
Given that $\mu_{\free}\in\Cal G$ (Proposition 2.9), it follows
immediately from (2.46) that $\mu_{\free}$ is extremal in
$\Cal G$ and hence also in $\Cal G_o$ (since
$\mu_{\free}$ is of course $H$-invariant).  Ergodicity then follows
from the fact that all extremal measures in $\Cal G_o$ are
$H$-ergodic (\cite{Pr}, Theorem 4.1).
\qed
\enddemo

\remark{Remarks}
\item{1)} {\it FKG Ordering of States}:
Using the fact that the wired state is the coarsest of all
states, we have analogues of (2.45) and (2.46) for the
wired measure, and thus
$$
\mu_{\free}
\underset\text{FKG} \to\leq\,\,\,
\mu
\underset\text{FKG} \to\leq\,\,\,
\mu_{\wir}
\qquad
\text{for all}\,\,\,\mu\in\Cal G \,\,.
\tag 2.47
$$
Note of course that this does not imply $\mu_{\wir}\in\Cal G$.
\item{2)} {\it The Size of $\Cal G$}:  By Proposition 2.9,
$\mu_{\free}\in\Cal G$ so that $|\Cal G| \geq 1$ for all
$q \geq 1$ and all inverse temperatures $\beta$.  Let
$P_\infty^{\wir}(\beta)$ denote the percolation probabilty
in the wired measure, which of course coincides with the
magnetization for integer $q$.  According to
a result of \cite{ACCN} (Theorem A.2), whenever
$P_\infty^{\wir}(\beta) = 0$ (i\.e. $\beta \leq \beta_t$ for
systems with second-order transitions and $\beta < \beta_t$
for those with first-order transitions) $\mu_{\free} = \mu_{\wir}$,
so that by (2.47) and Proposition 2.9, $|\Cal G| = 1$.
It is expected that $|\Cal G| = 1$ also for $\beta > \beta_t$,
but there are only incomplete results for $d=2$:  The
two-dimensional dual of the \cite{ACCN} result says
$P_\infty^{\wir}(\beta^\ast ) = 0$ implies
$|\Cal G| = 1$, i\.e. there is one state for
$\beta > \beta_t^\ast$, which presumably coincides with
$\beta_t$ (see also \cite{Grm}).  However,
one expects more states at the transition point
in systems with first-order transitions.  For
$q$ large enough and  $d=2$, convergent expansions
(\cite{KoS}, \cite{LMR}) can
be used to show that there are $q+1$ distinct
translation-invariant
spin states (which transform into two distinct
translation-invariant random cluster states -- the
free and the wired).  There are presumably no
non-translation-invariant states.  Thus we expect
$|\Cal G| = 2$ for $\beta = \beta_t$ and
$q$ large enough in $d=2$.  In three dimensions,
convergent expansions \cite{MMRS} can be used to show that for
$q$ large enough, in addition to the translation-invariant
states discussed above, there are infinitely many
non-translation-invariant ``Dobrushin-type'' states
corresponding here to states constructed from
wiring diagrams which coincide with $W_{\wir}$
above a certain hyperplane and with $W_{\free}$
below that plane.  We expect that these expansions
can also be used to show that these non-translation-invariant
states satisfy our DLR equation (2.34), so that at
$\beta = \beta_t$, $|\Cal G| = \infty$ for
$q$ large enough in $d \geq 3$, in contrast to
the conjecture of \cite{Grm}.
\item{3)}  {\it States with Constraints:}  In the remark
above, we mentioned ``Dobrushin-type'' states which we expect
to be in $\Cal G$; these states were
constructed from a combination of wired
and free boundary conditions.  There are, however, many
Dobrushin-type states in the spin system whose
transforms are not
in $\Cal G$ --- namely, mixed states in which various
components of the boundary have different
values of the spin.  In the random cluster model, these correspond
to states with constraints --- certain components cannot
be connected to other components.
Therefore, in order to formulate DLR equations for these
states, one has to supplement our wiring diagrams with
some notion of constraints.  While this is possible
for individual finite-volume states, it is not clear
how constraints should be induced by a given configuration
$\omega\in\Omega$, nor whether the resulting measures
would obey even finite-volume consistency conditions.
\endremark

\bigskip
\head{3. The Covariance Matrix}
\endhead

\medskip
\subhead{3.1. The Random Cluster Representation of the
Covariance Matrix}
\endsubhead
\smallskip

In this section, we rewrite the
covariance matrices
with free and constant boundary
conditions,
$$
G_{\free}^{mn}(x-y)
=
{\langle}q\delta(\sigma_x,m)\,;\,q\delta(\sigma_y,n){\rangle}_{\free}
\,
\tag 3.1
$$
and
$$
G_{c}^{mn}(x-y)
=
{\langle}q\delta(\sigma_x,m)\,;\,q\delta(\sigma_y,n){\rangle}_{c}
\,,
\tag 3.2
$$
in terms of
the random
cluster representation
of Fortuin and Kasteleyn
\cite{FK}.  We do this by first deriving
finite-volume expressions and then taking
infinite-volume limits.

\bigskip
\subsubhead{3.1.1. The Covariance Matrix in Finite Volume}
\endsubsubhead
\smallskip

Before deriving our representation
for the covariance  matrix, we recall
the corresponding result for the
(finite-volume)
magnetization
$$
M_x(\beta,\Lambda)=
{1\over q-1}{\langle}q \delta(\sigma_{x},0)-1{\rangle}_{0,\Lambda}
\,.
\tag 3.3
$$
Using the
symbol
$X\leftrightarrow Y$
for the event
that the set $X$ is connected to the set $Y$
by a finite path of occupied bonds,
the expression (2.15) almost immediately gives
$$
M_x(\beta,\Lambda)=
\mu_{\wir,\Lambda}
(x \leftrightarrow \partial\Lambda)
\,.
\tag 3.4
$$
For future reference, we note that this can
be easily generalized to the expectation
of $e^{ip\sigma_x}$ with
$p\in\hat {S}\setminus\{0\}=\{2\pi/q,\cdots 2\pi(q-1)/q\}$.
We obtain
$$
{\langle}e^{ip\sigma_x}{\rangle}_{0,\Lambda}=
\mu_{\wir,\Lambda}
(x \leftrightarrow \partial\Lambda)
\quad\text{if}\quad
p\in\hat S\setminus \{ 0 \}
\,.
\tag 3.5
$$

We begin by considering the finite-volume two-point
function with free boundary conditions,
$$
G_{\free,\Lambda}^{mn}(x,y)=
{\langle}q\delta(\sigma_x,m)
\,;\,
q\delta(\sigma_y,n){\rangle}_{\free,\Lambda}
\,.
\tag 3.6
$$
Using the fact that
${\langle}q\delta(\sigma_x,m){\rangle}_{\free,\Lambda}=1$
for all $m$ and all $x\in\Lambda$, we first rewrite
$G_{\free,\Lambda}^{mn}(x,y)$ as
an untruncated expectation value
$$
G_{\free,\Lambda}^{mn}(x,y)
=
{\langle}(q\delta(\sigma_x,m)-1)
(q\delta(\sigma_y,n)-1){\rangle}_{\free,\Lambda}
\,.
\tag 3.7
$$
Now observe that
$E_\free(
(q\delta(\sigma_x,m)-1)
(q\delta(\sigma_y,n)-1)
|\omega)=0$
if $x$ and $y$ are not connected in the configuration
$\omega$, while
$
E_\free(
(q\delta(\sigma_x,m)-1)
(q\delta(\sigma_y,n)-1)
|\omega)
=q\delta(m,n)-1$
if $x$ and $y$ are connected.
Thus, defining the connectivity in the
FK representation
$$
\tau_{\free,\Lambda}(x,y)=\mu_{\free,\Lambda}
(x \leftrightarrow y)
\,,
\tag 3.8
$$
we obtain the following:
\proclaim{Lemma 3.1}
The finite-volume covariance matrix
with free boundary conditions has
the representation
$$
G_{\free,\Lambda}^{mn}(x,y)=
(q\delta(m,n)-1)\tau_{\free,\Lambda}(x,y)
\,.
\tag 3.9
$$
\endproclaim
\remark {Remark}  The result (2.9) in \cite{ACCN} for
the usual two-point function,
$$
\frac{1}{q-1}{\langle} q\delta(\sigma_x,\sigma_y)-1
{\rangle} _{\free,\Lambda}
=\tau_{\free,\Lambda}(x,y) \,,
$$
is proportional to the
trace of our expression (3.9).
\endremark
\bigskip

Next, we rewrite the finite-volume
covariance matrix
with constant  boundary
conditions,
$$
G_{c,\Lambda}^{mn}(x-y)
=
{\langle}q\delta(\sigma_x,m)
\,;\,
q\delta(\sigma_y,n){\rangle}_{c,\Lambda}.
\tag 3.10
$$
To this end, we define the finite-cluster
connectivity
$$
\tau^{\fin}_{\wir,\Lambda}(x,y)
=\mu_{\wir,\Lambda}
(\{x \leftrightarrow y\}\cap
\{x\not\leftrightarrow\partial\Lambda\}),
\tag 3.11
$$
and the covariance of the events that
$x$ and $y$ are connected to the boundary
$\partial\Lambda$
$$
\align
C_{\wir,\Lambda}(x,y)=
\mu_{\wir,\Lambda}
(\{x\leftrightarrow\partial\Lambda\}
\cap
\{y\leftrightarrow\partial\Lambda)\}
-\mu_{\wir,\Lambda}
(x\leftrightarrow\partial\Lambda)
\mu_{\wir,\Lambda}
(y\leftrightarrow\partial\Lambda).
\tag 3.12
\endalign
$$
We have:
\proclaim{Lemma 3.2}
The finite-volume covariance matrix
with constant boundary conditions has the
representation
$$
\align
&G_{c,\Lambda}^{mn}(x,y)
=
(q\delta(m,n)-1)\,
\tau^{\fin}_{\wir,\Lambda}(x,y)
%\\&\quad
+
(q\delta(m,c)-1) (q\delta(n,c)-1)
C_{\wir,\Lambda}(x,y)
\,.
\tag 3.13
\endalign
$$
\endproclaim
\bigskip

\demo{Proof} By the symmetry of the
model, it is enough to establish the lemma for
$c=0$.
In a first step, we prove a similar relation
for
$
{\langle}e^{ip\sigma_x}
\,;\,
e^{-ip^\prime\sigma_y}{\rangle}_{0,\Lambda}
$,
namely
$$
\align
\langle e^{ip\sigma_x}
\,;\,
e^{-ip^\prime\sigma_y}\rangle_{0,\Lambda}
=
(1-\delta(p,0))(1-\delta(p^\prime,0))
\bigl(
C_{\wir,\Lambda}(x,y)
+
\delta(p,p^\prime)\tau^{\fin}_{\wir,\Lambda}(x,y)
\bigr)
\,.
\tag 3.14
\endalign
$$
Assume w.l.o.g\. that $p \neq 0$
and $p^{\prime} \neq 0$, since otherwise
${\langle} e^{ip\sigma_x}
\,;\,
e^{-ip^\prime\sigma_y}{\rangle}_{0,\Lambda}
=0$. Then recalling the definition of
truncated expectation values and
observing that by (3.5)
$$
{\langle} e^{ip\sigma_x}{\rangle}_{0,\Lambda}
{\langle} e^{-ip^\prime\sigma_y}{\rangle}_{0,\Lambda}
=
\mu_{\wir,\Lambda}
(x\leftrightarrow\partial\Lambda)
\mu_{\wir,\Lambda}
(y\leftrightarrow\partial\Lambda),
$$
the proof of (3.14)
reduces to showing that
$$
{\langle} e^{ip\sigma_x}
e^{-ip^\prime\sigma_y}{\rangle}_{0,\Lambda}
=\mu_{\wir,\Lambda}
(\{x\leftrightarrow\partial\Lambda\}
\cap
\{y\leftrightarrow\partial\Lambda\})
+\delta(p,p^\prime)\tau^{\fin}_{\wir,\Lambda}(x,y) \,.
\tag 3.15
$$

We consider the cases $p=p^\prime$ and
$p\ne p^\prime$ separately: If
$p\ne p^\prime$, the expectation
$E_0(e^{ip\sigma_x}
e^{-ip^\prime\sigma_y}|\omega)$
is zero unless both $x$ and $y$ are connected
to the boundary, in which case
$E_0(e^{ip\sigma_x}
e^{-ip^\prime\sigma_y}|\omega)=1$\,.
As a consequence,
$$
{\langle} e^{ip\sigma_x}
e^{-ip^\prime\sigma_y}{\rangle}_{0,\Lambda}
=\mu_{\wir,\Lambda}
(\{x\leftrightarrow\partial\Lambda\}
\cap
\{y\leftrightarrow\partial\Lambda\})
\quad \text{if} \quad p\ne p^{\prime}.
\tag 3.16
$$
If
$p= p^\prime$, we consider two cases:
either $x\leftrightarrow\partial\Lambda$
in the configuration $\omega$ or
$x\not\leftrightarrow\partial\Lambda$.
In the first case,
$E_0(e^{ip\sigma_x}
e^{-ip^\prime\sigma_y}|\omega)=1$
if
$y\leftrightarrow\partial\Lambda$
as well, and
$E_0(e^{ip\sigma_x}
e^{-ip^\prime\sigma_y}|\omega)=0$
if
$y\not\leftrightarrow\partial\Lambda$,
yielding a contribution of
$\mu_{\wir,\Lambda}
(x\leftrightarrow\partial\Lambda
\;\text{and}\;
y\leftrightarrow\partial\Lambda)$.
In the second case,
$E_0(e^{ip\sigma_x}
e^{-ip^\prime\sigma_y}|\omega)=1$
if
$x\leftrightarrow y$, and
$E_0(e^{ip\sigma_x}
e^{-ip^\prime\sigma_y}|\omega)=0$
if
$x\not\leftrightarrow y$,
yielding
$\tau^{\fin}_{\wir,\Lambda}(x,y)$.  Thus
$$
{\langle} e^{ip\sigma_x}
e^{-ip^\prime\sigma_y}{\rangle}_{0,\Lambda}
=\mu_{\wir,\Lambda}
(\{x\leftrightarrow\partial\Lambda\}
\cap
\{y\leftrightarrow\partial\Lambda\})
+\tau^{\fin}_{\wir,\Lambda}(x,y)
\quad \text{if} \quad p= p^{\prime}.
\tag 3.17
$$
Equations (3.16) and (3.17) establish
(3.15) and hence (3.14).

Given (3.14), the proof of the lemma is
an easy exercise: observing that
the delta functions
$q\delta(\sigma_x,m)$ and $q\delta(\sigma_y,n)$
can be rewritten as
$
\sum_{p\in \hat S} e^{ip(\sigma_x-m)}$
and
$
\sum_{p^{\prime}\in \hat S}e^{-ip^\prime(\sigma_y-n)}$,
respectively,
we multiply  both sides of (3.14) by
$e^{i(p^\prime n-pm)}$
and sum over $p$ and $p^\prime$ to obtain
(3.13) for $c=0$.
\hfill\qed
\enddemo

\bigskip
\noindent
\subsubhead{3.1.2. The Covariance Matrix in Infinite Volume}
\endsubsubhead
\smallskip

In this subsection, we extend our representations of
the covariance matrix with free and wired boundary
conditions to the infinite volume.  To this end, we
again denote by $C(x) = C(x;\omega)$ the set of occupied
bonds connected to $x$
in the configuration $\omega$, and define
the (translation-invariant) analogues of
expression (3.8) for the connectivity,
$$
\tau_{\free}(x-y)=\mu_\free(x\leftrightarrow y)
\,,
\tag 3.18
$$
expression (3.11) for the finite-cluster connectivity,
$$
\tau^{\fin}_{\wir}(x-y)=
\mu_{\wir}(x\leftrightarrow y
\;\text{and}\;
|C(x)| < \infty)
\,,
\tag 3.19
$$
and expression (3.12) for the covariance,
$$
C_{\wir}(x-y)=\text{\rm Cov}_{\mu_\wir}(|C(x)|=\infty,|C(y)|=\infty)
\,.
\tag 3.20
$$
where in general
$\text{Cov}_\mu(A,B)=
\mu(A\cap B)-\mu(A)\mu(B)$ is the
covariance of events $A$ and $B$ with
respect to a measure $\mu$.  We call
the function $C_{\wir}(x-y)$ defined in (3.20)
the {\it infinite-cluster covariance}.
Our infinite-volume representation is
contained in:

\proclaim{Theorem  3.3}
The covariance matrices
$G_{\free}^{mn}(x-y)$
and
$G_{c}^{mn}(x-y)$
can be expressed as
$$
G_{\free}^{mn}(x-y)=
(q\delta(m,n)-1)\tau(x,y)
\tag 3.21
$$
and
$$
\align
G_{c}^{mn}(x-y)
=
(q\delta(m,n)-1)\,
\tau^{\fin}_{\wir}(x-y)
+
(q\delta(m,c)-1) (q\delta(n,c)-1)
C_{\wir}(x-y)
\,.
\tag 3.22
\endalign
$$
\endproclaim

\bigskip

\demo{Proof}
Given the corresponding finite-volume statements
in Lemma 3.1 and Lemma 3.2,
the theorem is an immediate consequence of the
following proposition.
\enddemo

\bigskip

\proclaim{Proposition 3.4}
Let $q\geq 1$ be real,
let $\tau_{\free,\Lambda}(x,y)$,
$\tau_{\wir,\Lambda}^\fin(x,y)$
and $C_{\wir,\Lambda}(x,y)$
be the quantities defined in
equations {\rm (3.8), (3.11)} and {\rm (3.12)},
and let $\tau_{\free}(x-y)$, $\tau^\fin(x-y)$
and $C_{\wir}(x-y)$ be the corresponding
infinite-volume quantities,
defined in equations {\rm (3.18), (3.19)} and {\rm (3.20)}.
Then the infinite-volume limits of
$\tau_{\free,\Lambda}(x,y)$,
$\tau_{\wir,\Lambda}^\fin(x,y)$
and $C_{\wir,\Lambda}(x,y)$
exist, and
$$
\tau_{\free}(x-y)=
\lim_{\Lambda\to\Bbb Z^d}
\tau_{\free,\Lambda}(x,y)
\,,
\tag 3.23
$$
$$
\tau^\fin_{\wir}(x-y)=
\lim_{\Lambda\to\Bbb Z^d}
\tau^\fin_{\wir,\Lambda}(x,y)
\tag 3.24
$$
and
$$
C_{\wir}(x-y)=
\lim_{\Lambda\to\Bbb Z^d}
C_{\wir,\Lambda}(x,y)
\,.
\tag 3.25
$$
\endproclaim
\bigskip

\remark{Remark}
For local observables, the existence of
the thermodynamic limit follows immediately
from the FKG monotonicity properties (2.18)
and (2.19) --- see equations (2.20) and (2.21).
This, however, does not imply the relations
(3.23), (3.24) and (3.25), since the events
in question are nonlocal; the relations can
only be established after an interchange of
limits.  In the ordered phase, this interchange
is not merely technical --- it is related to
the question of how the infinite cluster
emerges from large clusters in a finite volume.
Thus it depends sensitively on boundary conditions.
For example, for a {\it free} boundary condition
analogue of the finite-cluster connectivity
(3.19), an infinite-volume statement like
(3.24) is actually false.
\endremark

\demo{Proof}
Introducing the event $R_{x,y}(\Lambda)$ that
$x$ and $y$ are connected in $ B(\Lambda)$,
the right hand side of (3.23) is
$$
\lim_{\Lambda\to\Bbb Z^d}
\mu_{\free,\Lambda}
(x \leftrightarrow y)=
\lim_{\Lambda\to\Bbb Z^d}
\mu_{\free,\Lambda}
(R_{x,y}(\Lambda))
$$
while the left hand side is
$$
\mu_\free(x \leftrightarrow y)=
\lim_{\Lambda^\prime\to\Bbb Z^d}
\mu_{\free}
(R_{x,y}(\Lambda^\prime))=
\lim_{\Lambda^\prime\to\Bbb Z^d}
\lim_{\Lambda\to\Bbb Z^d}
\mu_{\free,\Lambda}
(R_{x,y}(\Lambda^\prime))
\,.
$$
We therefore have to show
that
$$
\lim_{\Lambda^\prime\to\Bbb Z^d}
\lim_{\Lambda\to\Bbb Z^d}
\mu_{\free,\Lambda}
(R_{x,y}(\Lambda^\prime))
=
\lim_{\Lambda\to\Bbb Z^d}
\mu_{\free,\Lambda}
(R_{x,y}(\Lambda))
\,.
\tag 3.26
$$
In order to prove (3.26),
we combine the monotonicity property
(2.18) with the fact that
$R_{x,y}(\Lambda^\prime)\subset R_{x,y}(\Lambda)$
if
$\Lambda^\prime\subset\Lambda$ to get
$$
\mu_{\free,\Lambda^\prime}
(R_{x,y}(\Lambda^\prime))
\leq
\mu_{\free,\Lambda}
(R_{x,y}(\Lambda^\prime))
\leq
\mu_{\free,\Lambda}
(R_{x,y}(\Lambda))
\quad\text{if}\quad
\Lambda^\prime\subset\Lambda
\,.
\tag 3.27
$$
Taking the limits
${\Lambda\to\Bbb Z^d}$ and
${\Lambda^\prime\to\Bbb Z^d}$,
the inequality (3.27) implies
equation (3.26).

In order to prove (3.24), we
consider
the event
$R_{x,y}^{\fin}(\Lambda)$ that
$x$ and $y$ are connected by a
cluster $C(x) \subset B(\Lambda)$,
as introduced in Proposition 2.7.
Recalling that
$\partial \Lambda \equiv
\{x\notin\Lambda|\dist(x,\Lambda)=1\}$,
we see that $R_{x,y}^{\fin}(\Lambda)$
is the intersection of the event
$R_{x,y}(\Lambda)$ with the event that
$x$ is not connected to
$\partial\Lambda$.
We claim that
$$
\mu_{\wir,\Lambda^\prime}
(R_{x,y}^{\fin}(\Lambda^\prime))
\leq
\mu_{\wir,\Lambda}
(R_{x,y}^{\fin}(\Lambda^\prime))
\leq
\mu_{\wir,\Lambda}
(R_{x,y}^{\fin}(\Lambda))
\quad\text{if}\quad
\Lambda^\prime\subset\Lambda
\,.
\tag 3.28
$$
As before, the second
inequality follows from the
fact that
$R_{x,y}^{\fin}(\Lambda^\prime)
\subset R_{x,y}^{\fin}(\Lambda)$
if
$\Lambda^\prime\subset\Lambda$,
which implies that
$\mu_{\wir,\Lambda}(R_{x,y}^{\fin}(\Lambda^\prime))$
is monotone increasing in $\Lambda^\prime$.
However, the monotonicity of
$\mu_{\wir,\Lambda}(R_{x,y}^{\fin}(\Lambda^\prime))$
in $\Lambda$ is less obvious because
$R_{x,y}^{\fin}(\Lambda^\prime)$
is neither an increasing nor a decreasing
event.
It is, however, an event of the form (2.27) considered in
Proposition 2.6 and its corollary. Namely,
$$
R_{x,y}^{\fin}(\Lambda^\prime)
=\bigcup_{B}\{C(x)=B\}
=\bigcup_{B}\{\omega_B=1\}\cap
\{\omega_{\partial^\ast B}=0\}
\,,
\tag 3.29
$$
where the union goes over all connected
sets $B\subset B(\Lambda^\prime)$ that join
$x$ to $y$,
$\omega_B$ is the configuration $\omega$
restricted to the set $B$,  and
$\partial^\ast B$ is the set of all bonds in
$\Bbb B_d\setminus B$ which are connected to $B$,
as in the proof of Proposition 2.7.
Thus by the corollary to Proposition 2.6,
$\mu_{\wir,\Lambda}
(R_{x,y}^{\fin}(\Lambda^\prime))$
is
an increasing function of
$\Lambda^\prime\subset\Lambda$, which
is actually stronger than
the first inequality of
(3.28).  This completes the proof
of (3.24).

In order to prove (3.25),
we remark
that it has been already shown in
\cite{ACCN} (Theorem 2.3c)
that
$
\mu_\wir(|C(x)|=\infty)=
\lim_{\Lambda\to\Bbb Z^d}
\mu_{\wir,\Lambda}
(x \leftrightarrow \partial\Lambda)
\,.
$
The proof of (3.25) therefore
reduces to showing
$$
\mu_\wir(|C(x)|=\infty
\;\text{and}\;
|C(y)|=\infty))
=
\lim_{\Lambda\to\Bbb Z^d}
\mu_{\wir,\Lambda}
(x \leftrightarrow \partial\Lambda
\;\text{and}\;
y \leftrightarrow \partial\Lambda)
\,.
\tag 3.30
$$
Proceeding as before,
we now introduce
$R_{x,y}^{\partial\Lambda}$ as the event that
both $x$ and $y$ are connected to
$\partial\Lambda$. With this notation,
equation (3.30) can be rewritten as
$$
\lim_{\Lambda^\prime\to\Bbb Z^d}
\lim_{\Lambda\to\Bbb Z^d}
\mu_{\wir,\Lambda}
(R_{x,y}^{\partial\Lambda^\prime})
=
\lim_{\Lambda\to\Bbb Z^d}
\mu_{\wir,\Lambda}
(R_{x,y}^{\partial\Lambda})
\,.
\tag 3.31
$$
Using (2.19) instead of (2.18),
and observing that
$R_{x,y}^{\partial\Lambda^\prime}\supset R_{x,y}^{\partial\Lambda}$
if
$\Lambda^\prime\subset\Lambda$, we obtain
$$
\mu_{\wir,\Lambda^\prime}
(R_{x,y}^{\partial\Lambda^\prime})
\geq
\mu_{\wir,\Lambda}
(R_{x,y}^{\partial\Lambda^\prime})
\geq
\mu_{\wir,\Lambda}
(R_{x,y}^{\partial\Lambda})
\quad\text{if}\quad
\Lambda^\prime\subset\Lambda
\,.
\tag 3.32
$$
As before, the proof is completed by
taking the limits
${\Lambda\to\Bbb Z^d}$ and
${\Lambda^\prime\to\Bbb Z^d}$.
\qed
\enddemo

\bigskip
\subhead{3.2. The Covariance Matrix and its Eigenvalues}
\endsubhead
\smallskip

Here we analyze the
structure of the covariance  matrix
$$
G_b^{mn}(x-y)
=
{\langle} q\delta(\sigma_x,m)\,;\,q\delta(\sigma_y,n){\rangle}_b\,,
\tag 3.33
$$
with free and constant boundary conditions, summarizing our
results in Theorem 3.5 at the end
of the section.  Before discussing particular
boundary conditions, we note that in general
$$
\sum_m G_{b}^{mn}(x-y)
=\sum_n G_{b}^{mn}(x-y)
=0\,,
\tag 3.34
$$
which follows from the fact that any truncated expectation
${\langle} A;B{\rangle}_b$ vanishes if either $A$ or $B$ is
constant, and from the obvious relation
$\sum_{m\in S}\delta(\sigma_x,m)=1$.  In particular,
this implies that,
independent of boundary conditions, $G_b^{mn}$ always has
a trivial eigenvalue $0$,
corresponding to an eigenvector
$\vec v_0=(1,\cdots,1)\in\Bbb R^q$.

Now consider the matrix with
free boundary conditions
$$
G_{\free}^{mn}(x-y)
=
{\langle} q\delta(\sigma_x,m)\,;\,q\delta(\sigma_y,n){\rangle}_{\free}
\,.
\tag 3.35
$$
Due to the permutation symmetry of the Hamilton function
(2.1) and the symmetry of the boundary conditions,
all diagonal elements are equal, as are all
off-diagonal elements.
%$G_{\free}^{mm}(x-y)=G_{\free}^{00}(x-y)$
%for all $m\in S$, and
%$G_{\free}^{mn}(x-y)=
%G_{\free}^{01}(x-y)$
%for all $m\neq n$.
Combining this with the observation (3.34),
we conclude that
%$G_{\free}^{00}(x-y)
%+(q-1)G_{\free}^{01}(x-y)=0$
%and hence
$$
G_{\free}^{mn}(x-y)
=\left(
\delta(m,n)
-(1-\delta(m,n))\frac{1}{q-1}
\right)
G_{\free}^{00}(x-y)
\,.
\tag 3.36
$$
Given (3.36), the matrix $G^{mn}_{\free}(x-y)$
is easily diagonalized. We find one trivial eigenvalue
$0$, corresponding to an eigenvector
$\vec v_0=(1,\cdots,1)$, and one $(q-1)$-fold
degenerate eigenvalue
$$
G_{\free}(x-y)=
\frac{q}{q-1}
G_{\free}^{00}(x-y)
=\frac{q}{q-1}
{\langle} q\delta(\sigma_x,\sigma_y)-1{\rangle}_{\free}
\,,
\tag 3.37
$$
corresponding to the ($q-1$)-dimensional eigenspace
orthogonal to $\vec v_0$.  In the second equality
in (3.37), we have reexpressed
$G_{\free}^{00}(x-y)$ as the usual two-point
function.
\remark {Remark}  The above results imply that, in the
free boundary condition case, the covariance matrix of
the $q$-state Potts model does
not contain more information than the standard two-point
function.  As we will see below (and as should be clear
from the fact that $G^{mn}(x-y)$ always has one trivial
eigenvalue), the same is true of the covariance matrix
of the Ising model ($q=2$) with constant boundary conditions.
This may explain why the covariance matrix has not been
more widely studied previously.  However, as we shall see
below and in subsequent sections, the $q\geq 3$ state
matrix with constant boundary conditions does have
additional content, and this content has a clear stochastic
geometric interpretation.
\endremark
\bigskip

Next we analyze the covariance matrix with
constant boundary conditions,
$$
G_{c}^{mn}(x-y)
=
{\langle} q\delta(\sigma_x,m)\,;\,q\delta(\sigma_y,n){\rangle}_{c}
\,.
\tag 3.38
$$
Starting with the special case $q=2$, we use (3.34) to conclude that
$
G_{c}^{00}(x-y)
=G_{c}^{11}(x-y)
=-G_{c}^{01}(x-y)
=-G_{c}^{10}(x-y)$.
Combined with the fact that
$
G_{0}^{00}(x-y)
=G_{1}^{11}(x-y)
$
by the symmetry of the model,
we obtain
$$
G_{c}^{mn}(x-y)
=
\bigl(
\delta(m,n)-(1-\delta(n,m)
\bigr)
G_{0}^{00}(x-y)
\quad\text{for}\,\, q=2
\,.
\tag 3.39
$$
Observing that the matrix structure of
(3.39) is identical to that of (3.36) with
$q=2$, we see that we again obtain a trivial
eigenvalue of 0 and an eigenvalue
$$
G_{\wir}^{(1)}(x-y)=
G_{0}^{00}(x-y)
={\langle} s_x ; s_y {\rangle}_0
\quad\text{for}\,\, q=2
\,.
\tag 3.40
$$
Here we have rewritten $G_{0}^{00}(x-y)$
in terms
of standard
Ising spins $s_x = 2\delta(\sigma_x,0) - 1$.

For $q\neq 2$, the matrix structure of
$G_{c}^{mn}(x-y)$
is less trivial.
Using relation (3.34) and the fact that constant
boundary conditions $c \in S$ leave the symmetry
of permutations among elements of $S \setminus \{c\}$
unbroken, it is easy to show
that there are only two independent
matrix elements. Taking these to be
$G_{0}^{00}(x-y)$
and
$G_{0}^{11}(x-y)$,
we obtain
$$
\align
G_{c}^{cc}(x-y)
&=
G_{0}^{00}(x-y)
\quad\text{if}\quad c\in S
\,,
\\
G_{c}^{nn}(x-y)
&=
G_{0}^{11}(x-y)
\quad\text{if}\quad n\neq c
\,,
\\
G_{c}^{cn}(x-y)
&=
G_{c}^{nc}(x-y)
=
-\frac 1{q-1}G_{0}^{00}(x-y)
\quad\text{if}\quad n\neq c \, ,\quad\text{and}
\\
G_{c}^{mn}(x-y)
&=\frac 1{(q-1)(q-2)}G_{0}^{00}(x-y)
-\frac 1{q-2}G_{0}^{11}(x-y)
\quad\text{if}\quad n\neq m,\;n,m\neq c\,.
\\
\tag 3.41
\endalign
$$

In order to diagonalize
$G_{c}^{mn}(x-y)$, we begin by
observing that the expectation
${\langle} \cdot{\rangle}_{c}$ is invariant
under the group
$S_{q-1}$ of permutations
of $S \setminus \{c\}$.
Diagonalizing $G_{c}^{mn}$
first on the Hilbert space
corresponding to the trivial
representation of $S_{q-1}$,
we identify two eigenvectors:
$\vec v_0=(1,\cdots,1)$, corresponding
to the simple eigenvalue zero, and
$\vec v_1$, with components
$(v_1)_m=q\delta(m,c)-1$,
corresponding to the nontrivial
simple eigenvalue
$$
G_{\wir}^{(1)}(x-y)
=\frac q{q-1}
G_{0}^{00}(x-y)
=\frac q{q-1}
{\langle} q\delta(\sigma_x,0);q\delta(\sigma_y,0){\rangle}_0
\,.
\tag 3.42
$$
On the remaining ($q-2$)-dimensional subspace
orthogonal to $\vec v_0$ and $\vec v_1$, we finally
obtain the ($q-2$)-fold degenerate eigenvalue
$$
\align
G_{\wir}^{(2)}(x-y)
&=
\frac{q-1}{q-2}\,
G_{0}^{11}(x-y)
-\frac 1{(q-1)(q-2)}\,
G_{0}^{00}(x-y)
\\
&=
G_{0}^{11}(x-y)
-G_{0}^{12}(x-y)
\,.
\tag 3.43
\endalign
$$
It is interesting to note
that, as in (3.37), it is possible to
express the eigenvalue
$
G_{\wir}^{(2)}(x-y)
$
in terms of an {\it un}truncated expectation.
Indeed, we may simply rewrite the second line
in (3.43) as
$$
\align
G_{\wir}^{(2)}(x-y)
&=\frac{1}{2}
{\langle} (q\delta(\sigma_x,1)-q\delta(\sigma_x,2))
\,;\,
(q\delta(\sigma_y,1)-q\delta(\sigma_y,2)){\rangle}_{0}
\\
&=\frac{1}{2}
{\langle} (q\delta(\sigma_x,1)-q\delta(\sigma_x,2))
\,
(q\delta(\sigma_y,1)-q\delta(\sigma_y,2)){\rangle}_{0}
\,.
\tag 3.44
\endalign
$$

We both summarize the results of this section and
establish their stochastic geometric significance
in the following:

\proclaim{Theorem 3.5}
Consider the $q$-state Potts model with
$q\geq 2$.  Then the free boundary condition
covariance matrix
$G_{\free}^{mn}(x-y)$
has the simple eigenvalue zero
corresponding to the eigenvector
$\vec v_0=(1,\cdots,1)$, and a ($q-1$)-fold degenerate eigenvalue
$$
G_{\free}(x-y)
=
\frac{q}{q-1}
{\langle} q\delta(\sigma_x,\sigma_y)-1{\rangle}_{\free}
\tag 3.45
$$
corresponding to the subspace orthogonal to $\vec v_0$.
For $q\geq 3$
and all $c\in S$, the
constant boundary condition covariance matrix
$
G_{c}^{mn}(x-y)
$
has the simple eigenvalue zero
corresponding to the eigenvector $\vec v_0=(1,\cdots,1)$,
a nontrivial simple eigenvalue
$$
G_{\wir}^{(1)}(x-y)
=\frac q{q-1}
{\langle} q\delta(\sigma_x,0)
\,;\,
q\delta(\sigma_y,0){\rangle}_{0}
\tag 3.46
$$
corresponding to the eigenvector
$\vec v_1$, with components
$(v_1)_m=q\delta(m,c)-1$,
where $\vec v_0$ and $\vec v_1$
belong to the trivial representation
of the unbroken subgroup $S_{q-1}$,
and a ($q-2$)-fold degenerate eigenvalue
$$
G_{\wir}^{(2)}(x-y)
=\frac{1}{2}
{\langle} (q\delta(\sigma_x,1)-q\delta(\sigma_x,2))
\,
(q\delta(\sigma_y,1)-q\delta(\sigma_y,2)){\rangle}_{0}
\tag 3.47
$$
corresponding to the subspace orthogonal to
$\vec v_0$ and $\vec v_1$.
For $q=2$, the matrix
$G_{c}^{mn}(x-y)$ has only
the trivial eigenvalue zero and
the eigenvalue
$G_{\wir}^{(1)}(x-y)$.

Moreover,
the eigenvalues
$G_{\free}(x-y)$,
$G_{\wir}^{(1)}(x-y)$ and
$G_{\wir}^{(2)}(x-y)$ can be expressed
in the random cluster representation as
$$
G_{\free}(x-y)=q\,\tau_{\free}(x-y)
\,,
\tag 3.48
$$
$$
G_{\wir}^{(1)}(x-y)=q\,\tau^{\fin}_{\wir}(x-y)+q(q-1)C_{\wir}(x-y)
\tag 3.49
$$
and
$$
G_{\wir}^{(2)}(x-y)=q\,\tau^{\fin}_{\wir}(x-y)
\,.
\tag 3.50
$$
\endproclaim

\demo{Proof}
It only remains to establish the random cluster representations
(3.48), (3.49) and (3.50) of the eigenvalues.  But these follow
immediately from expressions (3.37), (3.42) and (3.43) for the
eigenvalues in terms of the matrix elements, and expressions
(3.21) and (3.22) relating the matrix elements to the random
cluster connectivities and cluster covariance.
\qed
\enddemo

\bigskip
\head{4. The Correlation Lengths}
\endhead

\medskip
\subhead{4.1. Existence of the Lengths
$\xi_\free$, ${\xi}^{(1)}_\wir$ and $\xi^{(2)}_\wir$}
\endsubhead
\smallskip

In this subsection, we
establish the existence of the limits
(1.7), (1.8) and (1.9) using standard reflection positivity
arguments. Namely, introducing the
unit lattice vector
$\hat e_1=(1,0,\cdots,0)\in \Bbb Z^d$,
we prove the following:

\proclaim{Theorem 4.1}Let $q\geq 2$ be an integer,
and let
$G(t)$ denote $G_\free(t\hat e_1)$,
$G_\wir^{(1)}(t\hat e_1)$ or (for $q\geq 3$)
$G_\wir^{(2)}(t\hat e_1)$, see Theorem 3.5.
Then $G(t)$ is a non-negative,
monotone decreasing
and log convex function of $t$, so
that
$(G(t)/G(0))^{1/t}$ is monotone
increasing in $t$, the limit
$$
1/\xi\equiv -\lim_{t\to\infty}\frac{\log G(t)}{t}
\tag 4.1
$$
exists and the function $G(t)$ obeys the {\rm a priori}
bound
$$
G(t)\leq G(0) e^{-t/\xi}
{}.
\tag 4.2
$$
Furthermore, denoting the correlation
lengths defined in (4.1)
by $\xi_\free$, $\xi_\wir^{(1)}$ and
$\xi_\wir^{(2)}$, we have
$$
\xi_\wir^{(1)}\,\geq
\,\xi_\wir^{(2)}.
\tag 4.3
$$

\endproclaim

\demo{Proof}
We start with the observation that each $G(t)$
can be written as a truncated (infinite-volume)
expectation of the form
$$
G(t)=\langle A;T^tA\rangle_b,
\tag 4.4
$$
where $b$ denotes either free or constant boundary conditions,
$A=A(\sigma_x)$ is an observable which depends
on the spin variable $\sigma_x$ of a single point
$x\in\Bbb Z^d$, and $T^tA$ is the translation of
the observable $A$ by $t\hat e_1$.
Equation (4.4) follows from  (3.37) and (3.35) for
$G_\free$, from (3.42) for
$G_\wir^{(1)}$ and from (3.44) for
$G_\wir^{(2)}$.

Due to the reflection
positivity of the model (see Appendix A for a review
of the basic ideas), $T$ can be represented as a
non-negative contraction ($0\leq T\leq 1$)
on a  Hilbert space $\Cal H$,
and
$$
G(t)=(\psi,T^t\psi)
$$
for a suitable vector $\psi\in\Cal H$.
Obviously, this implies that $G(t)$
is a monotone decreasing, non-negative function of $t$.
By the Cauchy-Schwarz inequality,
$$
G(\half(t_1+t_2))=
(T^{t_1/2}\psi,T^{t_2/2}\psi)
\leq\sqrt{G(t_1)G(t_2)} \,,
$$
so that $G(t)$
%and hence also  $G(t)/G(0)$
is log convex. Noting
{$0<G(0)<\infty$},
this implies that
$(G(t)/G(0))^{1/t}$
is monotone increasing in $t$, which
in turn immediately implies existence of
the limit and the {\it a priori} bound.
Finally, (4.3) follows immediately
from the representation in Theorem 3.5
and existence of the limits.
\qed
\enddemo

\remark{Remark} For
$G_\free$ and
$G_\wir^{(2)}$, the existence of the corresponding
correlation lengths can also be
established by subadditivity
arguments (see Section 4.2 below).
While these arguments are more involved
than the reflection positivity
proof presented
above, they have the advantage that they
give existence of limits analogous to
(1.7) and (1.9) for non-integer values of $q$,
defined directly in terms of $\tau_{\free}$
and $\tau_{\wir}^{\fin}$ (c.f. (equations (3.48) and
(3.50)). Moreover, a slight
variation of these arguments can be used to establish
left continuity of the inverse correlation length
$1/\xi_\free(\beta)$ (Theorem 4.2)
and upper semicontinuity of $1/\xi^{(2)}_\wir$ (Theorem 4.3).
On the other
hand, subadditivity does not establish log convexity of $G(t)$ and
hence monotonicity of the full sequence $(G(t)/G(0))^{1/t}$, as
we have from the above theorem for integer $q$.
Furthermore, we are not aware of any proof of the existence
of $\xi^{(1)}_\wir$ which does not involve
reflection positivity.
\endremark

\bigskip
\subhead{4.2. Equivalent Characterizations of
$\xi_\free$, ${\xi}^{(1)}_\wir$ and $\xi^{(2)}_\wir$}
\endsubhead
\smallskip

We already have
stochastic geometric representations for the correlation lengths
$\xi_\free$ and $\xi_\wir^{(2)}$
as the decay rates of
of $\tau_\free$ and $\tau_\wir^\fin$
-- see Theorem 3.5.
In this subsection,
we provide a stochastic geometric representation
for $\xi_\wir^{(1)}$ (Theorem 4.4) and give
alternative
representations for $\xi_\free$ and $\xi_\wir^{(2)}$
(Theorem 4.3, Lemmas 4.6, 4.7 and 4.8).
On the one hand, these alternative representations allow
us to prove several results on the behaviour of
$\xi_\free$ and $\xi_\wir^{(2)}$,
in particular left continuity of
$1/\xi_\free(\beta)$ (Theorem 4.2),
 upper semicontinuity of
$1/\xi_\wir^{(2)}$ (Theorem 4.3),
and the two-dimensional dichotomy
(1.11) and (1.12) involving
$\xi_\free$ and $\xi_\wir^{(2)}$ discussed in the introduction.
On the other hand, the alternative representations
may be of interest for numerical determinations --
in particular the representation of
$\xi_\wir^{(2)}$ in terms of the probability
$\tau_\wir^\diam(n)$ that the diameter of the
cluster $C(0)$ is $n$ (Theorem 4.3).
The representation of $\xi^{(1)}_\wir$ as
the decay rate of the covariance
$C_\wir$,
provided the magnetization $M(\beta)>0$,
may be of interest both to mathematicians and
numerical physicists.  It is worth noting that
many of the results of this subsection are generalizations
of corresponding percolation results of \cite{CCGKS} to
$q \geq 1$, but the proofs are quite different due
to the lack of independence, the lack of
a BK inequality,
and the presence of boundary conditions.

Some of the results in this section (and most of the
proofs) are of a rather technical nature.  In particular,
we introduce many connectivity functions and ultimately
show that they have only a few independent decay rates.
However, in the process, the notation and the arguments
become rather cumbersome.  In order to simplify matters,
we first introduce only a few ``physical'' connectivity
functions and summarize the results of independent interest
on $\xi_\free$, $\xi_\wir^{(2)}$ and $\xi_\wir^{(1)}$ in
Theorems 4.2, 4.3 and 4.4, respectively.  The remainder
of the subsection is devoted both to the proof of these
results and to the statement and proof of several
more techncial results which we will need for our proof
of the two-dimensional dichotomy in Section 5.

We start with a few definitions.
For $b=$ ``$\wir$'' or ``$\free$'',
we introduce
the on-axis connectivity function
$$
\tau_b(L\hat e_1)
=
\mu_b(0\leftrightarrow L\hat e_1),
\tag 4.5
$$
the on-axis finite-cluster connectivity
$$
\tau_b^\fin(L\hat e_1)=
\mu_b(\{0\leftrightarrow L\hat e_1\}\cap\{|C(0)|<\infty\}),
\tag 4.6
$$
the diameter function
$$
\tau_b^{\diam}(L)
=
\mu_b(\diam\, C(0)=L)
,
\tag 4.7
$$
where $\diam\, C(0)$ denotes the diameter of the
cluster $C(0)$ in the $\ell_\infty$ norm,
i\. e\. the maximum diameter in any of the
$d$
coordinate directions,
and the covariance
$$
C_b(x-y)=
\mu_b(
\{|C(x)|=\infty\}
\cap\{|C(y)|=\infty\}
) - P_\infty^b(\beta)^2,
\tag 4.8
$$
where
$$
P_\infty^b(\beta)=\mu_b(|C(0)|=\infty).
\tag 4.9
$$
We denote the corresponding correlation lengths
-- whenever they exist --
by
$\xi_b$, $\xi_b^\fin$, $\xi_b^\diam$ and
$\xi_b^C$.

\proclaim{Theorem 4.2}
Let
$0\leq \beta\leq \infty$
and
$q\geq 1$.  Then
the
correlation lenths
$\xi_\wir$ and $\xi_\free$ exist,
$\xi_\free\leq\xi_\wir$,
and
$1/\xi_\free$
is a left continuous function of $\beta$.
\endproclaim

\proclaim{Theorem 4.3}
Let $0<\beta<\infty$
and
$q\geq 1$.
Then for $b=$ ``$\wir$''  or  ``$\free$'',
the correlation
lengths $\xi_b^\fin$
and $\xi_b^\diam$
exist and are equal:
$
{\xi_b^\fin}
=
{\xi_b^\diam}
$.  Also
${\xi_\wir^\fin}\leq {\xi_\free^\fin}$,
and
${1}/{\xi_\wir^\fin}$ is an upper semicontinuous
function of $\beta$.
If
$q \geq 3$ is an integer,
then in addition
$$
\xi^{(2)}_\wir=\xi_\wir^\fin={\xi_\wir^\diam}
{}.
\tag 4.10
$$
\endproclaim

\remark{Remark}
Combined with the obvious inequality
$\xi_\free^\fin\leq\xi_\free$, the inequalities
from Theorem 4.2 and Theorem 4.3 give
$$
{\xi_\wir^\fin}\leq {\xi_\free^\fin}
\leq
\xi_\free\leq\xi_\wir
$$
provided
$0<\beta<\infty$ and $q\geq 1$.
\endremark

\proclaim{Theorem 4.4}
Let
$0\leq \beta\leq\infty$
and $q\geq 2$ be an integer.
Then the
 correlation length
$\xi_\wir^{C}$
exists and
$$
\xi_\wir^{C}=\xi_\wir^{(1)}
\qquad\text{if}\qquad
M(\beta)>0
\tag 4.11
$$
while
$$
\xi_\wir^{C}=0
\qquad\text{and}\qquad
\xi_\wir^{(1)}=\xi_\wir^{(2)}=\xi_\free
\qquad\text{if}\qquad
M(\beta)=0.
\tag 4.12
$$
\endproclaim
\medskip

In order to prove Theorem 4.4, we use  a proposition
which may be of independent interest and is stated
next.

\proclaim{Proposition 4.5} Let
$0\leq \beta\leq\infty$
and
$q\geq 1$. Then
$$
C_b(x-y)\geq \tau_b^\fin(x-y)P_\infty^b(\beta).
\tag 4.13
$$
\endproclaim

\demo{Proof of Proposition 4.5}
The proof of this proposition is
an easy generalization of the
proof of the corresponding
statement in \cite{CCGKS}.
For a set $B\subset \Bbb B_d$,
let $P(B)$ be the set of points $x$ such that
$x\in \partial b$ for some bond $b\in B$.
Denoting by $\Cal B_x$ the family of finite
connected subsets $B\subset \Bbb B_d$
for which $x\in P(B)$, we have
$$
\align
C_b&(x-y)
=
\mu_b(|C(x)|<\infty)\mu_b(|C(y)|=\infty)
-
\mu_b(\{|C(x)|<\infty)\}\cap\{|C(y)|=\infty\})
\\
&=
\sum_{B\in\Cal B_x}
\bigl(
\mu_b(C(x)=B)\mu_b(|C(y)|=\infty)
-
\mu_b(\{C(x)=B\}\cap\{|C(y)|=\infty\})
\bigr)
\\
&=
\sum_{B\in\Cal B_x}
\mu_b(C(x)=B)
\bigl(
\mu_b(|C(y)|=\infty)
-
\mu_b(|C(y)|=\infty\mid C(x)=B)
\bigr)
\\
&\geq
\sum_{{B\in\Cal B_x:}\atop {y\in P(B)}}
\mu_b(C(x)=B)
\bigl(
\mu_b(|C(y)|=\infty)
-
\mu_b(|C(y)|=\infty\mid C(x)=B)
\bigr)
\\
&=
\sum_{{B\in\Cal B_x:}\atop {y\in P(B)}}
\mu_b(C(x)=B)
\mu_b(|C(y)|=\infty)
=\tau_b^\fin(x-y)P_\infty^b(\beta)
\endalign
$$
where in the fourth step we have used that
for all $B\in\Cal B_x$
$$
\{C(x)=B\}=
\{\omega_{B}=1\}\cap\{\omega_{\partial^*B}=0\}
\equiv
A_2\cap D
$$
is an event of the form considered in the second
inequality (part 2i) of Proposition 2.6, and hence
$$
\mu_b(|C(y)|=\infty)
-
\mu_b(|C(y)|=\infty\mid C(x)=B)
\geq 0.
$$
Here, as in
the proof of Proposition 2.7,
$\omega_B$ is the configuration $\omega$ restricted
to $B$, and
the boundary
$\partial^*B$ of $B$
is the set of all bonds in $\Bbb B_d\setminus B$
which are connected to $B$.
\qed
\enddemo

\demo{Proof of Theorem 4.4}
Since $M(\beta)=P_\infty^\wir(\beta)$,
we have that for $M(\beta)>0$
$$
\align
(q-1)C_\wir(x-y)&
+\tau_\wir^\fin(x-y)
\leq
\biggl((q-1)+\frac 1{M(\beta)}\biggr)
C_\wir(x-y)\leq
\\
&\leq
\biggl(1+\frac 1{(q-1)M(\beta)}\biggr)
\bigl((q-1)C_\wir(x-y)
+\tau_\wir^\fin(x-y)
\bigr)
\endalign
$$
by Proposition 4.5 and the fact that
$\tau_\wir^\fin\geq 0$.
Combined with Theorem 4.1., which guarantees the existence of the
inverse correlation length
$$
1/\xi^{(1)}_{\wir}=-\lim_{L\to\infty}
\frac {\log ((q-1)C_\wir(L\hat e_1)+\tau_\wir^\fin(L\hat e_1))}L
$$
provided $q\geq 2$ is an integer, we obtain the statement
of Theorem 4.4 for $M(\beta)>0$. On the other hand,
if $M(\beta)=P_\infty^\wir(\beta)=0$,
then
$\tau^\fin_\wir=\tau_\wir$
and $C_\wir(x-y)=0$
which implies
$\xi_\wir^{(1)}=\xi_\wir^{(2)}$
and
$\xi^C_\wir=0$. Finally, $M(\beta)=0$
implies  $\mu_\wir=\mu_\free$ \cite{ACCN}
and hence $\xi^{(2)}_\wir=\xi_\free$.
\qed
\enddemo

In order to prove Theorems 4.2 and 4.3,
we will need several approximations to
the
connectivity functions
$
\tau_b(L\hat e_1)
$
and
$
\tau_b^\fin(L\hat e_1)
$.
Additional approximations will
be needed to prove the dichotomy
(1.11) and (1.12) discussed in
the introduction.
Rather than introducing them as they arise,
we define all of them here, so that the
reader may more easily refer back to the
definitions.
We will consider several subsets of $\Bbb Z^d$,
namely the
the ``cylinder''
$$
H(L)=
\bigl\{
x\in\Bbb Z^d
\mid
0\leq x_1\leq L
\bigr\}
,
\tag 4.14
$$
the ``tunnel''
$$
T(M)=
\bigl\{
x\in \Bbb Z^d
\mid
-M/2\leq x_i\leq (M+1)/2
\quad i=2,\cdots,d
\bigr\}
\tag 4.15
$$
and the  ``box''
$$
\Lambda(L,M)=T(M)\cap H(L)
{}.
\tag 4.16
$$
We then consider
the following approximation to
$\tau_b(L\hat e_1)$  in the cylinder,
tunnel and box:
$$
 \tau_b^{\text{cyl}}(L)
=
\mu_{b,H(L)}(0\leftrightarrow L\hat e_1),
\tag 4.17
$$
$$
 \tau_b^{\text{tun}}
(L,M)
=
\mu_{b,T(M)}(0\leftrightarrow L\hat e_1)
\tag 4.18
$$
and
$$
 \tau_b^{\text{box}}(L,M)
=
\mu_{b,\Lambda(L,M)}(0\leftrightarrow L\hat e_1).
\tag 4.19
$$
In all cases
$b=$
``$\wir$''
or
``$\free$''.
Assuming that they
exist, we denote the
corresponding correlation
lengths
 by
$ \xi_b^{\text{cyl}}$,
$ \xi_b^{\text{tun}}(M)$,
and
$ \xi_b^{\text{box}}(M)$.
We also consider the several approximations
to
$\tau_b^\fin(L\hat e_1)$, namely
$$
\hat\tau_b^{\text{cyl}}(L)
=
\mu_{b,H(L)}
(\{0\leftrightarrow L\hat e_1\}
\cap
\{|C(0)|<\infty\}
\cap
\{
0\not\leftrightarrow \partial H(L)
\}
),
\tag 4.20
$$
$$
\hat\tau_b^{\text{box}}(L,M)
=
\mu_{b,\Lambda(L,M)}
(\{0\leftrightarrow L\hat e_1\}
\cap
\{
0\not\leftrightarrow \partial\Lambda(L,M)
\}
),
\tag 4.21
$$
$$
\tilde\tau_b^{\text{cyl}}(L)
=
\mu_{b}
(
\{0\leftrightarrow L\hat e_1\}
\cap
\{|C(0)|<\infty\}
\cap
\{C(0)\subset B(H(L))\}
)
\tag 4.22
$$
and
$$
\tilde\tau_b^{\text{box}}(L,M)
=
\mu_{b}
(
\{0\leftrightarrow L\hat e_1\}
\cap
\{C(0)\subset B(\Lambda(L,M))\}
),
\tag 4.23
$$
denoting the corresponding correlation
lengths -- whenever they exist --
by
$\hat\xi_b^{\text{cyl}}$,
$\hat\xi_b^{\text{box}}(M)$,
$\tilde\xi_b^{\text{cyl}}$
and
$\tilde\xi_b^{\text{box}}(M)$.

We note that the distinction between (4.17) -- (4.21)
and (4.22), (4.23) is that in the former quantities
the probabilities are computed with
respect to  measures that live on the relevant
sets $\Lambda$,
while in the latter the probabilites are
computed with respect to the
full measures $\mu_{b}$,
but the events in question
occur in the relevant sets
$B(\Lambda)$.

Our first lemma gives the
equivalence of several definitions
of the correlation length $\xi_b$ and
will be used at the end of this section
to prove Theorem 4.2.

\proclaim{Lemma 4.6}
Let $0\leq\beta\leq\infty$
and
$q\geq 1$.
Let
$ \tau(L)$ denote $\tau_\wir(L\hat e_1)$,
$\tau_\free(L\hat e_1)$,
$ \tau_\free^{\text{tun}}(L,M)$,
$ \tau_\free^{\text{cyl}}(L)$,
or\,\,
$ \tau_\free^{\text{box}}(L,M)$.
Then
the correlation length $\xi$
corresponding to
$
\tau(L)
$
exists,
and
$
 \tau(L)\leq e^{-L/ \xi}
$.
Furthermore, the correlation
lengths
$ \xi_\free^{\text{tun}}(M)$
and
$ \xi_\free^{\text{box}}(M)$
are monotone decreasing in $M$,
$ \xi_\free^{\text{tun}}(M)
=
\xi_\free^{\text{box}}(M)$,
and
$$
\xi_\free=
 \xi_\free^{\text{cyl}}=
 \xi_\free^{\text{tun}}=
 \xi_\free^{\text{box}},
\tag 4.24
$$
where
$
 \xi_\free^{\text{tun}}=
\lim_{M\to\infty} \xi_\free^{\text{tun}}(M)
$
%\qquad{\text{and}}\qquad
and
$
 \xi_\free^{\text{box}}=
\lim_{M\to\infty} \xi_\free^{\text{box}}(M).
$
\endproclaim

\demo{Proof}
Considering an arbitrary subset $\Lambda\subset\Bbb Z^d$
and two points $x$ and $y$ in $\Lambda$,
we note that by the FKG inequality
(2.17)
$$
\mu_{b,\Lambda}(x\leftrightarrow y)
\geq
\mu_{b,\Lambda}(x\leftrightarrow z)
\mu_{b,\Lambda}(z\leftrightarrow y)
\tag 4.25
$$
for all $z\in\Lambda$; furthermore,
by the FKG monotonicity (2.18)
$$
\mu_{\free,\Lambda}(x\leftrightarrow y)
\geq
\mu_{\free,\Lambda^\prime}(x\leftrightarrow y)
\tag 4.26
$$
for all $\Lambda^\prime\subset\Lambda$ containing
$x$ and $y$. Using these inequalities, one obtains
subadditivity, and hence existence of the
corresponding correlation length $ \xi$, together with
the {\it a priori} bound
$ \tau(L)\leq e^{-L/ \xi}$ for
all five connectivity functions
$ \tau(L)$ considered in the theorem.
Observing that the monotonicity (4.26)
implies the monotonicity of
$\tau^{\text{box}}_\free(L,M)$ and
$\tau^{\text{tun}}_\free(L,M)$ in $M$, one obtains
the monotonicity of  $ \xi_\free^{\text{tun}}(M)$
and
$ \xi_\free^{\text{box}}(M)$ in $M$,
as well as the justification of the
interchange of limits
$$
\lim_{M\rightarrow\infty}\,\,
\lim_{L\rightarrow\infty}
\frac{\log\tau_\free^{\text{tun}}(L,M)}{L}
=
\lim_{L\rightarrow\infty}\,\,
\lim_{M\rightarrow\infty}
\frac{\log\tau_\free^{\text{tun}}(L,M)}{L}
$$
and similarly for $\tau_\free^{\text{box}}$.
The only additional ingredient
needed in the proof of the equalities
$$
 \xi_\free=
\lim_{M\to\infty}
\xi_\free^{\text{tun}}(M)
\qquad{\text{and}}\qquad
 \xi_\free^{\text{cyl}}=
\lim_{M\to\infty}
\xi_\free^{\text{box}}(M).
$$
is
that $\mu_{\free,T(M)}(x\leftrightarrow y)$ converges to
$\mu_{\free}(x\leftrightarrow y)$
(and similarly for $\tau_\free^{\text{box}}$),
which is established in
the same way
as (3.23).

We are left with the proof of the equalities
$ \xi_\free=\xi_\free^{\text{cyl}}$
and
$\xi_\free^{\text{tun}}(M)=
\xi_\free^{\text{box}}(M)$.
To this end,
we use
(4.25) and (4.26)
to get the bound
$$
\mu_{\free}
(0 \leftrightarrow nL\hat e_1)
\geq
\mu_{\free,H(nL)}
(0\leftrightarrow nL\hat e_1)
\geq
\prod_{i=0}^{n-1}
\mu_{\free,H(nL)}
(iL\hat e_1
\leftrightarrow
(i+1)L\hat e_1).
\tag 4.27
$$
Taking the limit $n\to\infty$,
and noting that all but say $\sqrt n$ terms on the
right hand side have arguments
which are sufficiently far
from the boundary, we obtain
$$
e^{-L/\xi_\free}\geq
e^{-L/\xi_\free^{\text{cyl}}}\geq
\tau_\free(L\hat e_1)
$$
which, in the limit $L\to\infty$,
implies that
$ \xi_\free=\xi_\free^{\text{cyl}}$.
The equality of
$\xi_\free^{\text{tun}}(M)$
and
$\xi_\free^{\text{box}}(M)$
is proved in the same way.
\qed
\enddemo

The next two lemmas give several
useful relations
between the correlation lengths
corresponding to (4.20) -- (4.23),
and are important ingredients for the
proof of Theorem 4.3 (see below) and for the
proof of the dichotomy (1.11) and (1.12)
(see Section 5). In order to state
the first of these two lemmas,
we introduce for each
$x_\perp\in\Bbb Z^{d-1}\cap[-M,M]^{d-1}$
the off-axis connectivity
function in the box,
$$
\tilde\tau_b^{\text{box}}(L,M;x_\perp)
=
\mu_{b}
(
\{0\leftrightarrow (L,x_\perp)\}
\cap
\{C(0)\subset B(\Lambda(L,M))\}
),
\tag 4.28
$$
and for each
$x_\perp\in\Bbb Z^{d-1}$
the off-axis connectivity
function in the cylinder
$$
\tilde\tau_b^{\text{cyl}}(L;x_\perp)
=
\mu_{b}
(
\{0\leftrightarrow (L,x_\perp)\}
\cap
\{|C(0)|<\infty\}
\cap
\{C(0)\subset B(H(L))\}
).
\tag 4.29
$$
We note that
$A_M=\{C(0)\subset B(\Lambda(L,M))\}$
is an increasing sequence
of events which converges to
the event
$\{|C(0)|<\infty\}
\cap
\{C(0)\subset B(H(L))\}$.
As a consequence,
$$
\tilde\tau_b^{\text{box}}(L,M;x_\perp)
\nearrow
\tilde\tau_b^{\text{cyl}}(L;x_\perp)
\qquad\text{as}\qquad
M\nearrow\infty.
\tag 4.30
$$

\proclaim{Lemma 4.7}
Let
$0<\beta<\infty$ and
$q\geq 1$.  Then for
$b=$ ``$\wir$''  or  ``$\free$'', the correlation lengths
$\tilde\xi^\cyl_b$ and
$\tilde\xi_b^{\text{box}}(M)$
corresponding to the connectivity functions
(4.22) and (4.23),
as well as the limit
$
{\tilde\xi_b^{\text{box}}}=
\lim_{M\to\infty}{\tilde\xi_b^{\text{box}}(M)},
$
exist and
$$
{\tilde\xi_b^{\text{cyl}}}
=
{\tilde\xi_b^{\text{box}}}.
\tag 4.31
$$
In addition,
$$
\tilde\tau_b^{\text{box}}(L,M;x_\perp)
\leq
C(\beta,q)
e^{-L/\tilde\xi_b^{\text{box}}(M)}
\leq
C(\beta,q)
e^{-L/\tilde\xi_b^{\text{box}}},
\tag 4.32
$$
and
$$
\tilde\tau_b^{\text{cyl}}(L;x_\perp)
\leq
C(\beta,q)
e^{-L/\tilde\xi_b^{\text{cyl}}}
\tag 4.33
$$
where $C(\beta,q)<\infty$ is continous as a function
of $\beta$ and independent of $L$ and $M$.
\endproclaim

\remark{Remark}  We will later show that
$\tilde\xi_b^{\text{box}}
=\xi_b^\fin$ (see equation (4.53)), so that
Lemma 4.7, together with Theorem 4.3, gives
us yet another characterization of
$\xi^{(2)}_\wir$.
\endremark
\demo{Proof}
In order to prove the lemma, we first establish the
existence of a constant
$C(\beta,q)<\infty$ such that for
all
$x_\perp\in\Bbb Z^{d-1}\cap[-M,M]^{d-1}$,
$$
\tilde\tau_b^{\text{box}}(L_1,M;x_\perp)
\,
\tilde\tau_b^{\text{box}}(L_2,M;x_\perp)\leq
C(\beta,q)
\,
\tilde\tau_b^{\text{box}}(L_1+L_2+1,M;0).
\tag 4.34
$$
To this end, we first rewrite the left
hand side of (4.34) in a form
which allows us to apply Proposition 2.7.
We consider the boxes
$
\Lambda_1=\{x\in T(M)\mid 0\leq x_1\leq L_1\}
$,
$
\Lambda_2=\{x\in T(M)\mid L_1+1\leq x_1\leq L_1+L_2+1\}
$
and
$
\Lambda=\Lambda(L_1+L_2+1,M)
$,
and the points
$
x_-=(L_1,x_\perp),
$
%\quad
$
x_+=(L_1+1,x_\perp)
$
%\quad\text{and}\quad
and
$
y=(L_1+L_2+1,0).
$
Using the translation and reflection invariance of the
measures $\mu_b$, we then rewrite
$$
\tilde\tau_b^{\text{box}}(L_1,M;x_\perp)
\,
\tilde\tau_b^{\text{box}}(L_2,M;x_\perp)
=\mu_b(R^\fin_{0,x_-}(\Lambda_1))\,\mu_b(R^\fin_{x_+,y}(\Lambda_2))
\tag 4.35
$$
where,
as in Proposition 2.7,
$R^\fin_{0,x_-}(\Lambda_1)$ is the event that
$0$ and $x_-$ are connected by a cluster
$C(0)\subset B(\Lambda_1)$,
and similary for  $R^\fin_{x_+,y}(\Lambda_2)$.
Observing that $B^{\splus}(\Lambda_1)
\cap
B(\Lambda_2)
=B(\Lambda_1)
\cap
B^{\splus}(\Lambda_2)
=\emptyset$,
we apply Proposition 2.7 to obtain
$$
\mu_b(R^\fin_{0,x_-}(\Lambda_1))\,\mu_b(R^\fin_{x_+,y}(\Lambda_2))
\leq
\mu_b(R^\fin_{0,x_-}(\Lambda_1)\cap R^\fin_{x_+,y}(\Lambda_2) ).
\tag 4.36
$$
Next we note that all configurations
$\omega\in R^\fin_{0,x_-}(\Lambda_1)\cap R^\fin_{x_+,y}(\Lambda_2)$
would contribute to
the event
$R^\fin_{0,y}(\Lambda)$
if the vacant bond $\langle x_-,x_+\rangle$
were occupied. Using finite energy in the form
(2.24), we therefore conclude that
$$
\mu_b(R^\fin_{0,x_-}(\Lambda_1)\cap R^\fin_{x_+,y}(\Lambda_2) )
\leq
C(\beta,q)\,
\mu_b(R^\fin_{0,y}(\Lambda))
\tag 4.37
$$
for a suitable constant $C(\beta,q)<\infty$.
Observing that
$$
\mu_b(R^\fin_{0,y}(\Lambda))
\equiv
\tilde\tau_b^{\text{box}}(L_1+L_2+1,M;0)\,,
$$
we obtain the subadditivity
bound (4.34).
By standard arguments,
the
bound (4.34),
together with the monotone convergence
(4.30),
implies the
existence of
the correlation lengths
${\tilde\xi_b^{\text{cyl}}}$ and
$\tilde\xi_b^{\text{box}}(M)$,
and the limit
${\tilde\xi_b^{\text{box}}}=\lim_{L\to\infty}\tilde\xi_b^{\text{box}}(M)$,
the {\it a priori}{ bounds}
(4.32) and (4.33),
and the equality of
${\tilde\xi_b^{\text{cyl}}}$
and ${\tilde\xi_b^{\text{box}}}$.
Finally, we note that by the finite
energy relation (2.24),
without loss of generality $C(\beta,q)$
may be chosen to be a continuous function
of $\beta$.
\qed
\enddemo

\proclaim{Lemma 4.8}
Let
$0<\beta<\infty$
and
$q\geq 1$.
Then the correlation lengths
$\hat\xi^\cyl_\wir$ and
$\hat\xi_\wir^{\text{box}}(M)$
corresponding to the connectivity functions
(4.20) and (4.21),
as well as the limit
$
{\hat\xi_{\wir}^{\text{box}}}=
\lim_{M\to\infty}
{\hat\xi_{\wir}^{\text{box}}(M)}
$,
exist and
$$
{\hat\xi_{\wir}^{\text{cyl}}}
=
{\hat\xi_{\wir}^{\text{box}}}
=
{\tilde\xi_{\wir}^{\text{box}}}.
\tag 4.38
$$
In addition,
$$
\hat\tau_{\wir}^{\text{box}}(L,M)
\leq
C(\beta,q)
e^{-L/\hat\xi_{\wir}^{\text{box}}(M)}
\leq
C(\beta,q)
e^{-L/\hat\xi_{\wir}^{\text{box}}},
\tag 4.39
$$
and
$$
\hat\tau_{\wir}^{\text{cyl}}(L)
\leq
C(\beta,q)
e^{-L/\hat\xi_{\wir}^{\text{cyl}}}
\tag 4.40
$$
where $C(\beta,q)<\infty$ is continuous as a function
of $\beta$ and independent of $L$ and $M$.
\endproclaim

\remark{Remark}
While Lemma 4.7 gives us alternative representations
of $\xi_b^{\fin}$ in terms of decay rates of infinite-volume
connectivities, this lemma -- together with equation (4.53)
and Theorem 4.3 -- gives us representations of
$\xi^{(2)}_\wir$ in terms of finite-volume quantities.
\endremark

\demo{Proof}
Following the proof of Lemma 4.6, we first
establish two inequalities analogous to
(4.25) and (4.26).
In order to state them,
we introduce the cylinders
$$
H(L_1,L_2)=
\{x\in\Bbb Z^d\mid L_1\leq x_1\leq L_2\},
\tag 4.41
$$
the boxes
$$
\Lambda(L_1,L_2,M)=
H(L_1,L_2)\cap T(M),
\tag 4.42
$$
(with $T(M)$ as defined in (4.15)), the
events
$$
R^\fin_{L_1,L_2}(\Lambda)
=
\{L_1 \hat e_1
\leftrightarrow
L_2\hat e_1\}
\cup\{C(0)\subset B(\Lambda)\}\,,
\tag 4.43
$$
and in particular
$$
\tilde R^\fin_{L_1,L_2}(M)
=R^\fin_{L_1,L_2}(\Lambda(L_1,L_2,M)).
\tag 4.44
$$
We then claim that for a suitable constant
$C(\beta,q)<\infty$,
$$
\mu_{\wir,\Lambda}
(\tilde R^\fin_{L_1,L_2}(M))
\mu_{\wir,\Lambda}
(\tilde R^\fin_{L_2+1,L_3}(M))
\leq
C(\beta,q)
\mu_{\wir,\Lambda}
(\tilde R^\fin_{L_1,L_3}(M))
\tag 4.45
$$
if $\Lambda\supset\Lambda(L_1,L_2,M)$,
$$
\mu_{\wir,\Lambda^\prime}
(R^\fin_{L_1,L_2}(\Lambda))
\leq
\mu_{\wir,\Lambda^{\prime\prime}}
(R^\fin_{L_1,L_2}(\Lambda))
\tag 4.46
$$
if
$
\Lambda^{\prime\prime}
\supset\Lambda^\prime
\supset\Lambda
$,
and
$$
\mu_{\wir,\Lambda}
(R^\fin_{L_1,L_2}(\Lambda^\prime))
\leq
\mu_{\wir,\Lambda}
(R^\fin_{L_1,L_2}(\Lambda^{\prime\prime}))
\tag 4.47
$$
if
$
\Lambda^\prime
\subset
\Lambda^{\prime\prime}
\subset\Lambda
$.
While the first of these three
inequalities is proved in
exactly the same way as (4.34)
using finite energy and Proposition 2.6,
the second follows from Proposition 2.6 (and its corollary)
alone
since $R^\fin_{L_1,L_2}(\Lambda)$ is an event of
the form (2.27), see proof of Proposition 2.7 and equation (3.29).
The last of the inequalities
follows from the fact that
$
R^\fin_{L_1,L_2}(\Lambda^\prime)
\subset
R^\fin_{L_1,L_2}(\Lambda^{\prime\prime})
$
if
$
\Lambda^\prime
\subset
\Lambda^{\prime\prime}
$.

Given (4.45) -- (4.47),
the proof of Lemma 4.8 is analogous to
that of Lemma 4.6, with
$$
\align
C(\beta,q)^{n-1}
&\mu_{\wir}
(\tilde R_{0,(nL+n-1)}(M))
\\
&\geq
C(\beta,q)^{n-1}
\mu_{\wir,\Lambda(0,nL+n-1,M)}
(\tilde R_{0,(nL+n-1)}(M))
\\
&\geq
\prod_{i=0}^{n-1}
\mu_{\wir,\Lambda(0,nL+n-1,M)}
(\tilde R_{i(L+1),L+i(L+1)}(M))
\tag 4.48
\endalign
$$
replacing the inequality (4.27).
\qed
\enddemo

We finally turn to the proofs of
Theorems 4.2 and 4.3.

\demo{Proof of Theorem 4.2}
The existence of the correlation
lengths $\xi_\free$ and $\xi_\wir$
has already been established
in Lemma 4.6,
and the inequality
$\xi_\free\leq\xi_\wir$
follows immediately from the FKG ordering
(2.23),
so all that remains to show is
left continuity of $1/\xi_\free(\beta)$.
Due to equation (4.24),
$1/\xi_\free(\beta)$
is a limit of finite-volume (and hence continuous)
quantities, namely
$$
1/\xi_\free(\beta)
=
-\lim_{M\to \infty}
\lim_{L\to\infty}
\frac{\log  \tau^{\text{box}}_\free(L,M)}L\,.
\tag 4.49
$$
As shown in the proof of Lemma 4.6,
the finite-volume connectivity
$ \tau^{\text{box}}_\free(L,M)$ is subadditive in $L$ and
monotone  increasing in $M$.  It is also monotone increasing
and continuous in
$\beta$. Choosing
suitable subsequences, e\. g\. $L=2^n$,
and noting the minus sign in (4.49),
this gives $1/\xi_\free(\beta)$ as
the limit of a decreasing sequence of continuous
decreasing functions,
and hence establishes the desired left continuity.
\qed
\enddemo

\demo{Proof of Theorem 4.3}
We start with the obvious bounds
$$
\tilde\tau_b^{\text{box}}(L,M)
\leq
\tilde\tau_b^\cyl(L)\leq\tau_b^\fin(L)
\leq
\sum_{n\geq L}\tau_b^\diam(n)
\tag 4.50
$$
and
$$
\tilde\tau_b^{\text{box}}(L,M)
\leq
\tau_b^\diam(L)
\qquad\text{for all}\quad M\leq L.
\tag 4.51
$$
Now consider the event
$\{C(0)=B\}$ where $B$ is some
given set of diameter $L$. Using  suitable
 rotations and  translations
by  vectors $t\in\Bbb Z^d$,
$|t|\leq L$,
each such cluster $B$ can be transformed into a
cluster $\tilde B \subset \Lambda(L,2L)$
 connecting the origin to a point
$x=(L,x_\perp)$ in the boundary of
$\Lambda(L,2L)$.
As  a consequence,
$$
\tau^\diam_b(L)
\leq
d(2L+1)^d
\sum_{{{x\in \Bbb Z^{d-1}}\atop{ |x_\perp|\leq L}}}
\tilde\tau_b^{\text{box}}(L,2L,x_\perp)
\leq d(2L+1)^{2d-1}e^{-L/\tilde\xi_b^{\text{box}}}\,,
\tag 4.52
$$
where we have used the {\it a priori} bound
(4.32) of Lemma 4.7
in the last step.

Combining the bounds (4.51) and (4.52)
with Lemma 4.7, we immediately
obtain the existence of the correlation length
$\xi^\diam_b$
and the equality
of $\xi^\diam_b$
and $\tilde\xi^{\text{box}}_b$.
Combining the bounds (4.50) and (4.52)
gives the existence of $\xi^\fin_b$
and the equality
of $\xi^\fin_b$
and
$\tilde\xi^{\text{box}}_b$,
provided $\tilde\xi^{\text{box}}_b<\infty$.
If, on the other hand,
$\tilde\xi^{\text{box}}_b=\tilde\xi_b^\cyl=\infty$,
we use the bound
$\tilde\tau_b^\cyl(L)\leq\tau_b^\fin(L)\leq 1$
to prove that the inverse correlation
length $1/{\xi_b^\fin}$ exists and is
equal to zero.
Thus we have the existence of
the correlation lengths
$\xi_b^\fin$ and $\xi_b^\diam$ and
the
equality
$$
\tilde\xi_b^{\text{box}}
=\xi_b^\fin=\xi_b^\diam\,.
\tag 4.53
$$
The final equivalence of Theorem 4.3, namely
$\xi_\wir^{(2)} = \xi_\wir^\fin$
for integer $q\geq 3$, follows immediately from
relation (3.50) of Theorem 3.5.

We are therefore left with the proofs
of the inequality
${\xi_\wir^\fin}\leq {\xi_\free^\fin}$
and the upper semicontinuity of
$1/\xi_\wir^\fin$.  Noting that
$\tau_b^\fin$ is the probability of
an event of the form considered in
Proposition 2.6, the inequality follows
immediately from the infinite-volume limit
of (2.31).
To prove the
upper semicontinuity,
we note that by Lemma 4.8 and equation (4.53) above,
$1/\xi_\wir^\fin$ can be written
as a limit of finite-volume
quantities,
namely
$$
1/\xi_\wir^\fin
=
1/\tilde\xi_\wir^{\text{box}}
=
1/\hat\xi_\wir^{\text{box}}
=-\lim_{M\to\infty}\lim_{L\to\infty}
\frac{\log\hat\tau_\wir^{\text{box}}(L,M)}L\,\,.
\tag 4.54
$$
Combined with  the
{\it a priori} bound (4.39) of
Lemma 4.8,
this implies
$$
e^{-1/\xi_\wir^\fin}
=\sup_{L,M}
\biggl\{
{\biggl(
\frac{\hat\tau_\wir^{\text{box}}(L,M)}
{C(\beta,q)}
\biggr)}^{1/L}
\biggr\}
$$
and hence
$$
1/\xi_\wir^\fin
=
\inf_{L,M}
\biggl\{
\frac{\log C(\beta,q)-
\log\hat\tau_\wir^{\text{box}}(L,M)}L
\biggr\}.
\tag 4.55
$$
Since both
$\log C(\beta,q)$ and
$-\log\hat\tau_\wir^{\text{box}}(L,M)$
are continuous  and hence upper
semicontinuous functions of $\beta$,
and since the infimum of upper semicontinuous
functions is upper semicontinuous, this establishes
the upper semicontinuity of
$1/\xi_\wir^\fin$.
\qed
\enddemo
In the above proof, we actually obtained one additional
equivalence which is not stated in Theorem 4.3, but which
will be necessary in the proof of our dichotomy.  Namely,
by equation (4.53), we have:
\proclaim{Corollary}  Let $0<\beta<\infty$ and $q\geq 1$.  Then
$\tilde\xi_b^{\text{box}}
=\xi_b^\fin$.
\endproclaim

\bigskip
\head{5. The Two-Dimensional Dichotomy and Related Results}
\endhead

\medskip
\subhead{5.1. Heuristics and Preliminaries}
\endsubhead
\smallskip

The goal of this section is a proof of the
two-dimensional dichotomy, the principal part
of which is the
duality relation
(1.11) for all $\beta$ in the low-temperature
regime.  In this subsection, we discuss
the heuristics of the relation, state our
results and briefly review two-dimensional
duality in the random cluster model.  In the next
two subsections, we derive upper and lower
bounds on
$\tau_{\wir}^{\fin}$ and its approximations in terms of
$\tau_{\free}$ and approximations to $\tau_{\free}^{\fin}$.
Finally in the fourth subsection, we put these
bounds together with the equivalence lemmas
of Section 4 and the ergodicity theorem of Section 2
to prove the
dichotomy.

In order to explain the heuristics
of the duality relation (1.11), let us consider
the representation of the random cluster
model in terms of the order-disorder contours
introduced in \cite{LMMRS},
see also \cite{BKM}.
In this representation, contours are defined as
(the connected components of) the
boundaries between regions of occupied
bonds, regarded as ordered regions,
and those of vacant bonds, regarded as
disordered regions.
Notice that in the wired measure, any finite
cluster of occupied bonds must be separated from
the infinite occupied cluster by a (disordered)
region of vacant bonds.  Thus all configurations
contributing to $\tau_\wir^\fin(x-y)$
have at least two contours surrounding
the points $x$ and $y$ --
one being the boundary between
the cluster connecting $x$ and $y$
and the disordered region, and the second being the
boundary between the disordered
region and the infinite cluster.

Let us begin by considering systems with first-order
transitions at the transition point $\beta_o$.
Since both the ordered and disordered phases
are stable at $\beta_o$, the two contours need
not remain near each other.  Indeed, under
similar circumstances,
\cite{MMRS} proved that two such order-disorder
interfaces tend to behave like independent
interfaces, leading to a surface
tension $\sigma_{oo}$ between two
ordered phases which is exactly
twice the surface tension
$\sigma_{od}$ between
an ordered and a disordered phase.
Now in our case, the minimal
combined area of the two
interfaces is $4|x-y|$.  Moreover,
we expect
that large  interfaces
are suppressed at a first-order
transition.  Thus we expect
$\tau_\wir^\fin(x-y)$
to decay exponentially
with a rate
$4\sigma_{od}=2\sigma_{oo}$,
which would imply
$$
1/\xi_\wir^\fin=2\sigma_{oo}\,.
\tag 5.1
$$
Obviously, this relation
should also be satisfied trivially at $\beta_o$ for
systems with second-order transitions
-- both sides should vanish.

Now consider the regime $\beta>\beta_o$
in a system with either a first- or
second-order transition.  In this regime, the disordered
phase is unstable, so that
large regions of vacant bonds are supressed.
Thus the two contours surrounding
$x$ and $y$ tend to
bind together, leading
to a single order-order
interface surrounding
the points of minimal
area $2|x-y|$. This leads to a exponential decay
with a rate
$2\sigma_{oo}$,
and hence again the
relation (5.1).

Note that due to the
duality relation
$
\sigma_{oo}(\beta)
=
1/\xi_\free(\beta^{\ast})
$,
equation (5.1) is equivalent
to the desired relation (1.11).

It would be interesting to make the
above heuristic arguments rigorous.
While this could presumably be
done for sufficiently large $q$,
a direct translation of these heursitics
into a proof for arbitrary
$q$ seems much more difficult. We therefore
follow a different route, based on our
inequalities involving decoupling events
(Proposition 2.6) and the equivalences
established in Section 4.

Before stating our main result, let us recall
that the dual inverse
temperature $\beta^{\ast}$ is defined by
$$
(e^\beta-1)(e^{\beta^{\ast}}-1)=q.
\tag 5.2
$$
Our main result is:

\proclaim{Theorem 5.1}
Let $d=2$, $q\geq 1$ real and $0<\beta<\infty$.
Then either
$$
P_\infty^\free(\beta^{\ast})=0
\qquad\text{and}\qquad
\xi_\wir^\fin(\beta)=\half\xi_\free(\beta^{\ast})
\tag 5.3
$$
or
$$
P_\infty^\free(\beta^{\ast})>0
\qquad\text{and}\qquad
\xi_\wir^\fin(\beta)=\xi_\free(\beta)\,.
\tag 5.4
$$
\endproclaim

\remark{Remarks}
\item{1)}
If, in addition, $q\geq 3$ is an integer,
it follows easily from the results of the last
section and the duality relation (1.13)
(a proof of which is given in Section 5.4)
that (5.3) and (5.4) may be replaced
by the dichotomy:\,\,
Either
$$
P_\infty^\free(\beta^{\ast})=0
\qquad\text{and}\qquad
\xi_\wir^{(1)}(\beta)
\geq
\xi_\wir^{(2)}(\beta)
=\half\xi_\free(\beta^{\ast})
\tag 5.3$^\prime$
$$
or
$$
P_\infty^\free(\beta^{\ast})>0
\qquad\text{and}\qquad
\xi_\wir^{(1)}(\beta)
=
\xi_\wir^{(2)}(\beta)
=\xi_\free(\beta)\,.
\tag 5.4$^\prime$
$$

\item{2)}
It follows from the duality relation
(1.13) and the monotonicity of
$P_\infty^b$
($b=$ ``free'' or ``wir'')
as a function of $\beta$
that the first branch of the
dichotomy (i\. e\. (5.3) or (5.3$^\prime$))
occurs when $\beta\geq\beta_o$,
the self-dual point,
and that the second branch
(i\. e\. (5.4) or (5.4$^\prime$))
occurs whenever
$\beta<\beta_t=\inf\{\beta\mid M(\beta)>0\}$.
It is presumably the case that $\beta_o=\beta_t$,
but rigorously this is only known for
sufficiently large $q$
(see e\.g\. \cite{LMR} where this has been shown
for $q>25$).
\endremark

\bigskip
We close this subsection with a few remarks on duality.
As usual, the dual site lattice $(\Bbb Z^{{\ast}})^2$
is the set
of points $x^{\ast}=(x_1^{\ast},x_2^{\ast})\in(\Bbb Z+\half)^2$
with half-integer coordinates,
and the dual bond lattice $\Bbb B_2^{\ast}$
is the set of nearest neighbor
bonds in $(\Bbb Z^{{\ast}})^2$.  To each
bond $b\in\Bbb B_2$,
there corresponds a dual bond $b^{\ast}\in\Bbb B_2^{\ast}$
 which has the same midpoint as
$b$. Similarly, to each configuration $\omega$
on $B\subset\Bbb B_2$,
there corresponds a dual configuration
$\omega^{\ast}$ on
$B^{\ast}=\{b^{\ast}\mid{b\in B}\}$,
given by
$$
\omega^{\ast}(b^{\ast})=
\cases 0&\text{if}\quad \omega(b)=1\\
 1&\text{if}\quad \omega(b)=0\,.
\endcases
\tag 5.5
$$
We will sometimes refer to the bonds
$b^{\ast}\in\Bbb B_2^{\ast}$ for which
$\omega^{\ast}(b^{\ast})=1$ as occupied dual bonds.
Given a finite box
$$
\Lambda=
\{x\in\Bbb Z^2\mid 0\leq x_1\leq L,\,
-M/2\leq x_2\leq (M+1)/2\}
$$
and the corresponding set of bonds $B^+(\Lambda)$,
one defines the dual of $\Lambda$ as
$$
\Lambda^{\ast}=\{x\in (\Bbb Z^{\ast})^2\mid
\exists y\in (\Bbb Z^{\ast})^2
\quad\text{with}\quad \langle x,y\rangle\in
(B^+(\Lambda))^{\ast}\}.
\tag 5.6
$$
Note that
in general $\Lambda$  and $\Lambda^{\ast}$
are not of the same cardinality.
Using the appropriate Euler relation
to relate $\#(\omega)$ to $\#(\omega^{\ast})$,
it is straightforward to check that
for a given configuration $\omega$
on $B^+(\Lambda)$ and its dual
$\omega^{\ast}$ on $B(\Lambda^{\ast})$,
$$
G_{\wir,\beta,\Lambda}(\omega)
=G_{\free,\beta^{\ast}\hskip -.09cm,\Lambda^{\ast}}(\omega^{\ast})\,,
\tag 5.7
$$
where we have explicitly indicated the temperature
dependence of the weights (2.14) and (2.16).
Thus for
each $A\in\Cal F_{B^+(\Lambda)}$,
$$
\mu_{\wir,\beta,\Lambda}(A)
=
\mu_{\free,\beta^{\ast}\hskip -.09cm,\Lambda^{\ast}}(A^{\ast}),
\tag 5.8
$$
where $A^{\ast}$ is the event
$A^{\ast}=\{\omega^{\ast}\mid\omega\in A\}$.
In the next two subsections, we will often
characterize events
$A\in\Cal F_{B^+(\Lambda)}$
in terms of the corresponding dual events.
A typical example is the event that
the cluster of the origin does not touch
the boundary $\partial\Lambda$, which
is equivalent to the
existence of a dual cluster in
$B(\Lambda^{\ast})$ containing a
closed loop which  surrounds the origin.

\bigskip
\subhead{5.2. The Upper Bound}
\endsubhead
\smallskip

Our upper bound is stated in terms of
the finite-volume approximation
$\hat\tau_\wir^{\text{box}}$
to $\tau_\wir^{\fin}$, see equation
(4.21). As in the last subsection,
we will often explicitly indicate the
$\beta$-dependence of the relevant
quantities.

\proclaim{Theorem 5.2}
Let $d=2$, $0<\beta<\infty$ and  $q\geq 1$.
Then there exists a constant
$C_1(\beta,q)<\infty$ such that
$$
\hat\tau_{\wir,\beta}^{\text{box}}(L,M)
\leq
C_1(\beta,q)\,\,
\bigl(\tau_{\free,\beta^\ast}((L-1)\hat e_1)\bigr)^2\,.
\tag 5.9
$$
\endproclaim

\demo{Proof}
Let $\Lambda$ denote the box
$\Lambda(L,M)$, see equation (4.16).
By its
definition (4.21), the connectivity function
$\hat\tau_{\wir,\beta}^{\text{box}}(L,M)$
is the probability,
in the measure
$\mu_{\wir,\beta,\Lambda}$,
of the event
$
R_{0,L}^\fin(\Lambda)=\{0\leftrightarrow L\hat e_1\}
\cap
\{0\not\leftrightarrow \partial\Lambda\}
$. Equivalently, $R_{0,L}^\fin(\Lambda)$
can be defined as the intersection of the
event $\{0\leftrightarrow L\hat e_1\}$
and the event that there is a closed
loop $\gamma^{\ast}$
of occupied dual bonds surrounding the points
$0$ and $L\hat e_1$.
Consider the points
$$
x_\pm^{\ast}=(-1/2,\pm 1/2)
\qquad\text{and}\qquad
y_\pm^{\ast}=(L+1/2,\pm 1/2)
$$
in $\Lambda^{\ast}$.
Given the fact that the connection
from $0$ to $L\hat e_1$ must occur
without touching $\partial\Lambda$, it is
clear that the dual loop
$\gamma^{\ast}$ must  consist of four pieces:
the bond $\langle x_-^{\ast},x_+^{\ast}\rangle$,
a path $\gamma_+^{\ast}$ connecting the
point $x_+^{\ast}$ to the point $y_+^{\ast}$, the bond
$\langle y_-^{\ast},y_+^{\ast}\rangle$,
and a path $\gamma_-^{\ast}$ connecting the
point $y_-^{\ast}$ to the point $x_-^{\ast}$.
Moreover, the two paths
$\gamma_\pm^{\ast}:x_\pm^{\ast}\to y_\pm^{\ast}$ must occur in
$
B(\Lambda^{\ast})\setminus
\{\langle x_-^{\ast},x_+^{\ast}\rangle,
\langle y_-^{\ast},y_+^{\ast}\rangle\}
$.
Let us denote by
$R^{\ast}_L$,
$R^{\ast}_R$,
$R^{\ast}_-$ and
$R^{\ast}_+$ the four events described above,
namely (dual) occupation of the bond
$\langle x_-^{\ast},x_+^{\ast}\rangle$, the
bond $\langle y_-^{\ast},y_+^{\ast}\rangle$
and some
paths $\gamma_\pm^{\ast}:x_\pm^{\ast}\to y_\pm^{\ast}$ in
$
B(\Lambda^{\ast})\setminus
\{\langle x_-^{\ast},x_+^{\ast}\rangle,
\langle y_-^{\ast},y_+^{\ast}\rangle\}
$, respectively.
Then $R_{0,L}^\fin(\Lambda)
=
\{0\leftrightarrow L\hat e_1\}
\cap R^{\ast}_L
\cap R^{\ast}_R
\cap R^{\ast}_-
\cap R^{\ast}_+
$.

Consider now a configuration
$\omega\in R^{\ast}_L \cap R^{\ast}_R
\cap R^{\ast}_- \cap R^{\ast}_+$. It is an
easily verified geometrical fact that
$\omega\in R_{0,L}^\fin(\Lambda)$
if and only if
the dual cluster joining $x_+^{\ast}$
to $y_+^{\ast}$ and the dual cluster
joining $x_-^{\ast}$ to $y_-^{\ast}$ are
connected only via the two bonds
$\langle x_-^{\ast},x_+^{\ast}\rangle$
and
$\langle y_-^{\ast},y_+^{\ast}\rangle$,
i\.e\. if and only if there is no
dual connection between
$x_-^{\ast}$ and $x_+^{\ast}$ in the
set
$
B(\Lambda^{\ast})\setminus
\{\langle x_-^{\ast},x_+^{\ast}\rangle,
\langle y_-^{\ast},y_+^{\ast}\rangle\}
$.
Using finite energy in the form
(2.24) to convert the two occupation events
$R^{\ast}_L$ and $R^{\ast}_R$ to
the events that the bonds $\langle x_-^{\ast},x_+^{\ast}\rangle$ and
$\langle y_-^{\ast},y_+^{\ast}\rangle$
are {\it vacant},
and the duality relation (5.8) to
transform the measure $\mu_{\wir,\beta,\Lambda}$ into the
free measure
$\mu_{\free,\beta^{\ast}\hskip -.09cm,\Lambda^{\ast}}$,
we therefore obtain
$$
\hat\tau_{\wir,\beta}^{\text{box}}(L,M)
\leq
C_1(\beta,q)
\,\mu_{\free,\beta^{\ast}\hskip -.09cm,\Lambda^{\ast}}
(
\{x_-^{\ast}\leftrightarrow y_-^{\ast}\}
\cap
\{x_+^{\ast}\leftrightarrow y_+^{\ast}\}
\cap
\{x_-^{\ast}\not\leftrightarrow x_+^{\ast}\}
)
\tag 5.10
$$
with $C_1(\beta,q)<\infty$ if
$0<\beta<\infty$.

Next, we note that by Proposition 2.8,
$$
\align
\mu_{\free,\beta^{\ast}\hskip -.09cm,\Lambda^{\ast}}
&(
\{x_-^{\ast}\leftrightarrow y_-^{\ast}\}
\cap
\{x_+^{\ast}\leftrightarrow y_+^{\ast}\}
\cap
\{x_-^{\ast}\not\leftrightarrow x_+^{\ast}\}
)
\\
&\leq
\mu_{\free,\beta^{\ast}\hskip -.09cm,\Lambda^{\ast}}
(x_+^{\ast}\leftrightarrow y_+^{\ast})\,\,
\mu_{\free,\beta^{\ast}\hskip -.09cm,\Lambda^{\ast}}
(x_-^{\ast}\leftrightarrow y_-^{\ast})\,.
\tag 5.11
\endalign
$$
Using the monotonicity (2.18), we obtain
$$
\mu_{\free,\beta^{\ast}\hskip -.09cm,\Lambda^{\ast}}
(x_\pm^{\ast}\leftrightarrow y_\pm^{\ast})\,
\leq\,
\mu_{\free,\beta^{\ast}}
(x_\pm^{\ast}\leftrightarrow y_\pm^{\ast})\,
=
\tau_{\free,\beta^{\ast}}((L+1)\hat e_1)\,,
\tag 5.12
$$
which, combined with (5.10) and (5.11), proves
the theorem.
\qed
\enddemo

\bigskip
\subhead{5.3. The Lower Bound}
\endsubhead
\smallskip

Our lower bound is given in terms of the approximation
$\tilde\tau_{\free}^{\text{box}}$ to
$\tau^{\free}_{\fin}$, see equation (4.23).

\proclaim{Theorem 5.3}
Let $d=2$, $0<\beta<\infty$ and $q\geq 1$.
Then there exists a constant
$C_2(\beta,q)>0$ such that
for all positive integers $M$ and $L$
$$
\tau_{\wir,\beta}^\fin(L\hat e_1)
\geq
C_2(\beta,q)^M
\bigl(
\tilde\tau_{\free,\beta^{\ast}}^{\text{box}}
(L-1,M)
\bigr)^2.
\tag 5.13
$$
\endproclaim

\demo{Proof}
In order to prove the theorem, we introduce
several sets in both $\Bbb Z^2$ and in its
dual $(\Bbb Z^{\ast})^2$. Consider the dual boxes
$$
\align
\Lambda_+^{\ast}
=\{x^{\ast}\in (\Bbb Z^{\ast})^2
&\mid
\half\leq x_1^{\ast}\leq L-\half
,\,\,
\half\leq x_2^{\ast}\leq 2M+\half
\},
\\
\Lambda_-^{\ast}
=\{x^{\ast}\in (\Bbb Z^{\ast})^2
&\mid
\half\leq x_1^{\ast}\leq L-\half
,\,\,
-(2M+\half)\leq x_2^{\ast}\leq -\half
\},
\endalign
$$
dual points
%$x^{\ast}_\pm$ and $y^{\ast}_\pm$ in
%$\partial\Lambda^{\ast}_\pm$:
$$
x_\pm^{\ast}=\half\hat e_1\pm (M+\half)\hat e_2,
\qquad
x_\pm^{\ast}=(L-\half)\hat e_1\pm (M+\half)\hat e_2,
$$
dual bonds
$$
b_{\pm,l}^{\ast}=\langle x^{\ast}_\pm,x^{\ast}_\pm-\hat e_1\rangle
,\qquad
b_{\pm,r}^{\ast}=\langle y^{\ast}_\pm,y^{\ast}_\pm+\hat e_1\rangle
,
$$
and dual vertical
lines $\gamma^{\ast}_l$
and $\gamma^{\ast}_r$ joining the
point $x^{\ast}_--\hat e_1$
to the point $x^{\ast}_+-\hat e_1$,
and the point
$y^{\ast}_-+\hat e_1$ to the point
 $y^{\ast}_++\hat e_1$,
respectively.
In addition to these sets
in $(\Bbb Z^{\ast})^2$ and $\Bbb B_2^{\ast}$,
consider the points
$$
\tilde x_\pm=\pm M\hat e_2
\qquad\text{and}\qquad
\tilde y_\pm=L\hat e_1 \pm M\hat e_2
,
$$
in $\Bbb Z^2$
and the vertical lines $\tilde\gamma_l$
and $\tilde\gamma_r$
in $\Bbb B_2$ that join
the point $\tilde x_-$
to the point $\tilde x_+$,
and the point $\tilde y_-$
to the point $\tilde y_+$,
respectively.

Consider now the events
$R^{\ast}_\pm$ that the points
$x^{\ast}_\pm$ and $y_\pm^{\ast}$
are connected by a path of dual
occupied bonds in $\Lambda^{\ast}_\pm$,
and the events $R_\alpha^{\ast}$
that all bonds in
$\gamma^{\ast}_\alpha
\cup b^{\ast}_{+,\alpha}
\cup b^{\ast}_{-,\alpha}$
($\alpha=l,r$) are occupied.
The event
$
R^{\ast}_+\cap R^{\ast}_-
\cap
R^{\ast}_l\cap R^{\ast}_r
$
then clearly implies the existence of an occupied dual
path surrounding the points
$0$ and $L\hat e_1$, so that
the event
$
\{0\leftrightarrow L\hat e_1\}
\cap
R^{\ast}_+\cap R^{\ast}_-
\cap
R^{\ast}_l\cap R^{\ast}_r
$
is contained in the desired event
$
R^\fin_{0,L}=
\{0\leftrightarrow L\hat e_1\}
\cap
\{|C(0)|<\infty\}
$.
As a consequence
$$
\tau_{\wir,\beta}^\fin
(L\hat e_1)
=
\mu_{\wir,\beta}
(R^\fin_{0,L})
\geq
\mu_{\wir,\beta}
\biggl(
\{0\leftrightarrow L\hat e_1\}
\cap
R^{\ast}_+\cap R^{\ast}_-
\cap
R^{\ast}_l\cap R^{\ast}_r
\biggr)\,.
\tag 5.14
$$
Our goal is to modify the event in the
argument of (5.14) so that (1) the
event $\{0\leftrightarrow L\hat e_1\}$
is guaranteed to occur, and (2) the two
dual paths across $\Lambda_+^{\ast}$
and $\Lambda_-^{\ast}$ carry decoupling
events that allow us to apply Proposition
2.6 to factor their probabilities.
We begin by degrading our estimate
(5.14) by constructing vertical lines
of occupied (direct) bonds:
$$
\tau_{\wir,\beta}^\fin
(L\hat e_1)
\geq
\mu_{\wir,\beta}
\biggl(
\{\omega_{\tilde\gamma_l\cup\tilde\gamma_r}=1\}
\cap
\{0\leftrightarrow L\hat e_1\}
\cap
R^{\ast}_+\cap R^{\ast}_-
\cap
R^{\ast}_l\cap R^{\ast}_r
\biggr).
$$
Here, as usual, $\omega_B$ denotes the configuration
$\omega$ restricted to $B$.
Using finite energy in
the form (2.24)
to flip
the $4M+6$ bonds
in
$\gamma^{\ast}_l
\cup b^{\ast}_{+,l}
\cup b^{\ast}_{-,l}
\cup\gamma^{\ast}_r
\cup b^{\ast}_{+,r}
\cup b^{\ast}_{-,r}$,
we then obtain
$$
\tau_{\wir,\beta}^\fin
(L\hat e_1)
\geq
C(\beta,q)^{4M+6}
\mu_{\wir,\beta}
\biggl(
\{\omega_{\tilde\gamma_l\cup\tilde\gamma_r}=1\}
\cap
\{0\leftrightarrow L\hat e_1\}
\cap
R^{\ast}_+\cap R^{\ast}_-
\biggr)
$$
with a suitable constant
$C(\beta,q)>0$.

Consider now the events
$(R^{\ast}_\pm)^\fin$
that $x_\pm^{\ast}$ and $y^{\ast}_\pm$
are  connected
by dual cluster
$C^{\ast}(x^{\ast}_\pm)\subset B(\Lambda^{\ast}_\pm)$,
i\.e\.
by clusters that lie entirely within $B(\Lambda^{\ast}_\pm)$,
and hence are surrounded by decoupling circuits of
occupied (direct) bonds in
$\bigl(B^+(\Lambda^{\ast}_\pm)\bigr)^{\ast}$.
Clearly
$R^{\ast}_\pm\supset (R^{\ast}_\pm)^\fin$
and thus
$$
\tau_{\wir,\beta}^\fin
(L\hat e_1)
\geq
C(\beta,q)^{4M+6}
\mu_{\wir,\beta}
\biggl(
\{\omega_{\tilde\gamma_l\cup\tilde\gamma_r}=1\}
\cap
\{0\leftrightarrow L\hat e_1\}
\cap
(R^{\ast}_+)^\fin\cap (R^{\ast}_-)^\fin
\biggr)\,.
\tag 5.15
$$
We now claim that
$$
\{\omega_{\tilde\gamma_l\cup\tilde\gamma_r}=1\}
\cap
\{0\leftrightarrow L\hat e_1\}
\cap
(R^{\ast}_+)^\fin\cap (R^{\ast}_-)^\fin
=
\{\omega_{\tilde\gamma_l\cup\tilde\gamma_r}=1\}
\cap
(R^{\ast}_+)^\fin\cap (R^{\ast}_-)^\fin\,.
\tag 5.16
$$
In order to see this, let
$\omega$ be a configuration
in $(R^{\ast}_+)^\fin$. As noted above, the condition
$C^{\ast}(x^{\ast}_\pm)\subset B(\Lambda^{\ast})$
 implies the existence of a
closed path
of occupied bonds in $\bigl(B^+(\Lambda^{\ast}_+)\bigr)^{\ast}$
surrounding $x^{\ast}_+$ and
$y^{\ast}_+$.
Given $\omega\in (R^{\ast}_+)^\fin$,
let $\gamma$ be
the innermost such path.
Since $\gamma$  surrounds
$x^{\ast}_+$ and
$y^{\ast}_+$,
but lies within
$\bigl(B^+(\Lambda^{\ast}_+)\bigr)^{\ast}$, it must visit
the points
$\tilde x_+$ and $\tilde y_+$; thus it provides
a connection
between $\tilde x_+$ and $\tilde y_+$
by a path of occupied bonds.
Observing that the vertical paths
$\tilde\gamma_l$
and
$\tilde\gamma_r$
connect the point
$0$ to the point $\tilde x_+$
and the point $L\hat e_1$
to the point $\tilde y_+$,
we see that there is automatically an occupied
path from $0$ to $L\hat e_1$,
which completes the proof of (5.16).

Using once more finite energy, the relations
(5.15) and (5.16) together
with the duality relation (5.8)
now imply that
$$
\align
\tau_{\wir,\beta}^\fin
(L\hat e_1)
&\geq
C(\beta,q)^{4M+6}
\mu_{\wir,\beta}
\biggl(
\{\omega_{\tilde\gamma_l\cup\tilde\gamma_r}=1\}
\cap
(R^{\ast}_+)^\fin\cap (R^{\ast}_-)^\fin
\biggr)
\\
&\geq
C(\beta,q)^{8M+6}
\mu_{\wir,\beta}
\bigl(
(R^{\ast}_+)^\fin\cap (R^{\ast}_-)^\fin
\bigr)
\\
&=
C(\beta,q)^{8M+6}
\mu_{\free,\beta^{\ast}}
\bigl(
R_+^\fin\cap R_-^\fin
\bigr),
\tag 5.17
\endalign
$$
where
$R_\pm^\fin$ are the events
dual to $(R_\pm^{\ast})^\fin$.

Finally, we use the
fact that $R_+^\fin$
and $R_-^\fin$
are events of the form
considered in
Proposition 2.7, so that
$$
\mu_{\free,\beta^{\ast}}
\bigl(R_+^\fin \cap R_-^\fin
\bigr)
\geq
\mu_{\free,\beta^{\ast}}
(R_+^\fin)\,\,
\mu_{\free,\beta^{\ast}}
(R_-^\fin)=
(\tilde\tau_{\free}^{\text{box}}(L,M))^2.
\tag 5.18
$$
This completes the proof of Theorem 5.3.
\qed
\enddemo

\smallskip

Notice that, in contrast to the proof of Theorem 5.2, the above
proof does not invoke monotonicity properties which depend on
boundary conditions.  Thus it can be used equally well to give
a lower bound on $\tau_{\free}^{\fin}$, namely:

\proclaim{Corollary}
Let $d=2$, $0<\beta<\infty$ and $q\geq 1$.
Then there exists a constant
$C_2(\beta,q)>0$ such that
for all positive integers $M$ and $L$
$$
\tau_{\free,\beta}^\fin(L\hat e_1)
\geq
C_2(\beta,q)^M
\bigl(
\tilde\tau_{\wir,\beta^{\ast}}^{\text{box}}
(L-1,M)
\bigr)^2.
\tag 5.19
$$
\endproclaim

\bigskip
\subhead{5.4. The Dichotomy and Percolation Probabilities}
\endsubhead
\smallskip

In this subsection we prove
Theorem 5.1. We start with a proposition
which is essentially a corollary to
the upper and lower bounds of
Theorems 5.2 and 5.3.

\proclaim{Proposition 5.4}
Let $d=2$, $0<\beta<\infty$ and  $q\geq 1$.
Then
$$
\xi_\wir^\fin(\beta)=\half\xi_\free(\beta^{\ast})
\qquad\text{if}\qquad
P_\infty^\free(\beta^{\ast})=0
\tag 5.20
$$
and
$$
\xi_\wir^\fin(\beta)=\xi_\free(\beta)\,
\qquad\text{if}\qquad
P_\infty^\wir(\beta)=0.
\tag 5.21
$$
\endproclaim

\demo{Proof}
Introducing the notation
$\hat\xi_\wir^{\text{box}}(M;\beta)$
and
$\tilde\xi_\free^{\text{box}}(M;\beta)$
to indicate the $\beta$-de\-pen\-dence of
the correlation lengths
$\hat\xi_\wir^{\text{box}}(M)$
and
$\tilde\xi_\free^{\text{box}}(M)$
corresponding to $\hat\tau_\wir^{\text{box}}$
and $\tilde\tau_\free^{\text{box}}$,
the upper and lower bounds of
Theorem 5.2 and 5.3 imply that
$$
\hat\xi_\wir^{\text{box}}(M;\beta)
\leq
\half\xi_\free(\beta^\ast)
\tag 5.22
$$
and
$$
\half\tilde\xi_\free^{\text{box}}(M;\beta^\ast)
\leq
\xi_\wir^\fin(\beta)
\tag 5.23
$$
Taking the limit $M\to\infty$,
and observing that the left hand
side of (5.22) goes to
$\xi^\fin_\wir(\beta)$
by Lemma 4.8 and the corollary at the
end of Section 4,
while the left hand side of
(5.23) goes to
$\xi^\fin_\free(\beta^*)$
by the same corollary and Lemma 4.7,
we conclude that
$$
\half\xi^\fin_\free(\beta^*)
\leq
\xi_\wir^\fin(\beta)
\leq
\half\xi_\free(\beta^*).
\tag 5.24
$$
Since
$\xi^\fin_\free(\beta^*)=\xi_\free(\beta^*)$
if $P_\infty^\free(\beta^*)=0$,
this implies the first part of the proposition.
If $P_\infty^\wir(\beta)=0$,
then
$\xi_\wir^\fin(\beta)=\xi_\wir(\beta)$.
In addition,
by the results of
\cite{ACCN},
$\mu_{\wir,\beta}=\mu_{\free,\beta}$ whenever
$P_\infty^\wir(\beta)=0$
and hence  $\xi_\free(\beta)=\xi_\wir(\beta)$.
This implies the second part of the proposition.
\qed
\enddemo

In order to complete the proof of Theorem 5.1,
we use the fact,
proven in Section 2.4, that the free measure
$\mu_{\free}$ is ergodic under any nontrivial
subgroup of the translation group (Theorem 2.10).
Since, in addition, $\mu_{\free}$ is an FKG measure
which is invariant under
horizontal and vertical translations
and axis reflections,
a bond percolation analogue of
the theorem of \cite{GKR} applies,
leading to

\proclaim{Theorem 5.5}
Let $d=2$, $0<\beta<\infty$ and $q\geq 1$,
and assume that $P_\infty^\free(\beta)>0$.
Then, with probability one with respect
to the free measure $\mu_{\free,\beta}$,
any finite set of sites in $\Bbb Z^2$
is surrounded by a circuit of occupied
bonds.
\endproclaim

\proclaim{Corollaries}
Let $d=2$, $0<\beta<\infty$ and $q\geq 1$.
Then
\item{1)}
$$
P_\infty^\free(\beta)\,P_\infty^\wir(\beta^\ast)=0\,.
\tag 5.25
$$
\medskip
\item{2)}
$P_\infty^\free(\beta) = 0 \quad \text{for all} \quad
\beta \geq \beta_o$\,.  In particular,
$P_\infty^\free(\beta)$
is left continuous at  $\beta_o$.
\medskip
\item{3)}  If
$P_\infty^\wir(\beta) = 0$ or $P_\infty^\free(\beta) > 0$,
then $\mu_{\free,\beta}(\cdot) = \mu_{\wir,\beta}(\cdot)$, and
in particular, $P_\infty^\free(\beta) = P_\infty^\wir(\beta)$.
\endproclaim

\demo{Proof}  As noted above, the theorem follows from
(a bond percolation analogue of) \cite{GKR} and Theorem 2.10.
Corollary 1 then follows immediately, and Corollary 2 also
follows easily -- see equation (1.14) and the paragraph
preceding it.  The first part of Corollary 3, namely that
$P_\infty^\wir(\beta) = 0$ implies
$\mu_{\free,\beta}(\cdot) = \mu_{\wir,\beta}(\cdot)$,
is a result of \cite{ACCN}.  That equality of the measures
is also implied by $P_\infty^\free(\beta) > 0$ follows from
(5.25), \cite{ACCN} and the self-duality of the model.
\qed
\enddemo

\remark{Remarks}
\item{1)}  Theorem 5.1 (the dichotomy) now
follows immediately from
Proposition 5.4 and equation (5.25).

\item{2)}  We expect that left continuity of
$P_\infty^\free(\beta)$
at  the transition point
holds in all dimensions provided $q\geq 1$.  However, we do
{\it not} expect  $P_\infty^\free(\beta)$ to be
right continuous at  the transition point if the system
has a first-order transition; indeed, for $q$
sufficiently large, this can be established using
Pirogov-Sinai theory,
as used e\.g\. in \cite{LMMRS}.
This is to be contrasted with the
behavior of $P_\infty^\wir(\beta)$.
By standard percolation arguments \cite{R},
namely expressing $P_\infty^\wir(\beta)$
as the decreasing limit of the finite-volume quantities
(3.4) (which are continuous and non-decreasing in
$\beta$),
$P_\infty^\wir(\beta)$ is
right continuous for all $\beta$ and all
$q \geq 1$ in dimension $d \geq 1$.  However, in dimension
$d \geq 2$, convergent expansions have
been used to show $P_\infty^\wir(\beta)$ is not
 left continuous
at the transition point
provided $q$ is
sufficiently large \cite{KoS}
(see also \cite{LMR} and \cite{LMMRS}).

\item{3)}  Corollary 3 implies that in two
dimensions the Gibbs state
is unique at all $\beta$ except those for which
$P_\infty^\free(\beta) = 0$ while $P_\infty^\wir(\beta) > 0$.
Presumably, this never occurs for systems with second-order
transitions ($q \leq 4$ in $d=2$), and occurs only at a single point
-- the transition point -- for systems with
first-order transitions ($q > 4$ in $d=2$).  Again, this can be proven
via expansion methods in
$d \geq 2$ for $q$ sufficiently large.
\endremark
\medskip
We conclude this section with a little result
which is an easy consequence of Proposition 5.4.
The result shows that continuity
of the magnetization at $\beta_o$ ensures criticality
of the transition, i\.e\. divergence of the
correlation length(s).

\proclaim{Proposition 5.6}  Let $d=2$ and $q \geq 1$.  Then
$M(\beta_o)\,\equiv P_\infty^\wir(\beta_o) = 0$ implies
$\xi_{\wir}^{\fin}(\beta_o) = \infty$ and hence also
$\xi_{\free}^{\fin}(\beta_o) = \xi_{\free}(\beta_o)
=\xi_{\wir}(\beta_o) = \infty$.
\endproclaim

\demo{Proof}
By the assumption $P_\infty^\wir(\beta_o) = 0$,
the FKG ordering of states (2.23) and the
definition of $\beta_o$, we have
$0= P_\infty^\free(\beta_o) = P_\infty^\free(\beta^{\ast}_o)$,
and hence by the
first branch (5.20) of Proposition 5.4:
$$
\xi_\wir^\fin(\beta_o)=\half\xi_\free(\beta^{\ast}_o)
=\half\xi_\free(\beta_o)\,.
\tag 5.26
$$
On the other hand, again by the assumption
$P_\infty^\wir(\beta_o) = 0$, we have the second branch
(5.21) of Proposition 5.4, namely:
$$
\xi_\wir^\fin(\beta_o)=\xi_\free(\beta_o)\,.
\tag 5.27
$$
{}From (5.26) and (5.27), we conclude that either
$\xi_\free(\beta_o) = 0$  or $\xi_\free(\beta_o) = \infty$.
The first case is easily ruled out by
considering e\.g\. $\tau_{\free}(\hat e_1)$.
That the other correlation lengths also diverge
is an immediate consequence of remark following
Theorem 4.3.
\qed
\enddemo
\smallskip

\bigskip
\head{Appendix: Reflection Positivity and the Transfer Matrix}
\endhead

The concept of reflection positivity and its consequences
are well-known tools in the context of field theory. For the
convenience of the reader we give a brief review in this
appendix.

We consider a (finite or infinite) lattice
$\Lambda\subset \Bbb Z^d$ which is invariant under
reflections at a plane $\Sigma$. Here $\Sigma$
is either a lattice plane or a plane which lies
halfway between two lattice planes.
Denoting the reflection at $\Sigma$ by $r$, we then decompose
$\Lambda$ as $\Lambda=\Lambda_+\cup\Lambda_-$,
where $\Lambda_+$ are the points on one side
of $\Sigma$, $\Lambda_-=r(\Lambda_+)$
are the points on the other
side of $\Sigma$, and
 $\Lambda_-\cap\Lambda_+=\Sigma\cap\Lambda$ (which is of
course empty if $\Sigma$ lies between two lattice planes).

For a local observable $A$ with support
$\supp A\subset\Lambda_+$, one introduces the
reflected observable $r(A)$ as
$$
(r(A))(\sigma)=A(r(\sigma))
\,
\tag A.1
$$
where
$r(\sigma)_x=\sigma_{r(x)}$.
Reflection positivity of the Potts model is
the statement that
$$
{\langle}\overline{r(A)}\,A{\rangle}_{b,\Lambda}\geq 0
\,.
\tag A.2
$$
The proof of (A.2) is standard; for the strategy,
see e.g. \cite{FILS}, \cite{Se}.

The inequality (A.2) has several important consequences.
Here we are mainly interested in the representation of
truncated expectation values as matrix elements of a
suitably defined transfer matrix $T$. In order to define
the transfer matrix $T$ in the setting considered here,
we need, in addition to the reflection invariance of
$\Lambda$, that $\Lambda$ is invariant under translations
perpendicular to $\Sigma$. We therefore assume that
$\Lambda$ is of the form
$$
\Lambda=\Bbb Z\times \Lambda_1
$$
where $\Lambda_1$ is a (finite or infinite) sublattice of
$\Bbb Z^{d-1}$. We then consider
the algebra $\Cal A_+$ of local observables with support
in $\Lambda_+$, where from now on
$$
\Lambda_+=\{(x, \vec x)\in \Bbb Z^d
|x\geq 0,\vec x\in \Lambda_1\}
\,.
$$
while
$
\Sigma=\{(x, \vec x)\in \Bbb Z^d
|x= 0,\vec x\in \Lambda_1\}
$.
Due to (A.2),
the equation
$$
{\langle}A\,,B{\rangle}
:={\langle}\overline{r(A)} B{\rangle}_{b,\Lambda}
\tag A.3
$$
defines a positive semi-definite scalar product
over $\Cal A_+$. Dividing out the corresponding
null space $\Cal N$ and completing the resulting
space in the usual way, this leads to the definition
of a Hilbert space $\Cal H=\overline{\Cal A_+/\Cal N}$.

Next, we introduce, for each local
observable $A\in \Cal A_+$, the observable $TA$
which is obtained from $A$ by translation by one
lattice unit in the positive direction perpendicular
to $\Sigma$. It is an easy consequence of the Cauchy-Schwarz
inequality for the scalar product (A.3), see \cite{Se}
for details, that
$T$ obeys the inequalities
$$
0\leq {\langle}A,TA{\rangle}\leq {\langle}A,A{\rangle}
\,.
\tag A.4
$$
The operator $T$ therefore defines a positive transfer
matrix, which obeys the inequalities
$$
0\leq T\leq 1
$$
as an operator on $\Cal H$. Observing that the
vector $\Omega$ corresponding to the constant
function $1\in\Cal A_+$ is an eigenvector of $T$ with
eigenvalue $1$, we note that the norm of $T$ is one.

We finally consider the interpretation of truncated
expectation values in the above Hilbert space representation.
Since
$$
{\langle}\overline{r(A)}{\rangle}_{b,\Lambda}=
{\langle}\overline{r(A)}\cdot 1{\rangle}_{b,\Lambda}=
{\langle}A,\Omega{\rangle}
\tag A.5a
$$
while
$$
{\langle}T^nA{\rangle}_{b,\Lambda}={\langle}\Omega,T^nA{\rangle}
={\langle}T^n\Omega,A{\rangle}
={\langle}\Omega,A{\rangle}
\,,
\tag A.5b
$$
one immediately obtains
$$
{\langle}\overline{r(A)};T^nA{\rangle}_{b,\Lambda}
={\langle}A,T^nA{\rangle}
-{\langle}A,\Omega{\rangle}{\langle}\Omega,A{\rangle}
\,.
\tag A.6
$$
Introducing the projection operator $P_\perp$ onto
the Hilbert space orthogonal to $\Omega$,
equation (A.6) becomes
$$
{\langle}\overline{r(A)};T^nA{\rangle}_{b,\Lambda}
={\langle}A_\perp,T^nA_\perp{\rangle}
\,,
\tag A.7
$$
where
$$
A_\perp=P_\perp A=A-{\langle}\Omega,A{\rangle}\Omega
\,.
\tag A.8
$$
If the support of $A$ is a subset the lattice plane
$\Sigma$, $r(A)=A$, and equation (A.7) reduces to
$$
{\langle}\overline{A};T^nA{\rangle}_{b,\Lambda}
={\langle}A_\perp,T^nA_\perp{\rangle}
\,.
\tag A.9
$$
Equation (A.9) is  an important technical tool
in the proof of the existence
of the correlation lenghts $\xi_{\wir}^{(1)}$ and
$\xi_{\wir}^{(2)}$.

\remark{Remark:}
In the context of Euclidean field theory,
the direction perpendicular to $\Sigma$ is
often interpretated as the Euclidean time.
The Hilbert space
$\Cal H=\overline{\Cal A_+/\Cal N}$
is then nothing but the quantum mechanical
Hilbert space of the considered model, and
$T$ is the generator of the Euclidean time
translations, i.e. $T=e^{-\epsilon H}$,
where $\epsilon$ is the lattice spacing and
$H$ is the Hamilton operator of the theory.

However, $\Cal H$ and $T$
have no such interpretation for the classical
Potts
model. This is due to the fact that here
$\Lambda$ is the lattice of a
{\it classical} system, and not a lattice approximation
to Euclidean space-time.

\endremark

\enddocument

\end
\bye